\documentclass[review]{elsarticle}  

\usepackage{graphicx}
\usepackage[caption=false,font=footnotesize]{subfig}
\usepackage{array}
\usepackage{booktabs}
\usepackage[hidelinks]{hyperref}
\usepackage{algorithm}
\usepackage[noend]{algorithmic}
\usepackage{wrapfig}

\usepackage{multirow}
\usepackage{threeparttable}

\usepackage{amssymb, amsmath}
\usepackage{mathtools}
\usepackage{bm}
\usepackage{siunitx}
\usepackage{tikz}
\usetikzlibrary{shapes,fit,positioning}

\DeclareMathOperator*{\argmin}{argmin}
\DeclareMathOperator*{\sign}{sign}

\begin{document}
	\title{Semi-Blind Source Separation with Learned Constraints}
	
	\author[1]{R\'emi~Carloni~Gertosio\corref{cor1}} \ead{remi.carlonigertosio@cea.fr}
	\author[1]{J\'er\^ome~Bobin} \ead{jerome.bobin@cea.fr}
	\author[1]{Fabio Acero} \ead{fabio.acero@cea.fr}
	\cortext[cor1]{Corresponding author}
	\affiliation[1]{organization={IRFU, CEA, Universite Paris-Saclay}, 
		postcode={F-91191}, city={Gif-sur-Yvette}, country={France}}
	
	\begin{abstract}
		Blind source separation (BSS) algorithms are unsupervised methods, which are the cornerstone of hyperspectral data analysis by allowing for physically meaningful data decompositions. BSS problems being ill-posed, the resolution requires efficient regularization schemes to better distinguish between the sources and yield interpretable solutions. For that purpose, we investigate a semi-supervised source separation approach in which we combine a projected alternating least-square algorithm with a learning-based regularization scheme. In this article, we focus on constraining the mixing matrix to belong to a learned manifold by making use of generative models. 
		Altogether, we show that this allows for an innovative BSS algorithm, with improved accuracy, which provides physically interpretable solutions. The proposed method, coined sGMCA, is tested on realistic hyperspectral astrophysical data in challenging scenarios involving strong noise, highly correlated spectra and unbalanced sources. The results highlight the significant benefit of the learned prior to reduce the leakages between the sources, which allows an overall better disentanglement.
	\end{abstract}
	

	\maketitle
	
	\paragraph*{Notations} 
	A vector is written in bold lowercase; $\mathbf{x}_n$ denotes the $n$th entry of vector $\mathbf{x}$. A matrix is written in bold uppercase; $\mathbf{X}_{n:}$ and $\mathbf{X}_{:m}$ are (column) vectors constituted respectively of the $n$th row and the $m$th column of matrix $\mathbf{X}$, while $\mathbf{X}_{nm}$ is the $(n,m)$th entry of $\mathbf{X}$. The $i$th element of a sequence of variables (\textit{e.g.}~of an iterative scheme, of a set) is noted with a superscript between parentheses, such as $\mathbf{X}^{(i)}$. An equality which stands for a definition is introduced with the symbol $\coloneqq$. $\left\lVert\cdot\right\rVert_p$ designates the $p$-norm for vectors and the entrywise $p$-norm for matrices. $\odot$ denotes the entrywise multiplication. $\mathbf{X}^\top$ and $\mathbf{X}^+$ designate respectively the transpose and the Moore–Penrose inverse of $\mathbf{X}$. $\iota_{\mathcal{X}}(\cdot)$ denotes the characteristic function of $\mathcal{X}$: $\iota_{\mathcal{X}}(x) = 0$ if $x\in\mathcal{X}$, $+\infty$ otherwise. Finally, $\Pi_{\mathcal{X}}(\cdot)$ designates the orthogonal projection on $\mathcal{X}$. 
	
	\section{Introduction}
	Hyperspectral data are generically constituted of several observations of a single region at different wavelengths across the electromagnetic spectrum. More precisely, let $\mathbf{X}_{j:} \in \mathbb{R}^P$ be a measurement made of $P$ samples or pixels at a channel $j \in [1\dots J]$, which corresponds to a given wavelength; $\mathbf{X}_{j:}$ may represent a 2D image, which is flattened into a vector. Under the linear mixture model, $\mathbf{X}_{j:}$ can be expressed as the weighted sum of $I$ elementary sources $\{\mathbf{S}_{i:} \in \mathbb{R}^P\}_{i \in [1\dots I]}$:
	\begin{equation}
		\mathbf{X}_{j:} = \sum_{i=1}^{I} \mathbf{A}_{ji} \mathbf{S}_{i:} + \mathbf{N}_{j:},
		\label{eq:model}
	\end{equation}
	where $\mathbf{A}_{ji}$ is the contribution of component $i$ at channel $j$ and $\mathbf{N}_{j:}$ is some additive noise. Equation \eqref{eq:model} can be rewritten with matrices, yielding:
	\begin{equation}
		\mathbf{X} =  \mathbf{A} \mathbf{S} + \mathbf{N},
		\label{eq:model2}
	\end{equation}
	where $\mathbf{X}\in\mathbb{R}^{J\times P}$, $\mathbf{A}\in\mathbb{R}^{J\times I}$ and $\mathbf{S}\in\mathbb{R}^{I\times P}$.
	With such notations, the columns $\left\{\mathbf{A}_{:i}\right\}_{i\in[1\dots I]}$ of the so-called mixing matrix $\mathbf{A}$ include the electromagnetic spectra of the elementary components. Note that the position of the source and mixing matrices is relative, in some fields the transposed model is considered instead. Hereafter, the overdetermined case is tackled, where there are more observations than sources ($J\geq I$). 
	
	Analyzing hyperspectral data can be tackled by factorizing the data $\mathbf{X}$ into a source matrix $\mathbf{S}$ and a mixing matrix $\mathbf{A}$. It is a very challenging ill-posed problem, which requires the use of prior information. So far, three kinds of approaches have been investigated in depth: \textit{(i)} the methods based on independent component analysis (ICA) \cite{HBSS}, which assume the statistical independence of the sources, \textit{(ii)}  the extensive class of nonnegative matrix factorization (NMF) algorithms \cite{Gillis20NMF}, which make the hypothesis that both the sources and the mixing matrix have nonnegative elements, and \textit{(iii)} the sparse blind source separation (BSS) methods, which exploit the sparsity of the sources in a given representation. The algorithm presented in this paper is part of the latter class. 
	
	When applied to real data, the first challenge of BSS is to unmix at best the sought-after elementary components by limiting "leakages", that is when the estimate of a component is contaminated by another component in the source ($\mathbf{S}$) and/or spectrum ($\mathbf{A}$) domains. In physical applications, this is fundamental, especially to disentangle complex mixtures such as those involving partially correlated sources \cite{Picquenot19}. To this end, it is essential to build efficient priors that better discriminate between the components. Secondly, the regularizations should provide physically relevant information to produce interpretable results. 
	
	The manifold hypothesis \cite{Fefferman13} states that natural signals, such as the sources and the spectra in the context of BSS, generally lie on low-dimensional manifolds embedded in larger spaces. 
	Physics sometimes provides analytical descriptions of such manifolds, with parametric models. For example, physical models of emission spectra have been integrated into a BSS scheme in astrophysics \cite{Irfan2019}. In practice, this solution nevertheless proves to be not very flexible and with a quality of separation that strongly depends on the precision of the analytical models employed. 
	Instead, machine learning techniques seem more adapted to model data manifolds. In order to tackle BSS problems, one way would amount to switch to end-to-end learning approaches, which directly provide an estimation of the sources and the mixing process \cite{brakel2017learning,Kameoka19,Jayaram20}.
	Nonetheless, such methods are not well adapted to account for the exact mixture model and the noise statistics, which is key in scientific applications. Consequently, hybrid techniques that combine standard variational approaches \cite{Scherzer09}, \textit{i.e.}~based on the minimization of a cost-function, and learned priors are better suited. In this framework, designing an effective BSS algorithm, with interpretable solutions, requires building regularizations for the mixing matrix and/or the sources that precisely account for the manifold properties.\\
	
	In this article, our main contributions are:
	\begin{itemize}
		\item We propose a source separation framework that associates a standard variational approach with a learned prior. To this effect, we choose to constrain the mixing matrix, that generally presents strong structures, to belong to the image space of generative models learned beforehand. The resulting regularization is expected to greatly constrain the components by efficiently rejecting the leakages, allowing for an improved separation.
		\item We introduce a new BSS algorithm that combines the algorithm GMCA (generalized morphological component analysis) with a learning-based modeling which is adapted to a low number of training samples, the interpolatory autoencoder (IAE). The resulting procedure is named {\it semi-blind} GMCA (sGMCA).
		\item We evaluate and compare the proposed method with several BSS algorithms on realistic hyperspectral astrophysical data in various challenging experimental settings. 
	\end{itemize}
	We characterize our method as \textit{semi-blind} because \textit{(i)} the regularization on the spectra is strong -- the IAE modeling implies a dimension reduction that notably reduces the solution space --  and \textit{(ii)} the components whose emission spectra are poorly or not known can still be identified blindly, which is essential in physical applications.

	\section{Towards a semi-blind source separation framework}
	
	\paragraph{\bf Associating a variational approach with a learning-based regularization}
	In the literature, combining variational approaches for inverse problems with regularization-learning procedures comes in different flavors. In \cite{Adler2017SolvingII,Adler2018LearnedPR,Gilton19,ADMMNet}, the proposed learning architecture builds upon a residual network to mimic standard proximal algorithms, in which regularization learning is one element of the learning scheme. So far, it is unclear whether this approach is well-suited for multiconvex problems (\textit{i.e.}~convex according to each variable when the other variables are fixed) such as BSS. A different way of solving inverse problems with learned regularizations consists in first inverting the observation operator (\textit{e.g.}~the mixing matrix, when the sources alone are estimated) with a fast and simple procedure (\textit{e.g.}~least-square solution) and then clean inversion artifacts thanks to a learned denoiser. This technique has been investigated both with standard convolutional networks denoisers \cite{doi:10.1002/mrm.26977,Jin17,Romano17,Sureau20} or with generative models with adversarial training \cite{NEURIPS2018_d903e960}. The underlying denoiser depends on the linear operator to be inverted, and in the case of BSS, this means that it should depend on the sources and the mixing matrix that are precisely the variables to be estimated.
	
	\paragraph{\bf Prior work in BSS} Incorporating learning-based priors in classical BSS methods is not novel. Research was mostly conducted in the specific framework of audio source separation. A neural-network-based denoiser has been used in \cite{Nugraha16} to refine the source estimate within a standard source separation scheme. Generative models built upon variational autoencoders have also been exploited in the context of single channel source separation \cite{Narayanaswamy20}. However, these methods are not easily transposable for the multichannel BSS problem that we propose to tackle. Moreover, we identify two major limitations: \textit{(i)} since the sought-after sources have a great variability, they would require a significant number of training samples, which might make it not well adapted to most physical applications, \textit{(ii)} the source denoising might not be constraining enough to efficiently discriminate between the sources and limit leakages of unknown morphologies or statistical distributions, especially in the presence of partial correlations. 
	
	\paragraph{\bf Constraining the spectra} Rather, we propose to make use of learned regularizations on the mixing matrix $\mathbf{A}$ (in the case of hyperspectral data, it generally corresponds to emission or reflectance spectra). Indeed, contrary to sources, the mixing matrix in hyperspectral images generally present a lower variability and exhibit stronger structures, which make them especially suitable for their modeling with machine learning. More particularly, we choose to constrain the sought-after spectra to belong to learned, low-dimensional manifolds. The resulting regularization is expected to greatly reject leakages, allowing for an improved disentanglement between the components. Also, enforcing the spectra to belong to learned and physically interpretable manifolds is of prime interest in the view of tackling physical applications. \\
	In order to learn spectrum models, we propose to use generative models. A generative model, whether based on standard autoencoders (AEs), variational autoencoders (VAEs), generative adversarial networks (GANs), \textit{etc.}, generically learns a mapping from a low-dimensional latent space to a data space, and, doing so, can parameterize a non-linear manifold in the data space. Thereafter, let us denote $g:\mathbb{R}^{N}  \rightarrow\mathbb{R}^{J}$ a generative model for the spectra; $g$ is a non-linear operator parameterized with a neural network. The image space of $g$ is supposed to approximate an underlying spectrum manifold $\mathcal{M}$:
	\begin{equation} 
		\label{eq:manifold}
		\mathcal{M} \approx \big\{\mathbf{a}\in\mathbb{R}^J, \exists \bm{\lambda} \in \mathbb{R}^{N}, \mathbf{a} = g\left(\bm{\lambda}\right) \big\}.
	\end{equation}
	
	\paragraph{\bf The interpolatory autoencoder as a generative model for the mixing matrix} In the present article, we use a particular AE to learn the generative models for spectra, the interpolatory autoencoder (IAE) \cite{BobinIAE2021}. The primary motivation is that, compared to classical AEs such as VAEs, the IAE requires much fewer training samples, which is convenient in a context where the available spectra are few. 
	Yet, we note that other AEs can be used in principle for the generative modeling; in this regard, the presentation of the method deliberately keeps a general formulation denoting $g$ some generative model.\\
	A description of the IAE can be found in \ref{app:iae}. Succinctly, the gist of the approeach is to generate samples on a manifold by linear interpolation between so-called "anchor points", which are known samples that belong to the manifold, in the latent space of an autoencoder.
	
	\section{The sGMCA algorithm}
	
	\subsection{Principle}
	
	The proposed source separation algorithm is based on a variational approach. In order to build the corresponding cost function, we first state the different assumptions on the sources, the mixing matrix and the noise.
	
	\subsubsection{Data-fidelity term} 
	In the following, the noise is assumed to be additive, Gaussian, independent and identically distributed\footnote{In low-count observations such as in X-ray imaging, the noise is strictly speaking Poisson. However, as the observation time increases, the noise can be approximated as additive, Gaussian, independent, but non-identically distributed -- the cost function can readily be adapted to account for this (weighted least squares).}. Thus, in accordance to the forward model in \eqref{eq:model2}, the data-fidelity term is chosen as being the Frobenius norm between the data $\mathbf{X}$ and the product $\mathbf{AS}$:
	\begin{equation}
		f\left(\mathbf{A}, \mathbf{S}\right) \coloneqq \frac{1}{2} \left\lVert\mathbf{X} - \mathbf{AS}\right\rVert^2_2.
	\end{equation}
	
	\subsubsection{Source regularization} The sources are assumed to be sparse in a given transformed domain, such as a wavelet representation. Let $\mathbf{W} \in \mathbb{R}^{P'\times P}$ be a dictionary of the considered sparsifying domain. 
	The sources are regularized by a $\ell_1$-penalization term in the transformed representation:
	\begin{equation}
		h_S\left(\mathbf{S}\right) \coloneqq \left\lVert\mathbf{\Lambda} \odot \left(\mathbf{S}\mathbf{W}^\top\right)\right\rVert_1.
	\end{equation}
	The hyperparameter $\mathbf{\Lambda}$ allows to tune the effect of the source regularization; being a matrix, it can depend on the source and the pixel.
	
	\subsubsection{Mixing matrix regularization} 
	The mixing matrix is regularized with the spectrum generative models. These are learned beforehand, depending on the expected components in the mixture.
	
	\paragraph{Application of the generative models}
	A first approach would amount to optimizing the spectra directly in the latent space of the generative models, leading to the following cost function:
	\begin{equation}
		\min\limits_{\mathbf{S}, \left\{\bm{\lambda}^{(i)}\right\}_{i\in[1\dots I]}} ~
		\frac{1}{2} \left\lVert\mathbf{X} - \mathbf{AS}\right\rVert^2_2 +
		\left\lVert\mathbf{\Lambda} \odot \left(\mathbf{S}\mathbf{W}^\top\right)\right\rVert_1
		\mathrm{~s.t.~} \forall i\in[1\dots I],~\mathbf{A}_{:i} = g^{(m_i)}\left(\bm{\lambda}^{(i)}\right),
	\end{equation}
	where $m_i$ is the index of the spectrum model which is associated to component $i$. 
	However, the optimization according to the $\{\bm{\lambda}^{(i)}\}_{i\in[1\dots I]}$ can be difficult to implement in practice. Above all, this formulation is problematic because the sources intervene in the application of the models; in particular, the impact of an estimation bias of $\mathbf{S}$ on the $\{\bm{\lambda}^{(i)}\}_{i\in[1\dots M]}$ is not necessarily well controlled.\\
	Rather, we take the spectrum models into consideration by constraining the spectra to belong to the image spaces of the generative models (which are supposed to approximate the underlying spectrum manifolds $\{\mathcal{M}^{(m_i)}\}_{i \in [1 \dots I]}$). To that end, the following constraint term on the mixing matrix can be considered: $\sum_{i \in[1\dots I]} \iota_{\mathcal{M}^{(m_i)}}\left(\mathbf{A}_{:i}\right)$, where $\iota_{\mathcal{X}}(\cdot)$ is recalled to be the characteristic function of $\mathcal{X}$: $\iota_{\mathcal{X}}(x) = 0$ if $x\in\mathcal{X}$, $+\infty$ otherwise. This makes it possible to decouple the application of the models from the sources, which facilitates the problem resolution.
	
	\paragraph{The semi-blind case} Until now, we supposed that all the spectra of the mixing matrix were known and modeled. However, in the more general {\it semi-blind} approach, we suppose that among the $I$ elementary components, $M$ are modeled and $I-M$ are fully unknown. Let $\mathbb{I} \subset [1\dots I]$ be the indices of the modeled components. \\
	If a spectrum is not modeled (\textit{i.e.}~$i\notin\mathbb{I}$), it is constrained to belong to the unit Euclidean ball $\mathcal{O} = \{\mathbf{a}\in\mathbb{R}^J, \left\lVert\mathbf{a}\right\rVert_2 \leq 1\}$; this standard constraint prevents scale degeneracies inherent to BSS such as $\|\mathbf{A}_{:i}\|_2 \rightarrow \infty$ and $\|\mathbf{S}_{i:}\|_2 \rightarrow 0$. \\
	The constraint term on the mixing matrix finally reads as:
	\begin{equation}
		h_A\left(\mathbf{A}\right) \coloneqq \sum_{i \in \mathbb{I}} \iota_{\mathcal{M}^{(m_i)}}\left(\mathbf{A}_{:i}\right) + \sum_{i \notin \mathbb{I}} \iota_{\mathcal{O}}\left(\mathbf{A}_{:i}\right).
	\end{equation}
	
	To summarize, sGMCA seeks the solution to the following problem:
	\begin{equation}
		\label{eq:obj}
		\min\limits_{\mathbf{A}, \mathbf{S}} ~
		\frac{1}{2} \left\lVert\mathbf{X} - \mathbf{AS}\right\rVert^2_2 +
		\left\lVert\mathbf{\Lambda} \odot \left(\mathbf{S}\mathbf{W}^\top\right)\right\rVert_1 
		+\sum_{i \in \mathbb{I}} \iota_{\mathcal{M}^{(m_i)}}\left(\mathbf{A}_{:i}\right) + \sum_{i \notin \mathbb{I}} \iota_{\mathcal{O}}\left(\mathbf{A}_{:i}\right).
	\end{equation}
	
	\subsection{Minimization scheme}
	
	\begin{algorithm}
		\caption{sGMCA}
		\label{alg:sgmca}
		\textbf{Inputs:} data $\mathbf{X}$, number of sources $I$, set of $M$ (already trained) generative models of spectra
		\begin{algorithmic}
			\STATE $\mathbf{A}, \mathbf{S} \leftarrow \mathrm{GMCA}(\mathbf{X}, I)$
			\WHILE{convergence not reached}
			\STATE \textit{Source update:}
			\STATE \textit{(1)} Least squares: $\mathbf{S} \leftarrow \mathbf{A}^+ \mathbf{X}$
			\STATE \textit{(2)} Determination of the thresholding parameters $\mathbf{\Lambda}$
			\STATE \textit{(3)} Thresholding: $\mathbf{S} \leftarrow \mathcal{T}_\mathbf{\Lambda}\left(\mathbf{S}\mathbf{W}^\top\right)\mathbf{W}$
			\STATE \textit{Mixing matrix update:}
			\STATE \textit{(1)} Least squares: $\mathbf{A} \leftarrow \mathbf{X} \mathbf{S}^+ $
			\STATE \textit{(2)} Spectrum identification \& model-to-spectrum mapping: determine $\mathbb{I}$ \& $\left\{m_i\right\}_{i\in\mathbb{I}}$ (see Alg.~\ref{alg:init})
			\STATE \textit{(3)} Constraint application:
			\FOR{$i\notin \mathbb{I}$}
			\STATE \textit{(3.a)} Projection on $\mathcal{O}$: $\mathbf{A}_{:i} \leftarrow \Pi_\mathcal{O}\left(\mathbf{A}_{:i}\right)$
			\ENDFOR
			\FOR {$i\in \mathbb{I}$}
			\STATE \textit{(3.b)} Manifold projection: $\mathbf{A}_{:i} \leftarrow \Pi_{\mathcal{M}^{(m_i)}}\left(\mathbf{A}_{:i}\right)$
			\ENDFOR
			\ENDWHILE
		\end{algorithmic}
		\textbf{Outputs:} mixing matrix $\mathbf{A}$, sources $\mathbf{S} $
	\end{algorithm}
	
	The sGMCA algorithm is based on GMCA \cite{bobin07a, KervazoPALM19}, which is a BSS algorithm built upon a projected alternating least squares (pALS) minimization procedure. Indeed, GMCA offers a flexible framework to tackle specific separation subproblems; it has incidentally be the subject of several extensions, \textit{e.g.}~DecGMCA to tackle joint deconvolution and separation when dealing with inhomogeneous observations \cite{Jiang2017}, including on the sphere \cite{Carloni2021}.\\
	The sGMCA method is described in Algorithm \ref{alg:sgmca}. 
	The sources and the mixing matrix are initialized with GMCA. At this point the solution is approximate; the mixing matrix and the sources are likely to be contaminated by residuals of other components (referred to as "leakages").
	$\mathbf{S}$ and $\mathbf{A}$ are updated alternatively and iteratively until convergence is reached. Each update comprises a least-square estimate, so as to minimize the data-fidelity term $f$, followed by the application of the proximal operator of the corresponding regularization term \cite{Boyd_Proximal14}.
	The procedure is stopped when either the estimated sources have stabilized ($\|\mathbf{S}^{(k)}-\mathbf{S}^{(k-1)}\|_2/\|\mathbf{S}^{(k)}\|_2 \leq \epsilon$ with $\epsilon = \num{1e-6}$ in practice) or a maximal number of iterations have been reached (50 in practice).
	
	Equation \eqref{eq:obj} is non-convex. It is nevertheless convex with respect to $\mathbf{S}$ when $\mathbf{A}$ is fixed. This is however not the case for $\mathbf{A}$ when $\mathbf{S}$ is fixed, since the manifolds $\{\mathcal{M}^{(m_i)}\}_{i\in\mathbb{I}}$ are likely not convex. Fortunately, if the first guess obtained by the least-square estimate is decent, and if the IAE generative models are accurate enough, the projections tend to be locally convex (see results in Section \ref{sec:model_learning}), which incidentally motivates the use of proximal tools. Being a non-convex problem, convergence to a critical point can at best be guaranteed. As far as we know, it is not proven that pALS converges, and in this respect, neither does sGMCA. However, in all tests performed, the algorithm always stabilized. We note that the stability is sometimes not strict, as minor cycles between the update of $\mathbf{S}$ and $\mathbf{A}$ can appear. This is most likely due to the convexity assumption of the spectrum update, which remains approximate. 
	
	\subsection{Source update}
	
	The proximal operator of the source constraint $h_S$ has an analytical form only if $\mathbf{W}$ is orthogonal, in which case it is a soft-thresholding in the transformed domain with thresholds $\mathbf{\Lambda}$, backprojected in the direct domain. Hereafter, we will keep this result as an approximation for non-orthogonal transformations, leading to the following source update:
		\begin{equation}
			\mathbf{S} \leftarrow \mathrm{prox}_{h_S}(\mathbf{A}^+\mathbf{X}) \approx  \mathcal{T}_\mathbf{\Lambda}\left(\left(\mathbf{A}^+\mathbf{X}\right)\mathbf{W}^\top\right)\mathbf{M}^\top,
		\end{equation}
		where $\mathbf{M}$ verifies $\mathbf{MW=\mathbf{I}}$ and $\mathcal{T}_\mathbf{\Lambda}(\cdot)=\max\left(|\cdot| - \mathbf{\Lambda}, \mathbf{0}\right) \odot \sign(\cdot)$, noting that the absolute, max and sign operators are element-wise and $\mathbf{0}$ is a matrix of zeros.
	
	The thresholding allows to refine the least-square estimate of the sources by removing the contaminating noise as well as the potential residuals of other sources, which could come from a misestimation of $\mathbf{A}$. 
	One of the major advantages of GMCA, which is adopted in the same way in sGMCA, is to propose an automatic selection strategy of the hyperparameters $\mathbf{\Lambda}$. In short, at each iteration \textit{(i)} the thresholds are set so to be three times greater than the noise level estimated in the current sources, and \textit{(ii)} they are further adapted to each sample by a $\ell_1$-reweighting scheme \cite{Candes_07_EnhancingSparsityby} in order to reduce the thresholding artifacts. More details and a thorough study on the impact of the choice of the hyperparameters can be found in \cite{KervazoPALM19}. 
	
	\subsection{Mixing matrix update}
	
	\subsubsection{Preliminary: projection on manifold}
	
	Let us firstly define the projection of a spectrum $\mathbf{a} \in \mathbb{R}^J$ on a manifold $\mathcal{M}$ associated to a generative model $g$:
	\begin{equation}
		\label{eq:proj}
		\Pi_\mathcal{M}\left(\mathbf{a}\right) \coloneqq g\left(\hat{\bm{\lambda}}\right),
	\end{equation}
	such that:
	\begin{equation}
		\label{eq:proj_lamda}
		\hat{\bm{\lambda}},~\hat{\rho} \coloneqq \argmin_{\bm{\lambda}\in\mathbb{R}^{N},\,\rho\in\mathbb{R^+}} \left\lVert\mathbf{a} - \rho\, g\left(\bm{\lambda}\right)\right\rVert^2_2.
	\end{equation}
	Equation \eqref{eq:proj_lamda} boils down to finding the latent parameters which minimize the Euclidean distance between the input spectrum $\mathbf{a}$ and the image space of the generative model. The coefficient $\rho$ allows for a rescaling and is necessary to take into account the scale indeterminacy on $\mathbf{a}$, which is inherent to BSS. Equation \eqref{eq:proj_lamda} is a constrained non-linear least-square problem, that does not admit a closed-form solution. It can nonetheless be estimated with a gradient-descent-based algorithm (\textit{e.g.} the Adam optimizer), since $g$ is differentiable as a neural-network based function. In this case, the latent parameters can be initialized using the encoders associated to the generative models.\\
	When the spectra are modeled by IAEs, we note that the latent parameters in Eq.~\eqref{eq:proj_lamda} are constrained to belong to \mbox{$\mathcal{T} \coloneqq \{\bm{\lambda}\in\mathbb{R}^{N}, \sum_{n=1}^N\bm{\lambda}_n=1\}$}, \textit{i.e.}~to sum to one, which is a particular requirement of the IAE. This requirement can be addressed by setting an element of $\bm{\lambda}$ as one minus the sum of the other elements in the gradient-descent-based minimization scheme.
	
	\subsubsection{Spectra identification and model-to-spectrum mapping}
	The least-square estimate, which constitutes the first step of the mixing matrix update, is equal to $\bar{\mathbf{A}} \coloneqq\mathbf{XS}^+$.
	Before applying the mixing matrix constraint, the following questions arise: \textit{(i)} among the $I$ spectra in $\bar{\mathbf{A}}$, which can be modeled by the provided generative models, and \textit{(ii)} more specifically, which model to associate with which identified spectrum? In other words, it raises the question of determining $\mathbb{I}$ and $\left\{m_i\right\}_{i\in\mathbb{I}}$ of Eq.~\eqref{eq:obj}.
	
	In physical applications where only one set of data is being worked on, identification can usually be made by hand (on the first iteration only, as the order does not change in the following iterations). Otherwise, we propose the spectrum identification procedure in Algorithm \ref{alg:init}. The gist is to associate iteratively the spectrum-model pair $(m,i)$ which minimizes the Euclidean distance between the spectra \textit{free of the interference} and their projections on manifolds, that is:
	\begin{equation}
		\epsilon^{(m,i)} \coloneqq \left\lVert\bar{\mathbf{A}}_{:i}-\mathbf{M}\bm{\mu}-{\rho}^{(m,i)}\,{\Pi_{\mathcal{M}^{(m)}}}\left(\bar{\mathbf{A}}_{:i}-\mathbf{M}\bm{\mu}\right)\right\rVert_2.
	\end{equation}
	Indeed, for robustness purposes, it is essential to account for the interferences that are likely to contaminate or even dominate the input spectra in $\bar{\mathbf{A}}$. This is done by subtracting the identified spectra up to the current iteration in $\mathbf{M}$, weighted by coefficients $\bm{\mu}$. These can be determined with a coarse grid-search, which is sufficient for the sake of identification\footnote{Assuming that the interfering spectra do not dominate the initial spectra, the search range of the coefficients can be limited between 0 and 1. In the performed tests, a grid step of 0.1 was enough to match the spectra with the correct models.}. Moreover, a simple forward pass in the encoder and decoder of the generative models can be employed as a fast spectrum projections $\{\Pi_{\mathcal{M}^{(m)}}\}_m$, without needing to resort to the more costly iterative scheme of Eq.~\eqref{eq:proj}. 
	\begin{algorithm}
		\caption{Spectrum identification and model-to-spectrum mapping}
		\label{alg:init}
		\begin{algorithmic}
			\STATE {\bfseries Input:} mixing matrix $\bar{\mathbf{A}}$, set of $M$ IAE generative models of spectra
			\STATE Initialize empty matrix $\mathbf{M}\in\mathbb{R}^{J\times 0}$
			\STATE $\mathbb{I} \leftarrow []$, $\mathbb{I}^C \leftarrow [1\dots I]$, $\mathbb{M} \leftarrow [1\dots M]$
			\FOR{ $k=1, \dots M$}
			\STATE $i, m, \bm{\mu} \leftarrow \argmin\limits_{i\in\mathbb{I}^C, m\in\mathbb{M}, \bm{\mu}\in [0,0.1, \dots 1]^{k-1}}\Big\lVert\bar{\mathbf{A}}_{:i}-\mathbf{M}\bm{\mu}$
			\STATE \hfill $-{\rho}^{(m,i)}~{\Pi_{\mathcal{M}^{(m)}}}\left(\bar{\mathbf{A}}_{:i}-\mathbf{M}\bm{\mu}\right)\Big\rVert^2$
			\STATE Append ${\rho}~{\Pi_{\mathcal{M}^{(m)}}}\left(\bar{\mathbf{A}}_{:i}-\mathbf{M}\bm{\mu}\right)$ to $\mathbf{M}$
			\STATE Append $i$ to $\mathbb{I}$, remove $i$ from $\mathbb{I}^C$, remove $m$ from $\mathbb{M}$
			\STATE $m_i \leftarrow m$
			\ENDFOR
			\STATE {\bfseries Output:} indices of the modeled spectra $\mathbb{I}$, model-to-spectrum map $\{m_i\}_{i\in\mathbb{I}}$
		\end{algorithmic}
	\end{algorithm}
	
	\subsubsection{Constraint application}
	It is recalled that the proximal operator of the characteristic function of a convex set is the orthogonal projection on the aforementioned set. Depending on whether a spectrum $\mathbf{A}_{:i}$ is modeled or not, the applied constraint differs:
	\begin{itemize}
		\item \textit{Unknown spectrum: projection on $\mathcal{O}$.} If a spectrum is not constrained (\textit{i.e.}~$i\notin\mathbb{I}$), the update reads as:
		\begin{equation}
			\mathbf{A}_{:i} \leftarrow \Pi_\mathcal{O}\left(\bar{\mathbf{A}}_{:i}\right) = \frac{\bar{\mathbf{A}}_{:i}}{\max\left(1, \left\lVert\bar{\mathbf{A}}_{:i}\right\rVert_2\right)}.
		\end{equation}
		\item \textit{Modeled spectrum:  projection on manifold.} If a spectrum is constrained (\textit{i.e.}~$i\in\mathbb{I}$), it is projected on its associated manifold:
		\begin{equation}
			\label{eq:projM}
			\mathbf{A}_{:i} \leftarrow \Pi_{\mathcal{M}^{(m_i)}}\left(\bar{\mathbf{A}}_{:i}\right).
		\end{equation}
	\end{itemize}
	
	\section{Numerical experiments}
	
\begin{figure}
		\begin{center}
			\subfloat[\label{fig:source_sync}Synchrotron]{\includegraphics[width=0.245\textwidth]{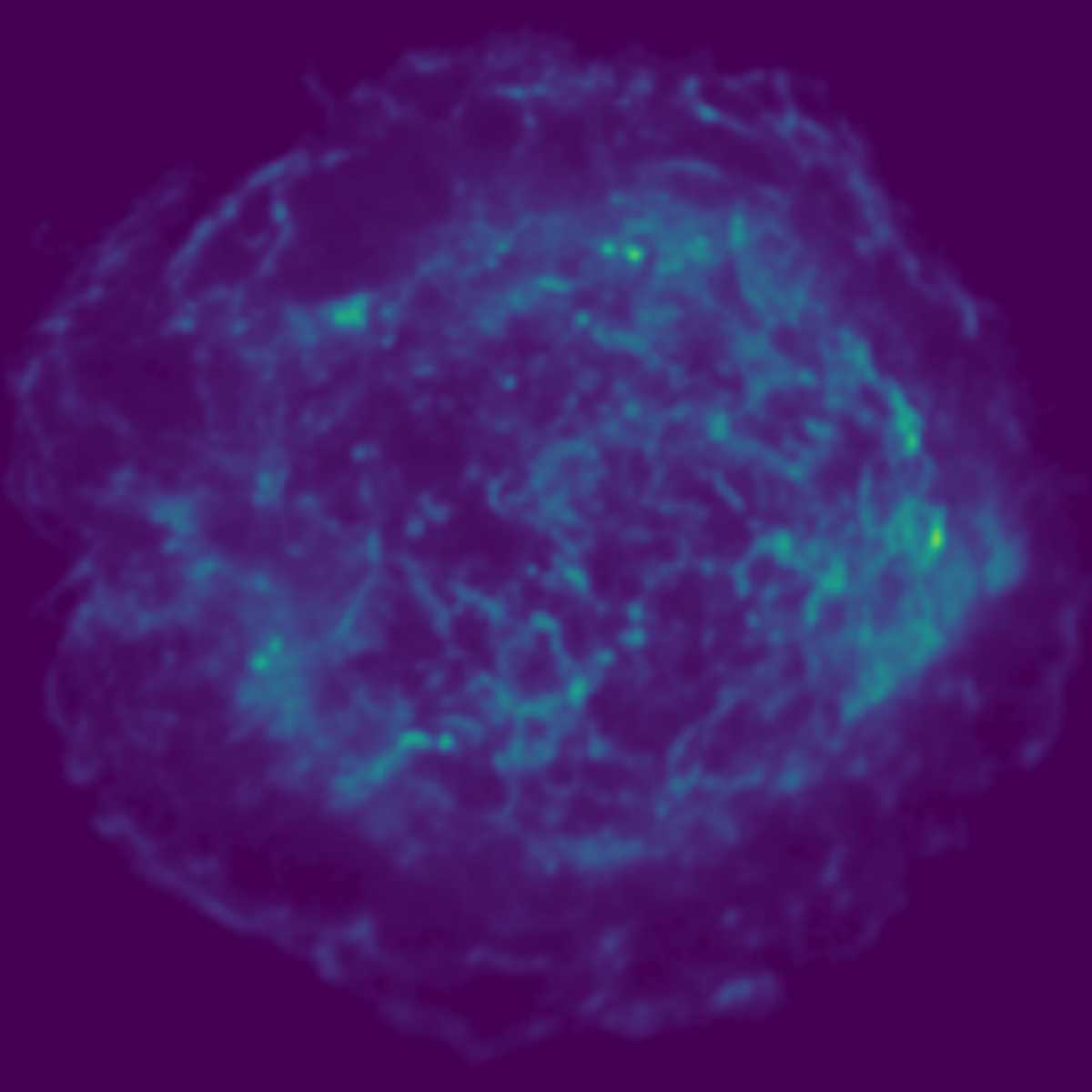}}
			\hfill
			\subfloat[\label{fig:source_therm}Thermal]{\includegraphics[width=0.245\textwidth]{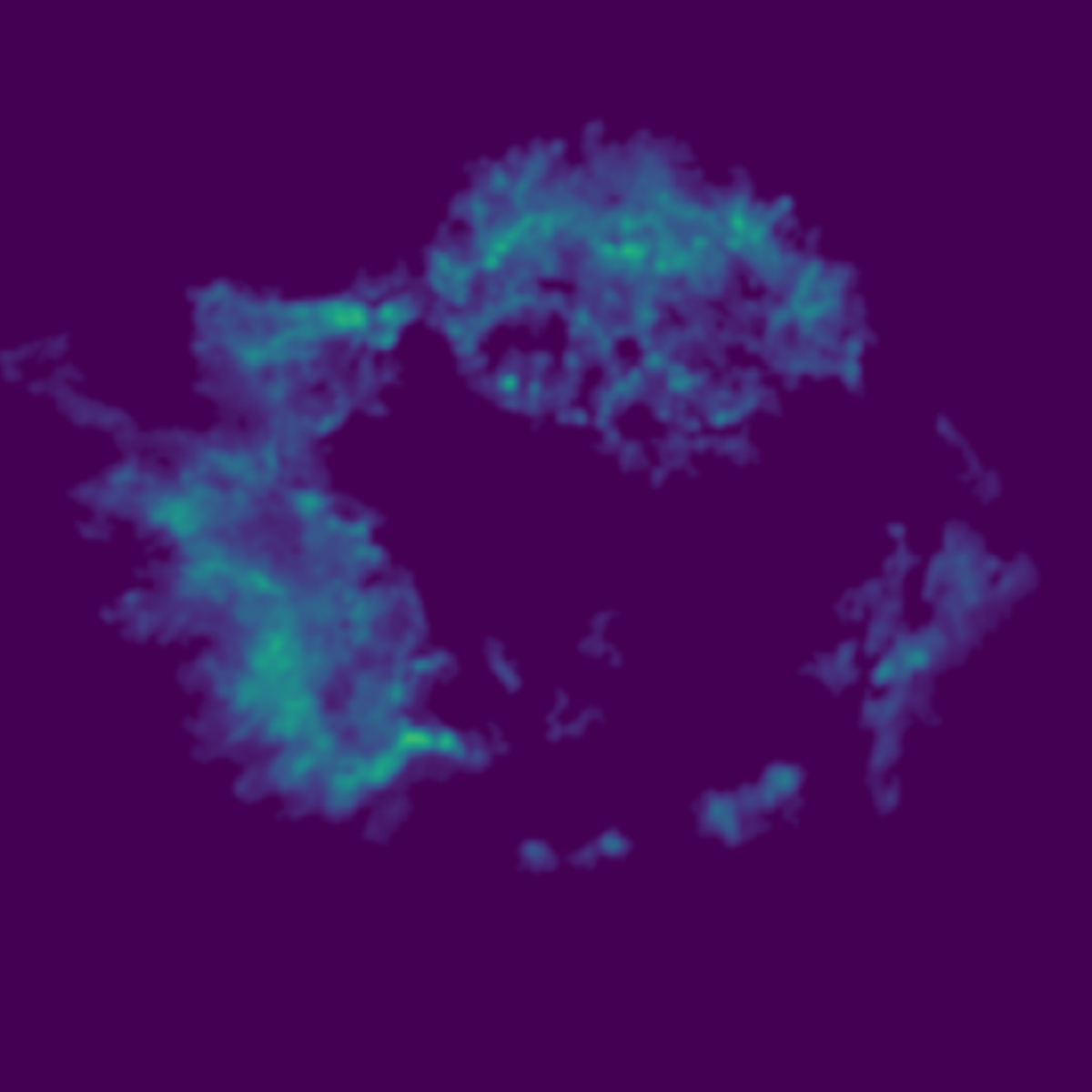}}
			\hfill
			\subfloat[\label{fig:source_iron1}Gaussian I]{\includegraphics[width=0.245\textwidth]{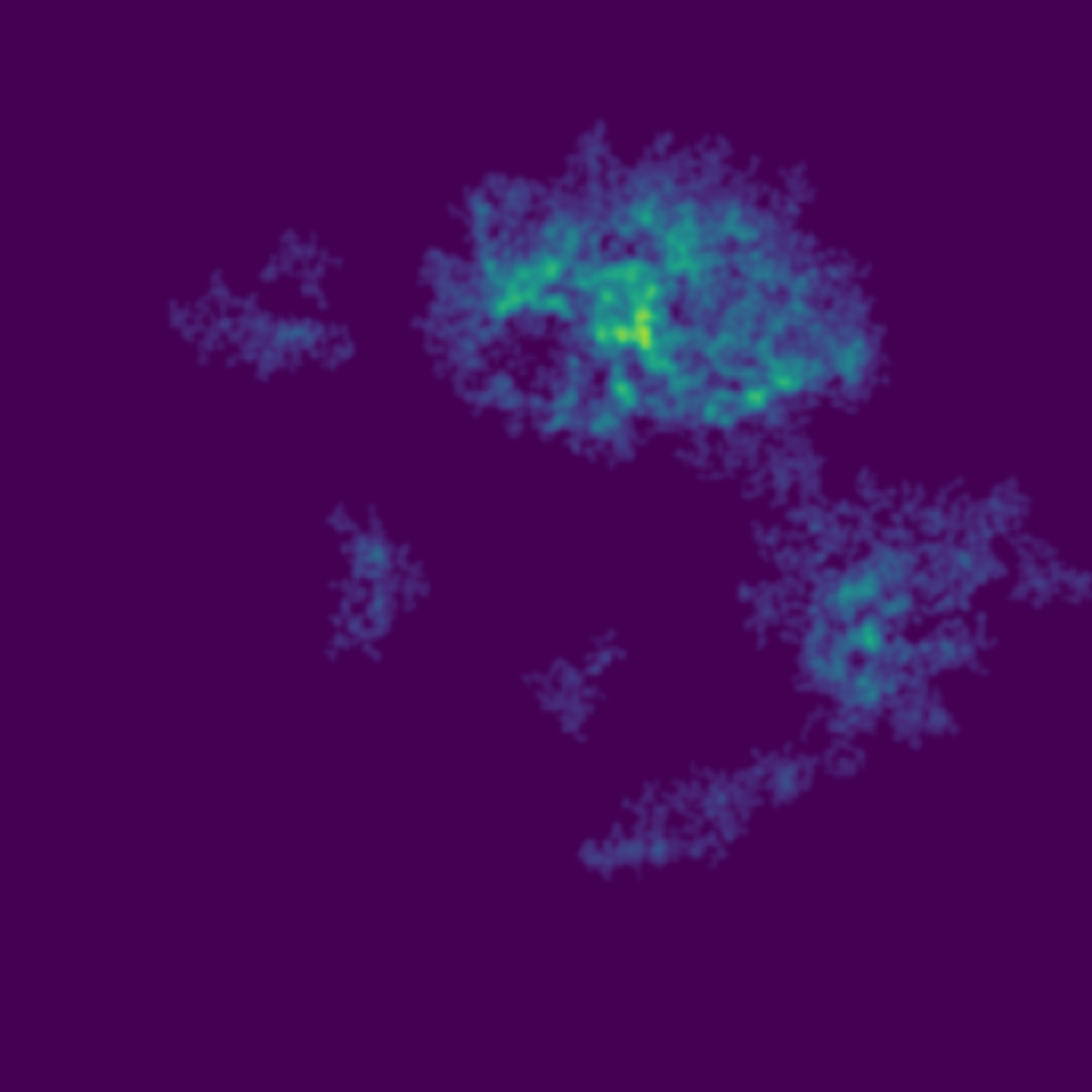}}
			\hfill
			\subfloat[\label{fig:source_iron2}Gaussian II]{\includegraphics[width=0.245\textwidth]{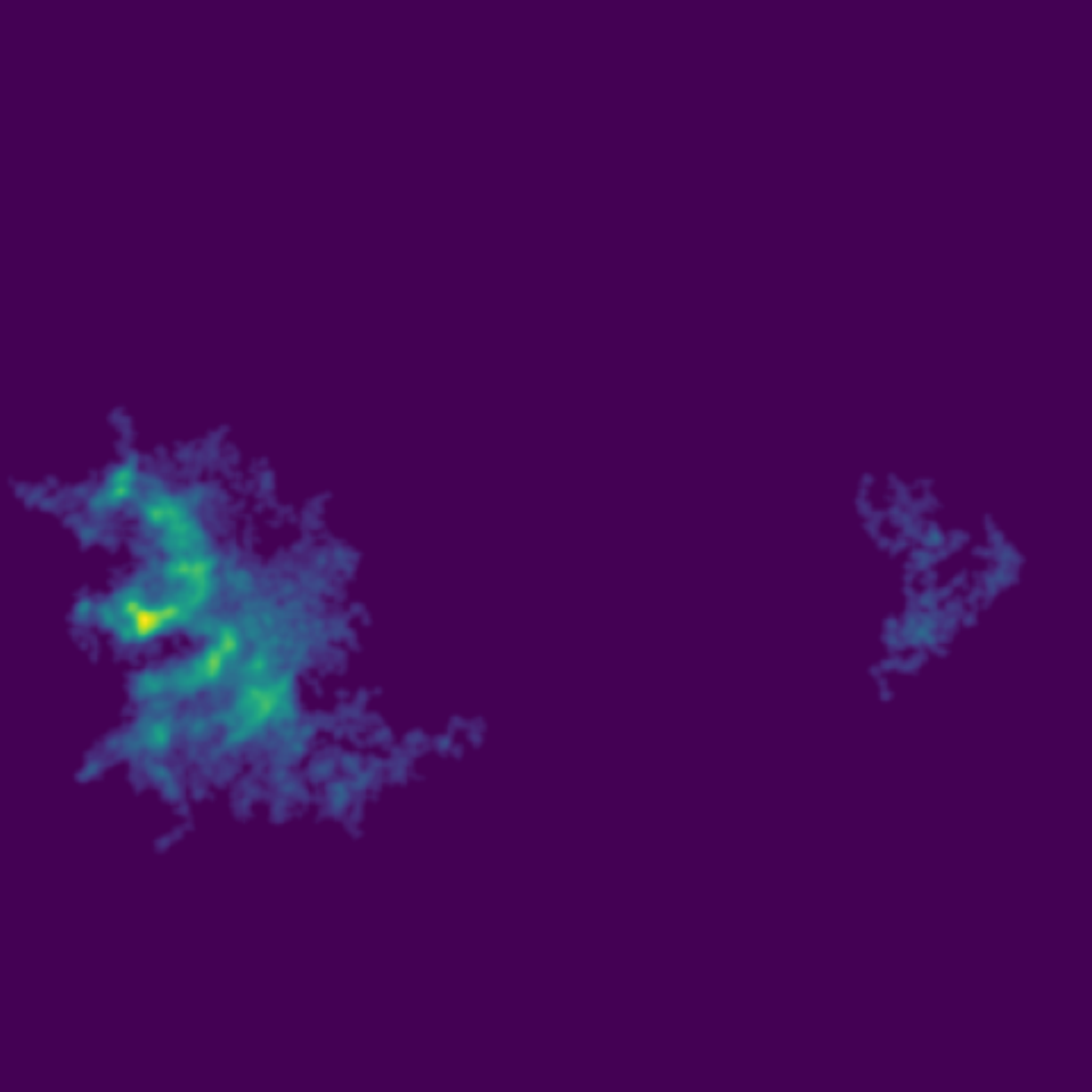}}
			\caption{Spatial templates obtained from Chandra X-ray observations of the Cassiopeia A supernova remnant (logarithmic scale).}
			\label{fig:sources}
		\end{center}
	\end{figure}

\begin{figure}
		\begin{center}
			\subfloat[\label{fig:spec_sync}Absorbed synchrotron power-law models]{\includegraphics[width=0.33\textwidth]{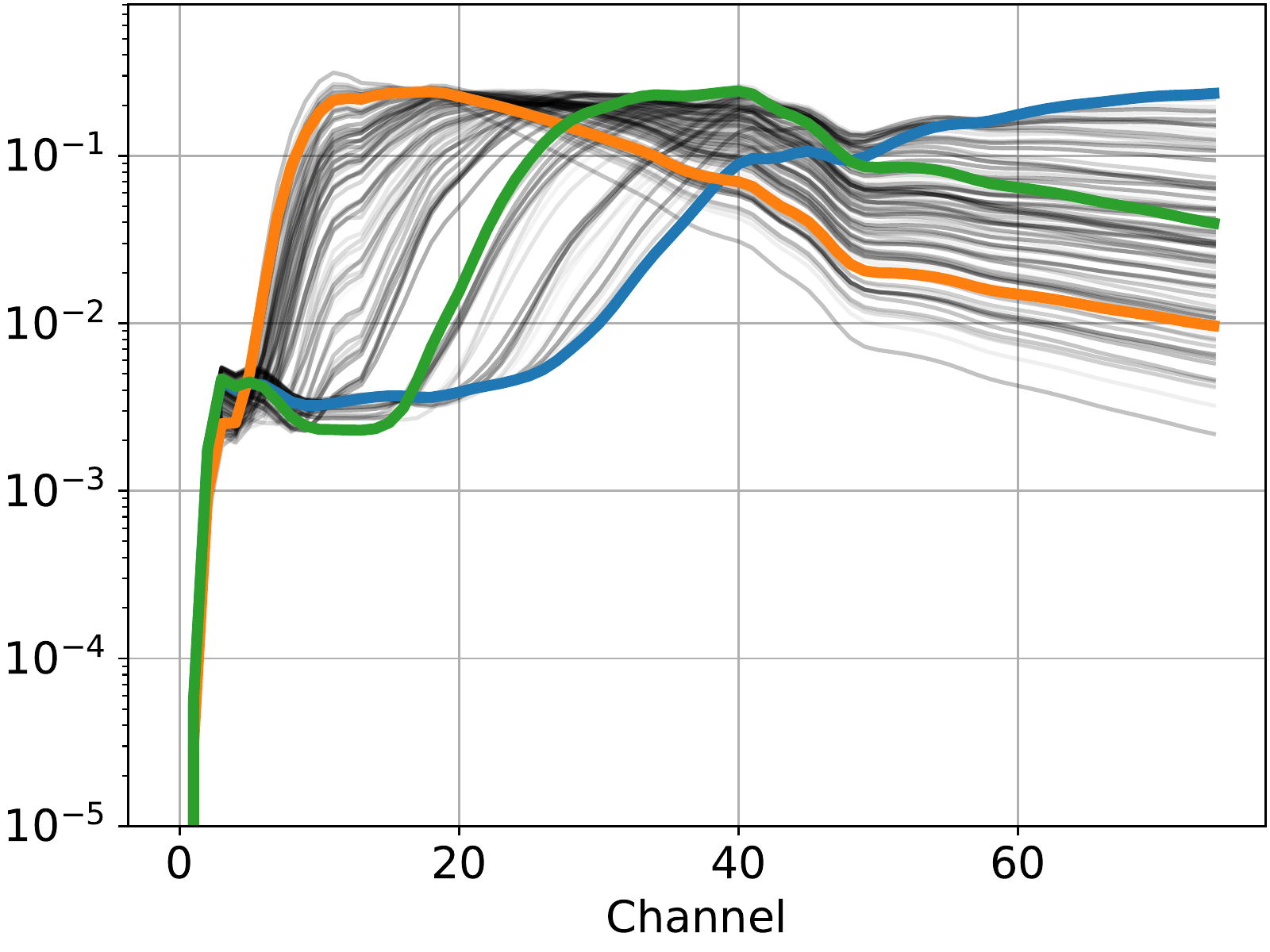}}
			\hfill
			\subfloat[\label{fig:spec_therm}Absorbed thermal models]{\includegraphics[width=0.33\textwidth]{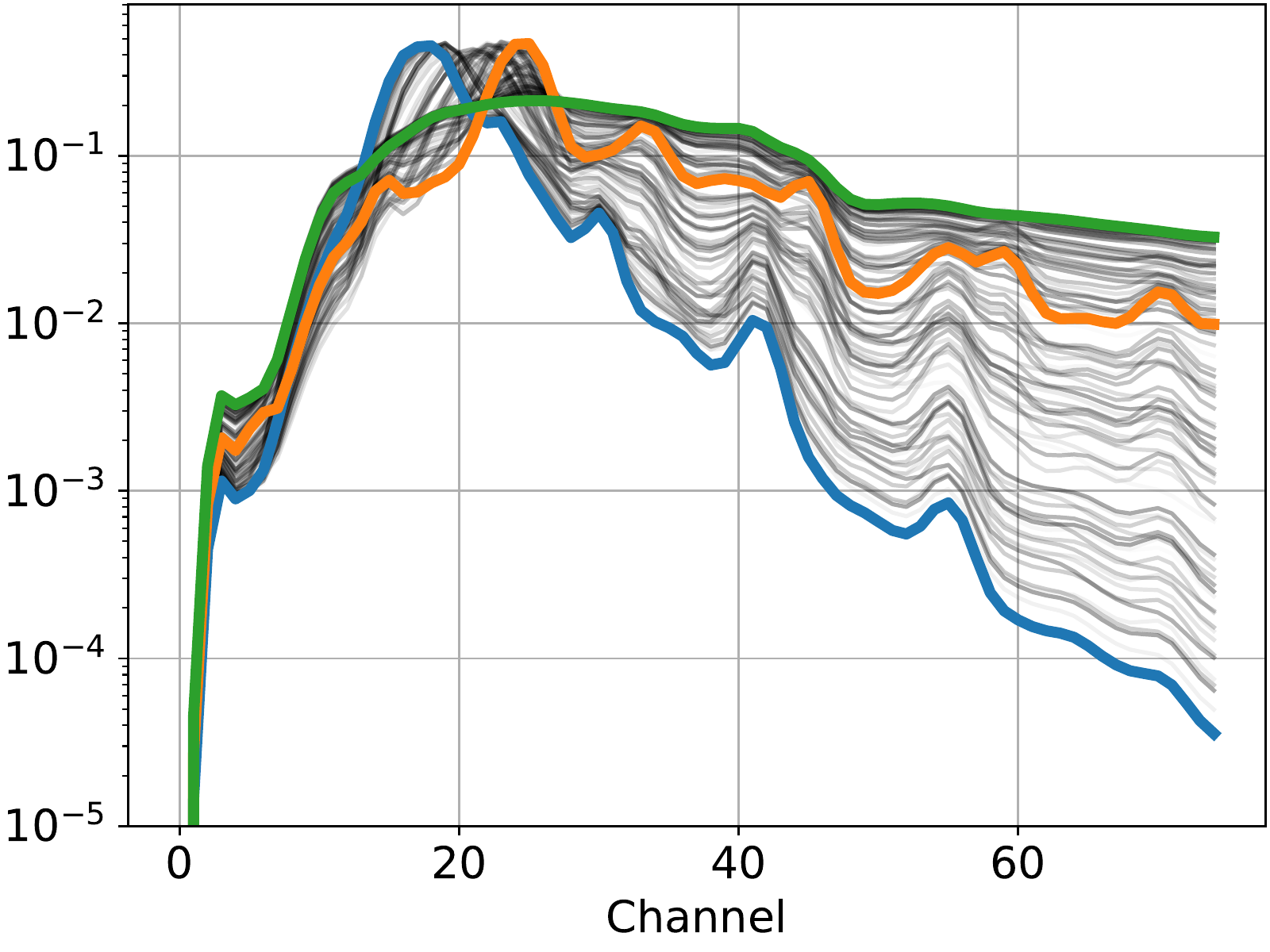}}
			\hfill
			\subfloat[\label{fig:spec_iron}Gaussian line models]{\includegraphics[width=0.33\textwidth]{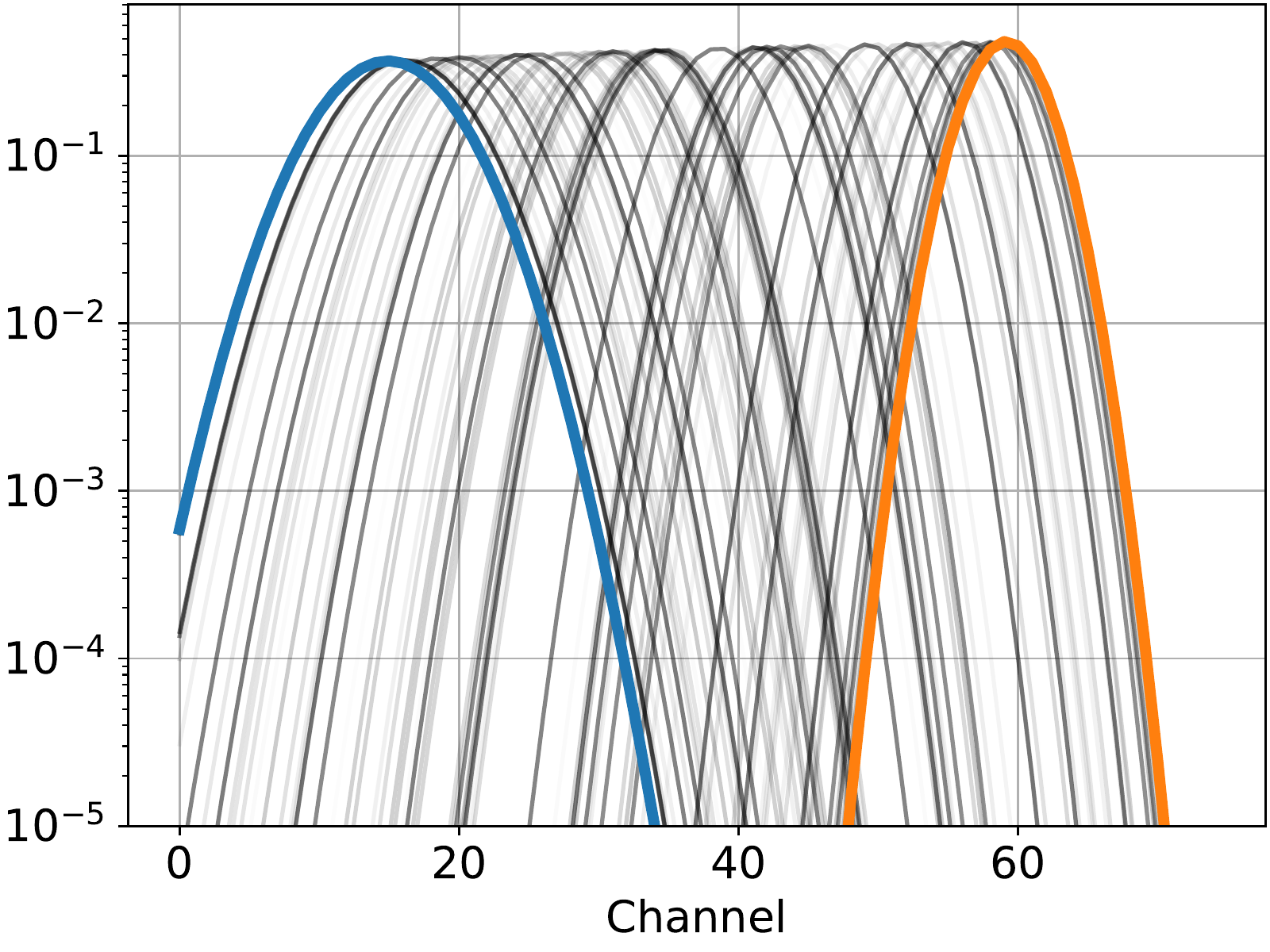}}
			\caption{Ensemble of three emission models. The colored thick lines are the chosen anchor points in the context of the IAE modeling.}
			\label{fig:spec}
		\end{center}
	\end{figure}

	In this section, the proposed method is evaluated on a realistic toy model of the Cassiopeia A supernova remnant observed by the X-ray space telescope Chandra\footnote{\href{https://chandra.harvard.edu}{\texttt{chandra.harvard.edu}}} (see \cite{Picquenot19} for more details about these data). The data are composed of $I=4$ sources of size $P=346\times346$ (see Fig.~\ref{fig:sources}), specifically one synchrotron source (radiation of the energetic charged particles in the supernova remnant by the synchrotron process), one thermal source (emission from a $10^7$~K plasma including continuum and lines emissions) and two Gaussian line emissions. The mixtures are observed over $J=75$ channels. An absorbed power-law model and a hot plasma emission model produced using the \textit{Astrophysical Plasma Emission Code} (APEC \cite{Forster20}) convolved with the spectral response of the Chandra telescope are used to generate the sets of synchrotron and thermal spectra (see Fig.~\ref{fig:spec_sync} and \ref{fig:spec_therm}). The emission line spectra are modeled as Gaussian kernels, with widths proportional to the center as shown in Fig.~\ref{fig:spec_iron}.
	
	We define three experimental parameters, namely:
	\begin{itemize}
		\item the signal-to-noise ratio \textit{SNR}, which is defined as the ratio of signal energy $\lVert\mathbf{AS}\rVert_2^2$ to the noise energy $\lVert\mathbf{N}\rVert_2^2$,
		\item the distance between the center of the two Gaussian line spectra $\delta$ (expressed in terms of spectral sample), to test how the proposed method separates two components whose spectra correlate -- this happens for example when the underlying physical processes are similar,
		\item the amplitude ratio $k$, which is a scalar by which the thermal and Gaussian sources are multiplied, so as to unbalance the sources and test how the algorithm recovers the least energetic components, which is a common situation in physical applications.
	\end{itemize}
	
	The sGMCA code that is used is open source (see \ref{app:code}). To the best of our knowledge, sGMCA is the first BSS method that makes use of a learned prior on the mixing matrix. We compare sGMCA with two benchmark experiments:
	\begin{itemize}
		\item an oracle version of GMCA, where $\mathbf{S}$ (respectively $\mathbf{A}$) is estimated with the ground-truth $\mathbf{A}$ (respectively $\mathbf{S}$); it provides an upper-bound for the reconstruction performances,
		\item an alternate version of sGMCA in which the spectra are regularized with a nearest-neighbor search among the spectra of the training sets; this experiment allows to better highlight the benefit of the IAE regularization.
	\end{itemize}
	The sGMCA algorithm is also compared to three standard BSS algorithms, which have no data-driven prior on the spectra, namely GMCA and:
	\begin{itemize}
		\item HALS \cite{Gillis_12_AcceleratedMultiplicativeUpdates}, which is a NMF algorithm designed to solve $\min_{\mathbf{A}\geq0, \mathbf{S}\geq0} \left\lVert\mathbf{X}-\mathbf{A}\mathbf{S}\right\rVert^2_2$ (the relationship $\geq$ is intended element-wise). The resolution is built upon a block-coordinate descent. The non-negativity of the solution is guaranteed thanks to multiplicative updates.
		\item SNMF \cite{LeRoux15}, which is a NMF algorithm enforcing the sparsity of the sources in the direct domain by solving $\min_{\mathbf{A}\geq0, \mathbf{S}\geq0} \left\lVert\mathbf{X}-\mathbf{A}\mathbf{S}\right\rVert^2_2 + \lambda \left\lVert\mathbf{S}\right\rVert_1$. Similarly to HALS, multiplicative updates ensure the non-negativity of the solution. To the best of our knowledge, there is no extension of SNMF in a transformed domain, which would better suit the data.
	\end{itemize}
	For the GMCA-based algorithms, the sparsity of the sources is enforced in the starlet (\textit{i.e.}~an isotropic undecimated wavelet) representation with two details scales \cite{Starck2015}.
	
	The estimated spectra are assessed with the spectral angular distance (SAD), recast in a logarithmic scale for the sake of precision:
	\begin{equation}
		\mathrm{SAD}^{(i)} = -10 \log_{10}\left( \arccos\left(\frac{{\mathbf{A}_{:i}}^\top{\mathbf{A}^*}_{:i}}{\left\lVert\mathbf{A}_{:i}\right\rVert_2\left\lVert{\mathbf{A}^*}_{:i}\right\rVert_2}\right)\right),
	\end{equation}
	with $\mathbf{A}^*$ the ground truth mixing matrix. Consequently, the greater the SAD, the more accurate the estimation. We will consider the overall SAD, that we define as the mean SAD over all spectra.\\
	The sources are evaluated with the signal-to-distortion, interference, noise and artifacts ratios (SDR, SIR, SNR\footnote{Note that \textit{SNR} in italics refers to the signal-to-noise ratio of the observed data, while SNR in roman is the criterion.}, SAR respectively), which are classical performance metrics in BSS \cite{Vincent06}. It is reminded that the SDR evaluates the overall reconstruction error, while the SIR, SNR and SAR assess more particularly the leakages from other sources, the noise contamination and other artifacts, respectively. Similarly to the SAD, we will use these metrics in the logarithmic scale.
	
	\subsection{Learning of the generative models} \label{sec:model_learning}
	
	Before tackling the source separation problem, the generative models of the three families of spectra need to be learned. 
	In the IAE framework \cite{BobinIAE2021}, the first step is to choose the anchor points and their numbers. The minimum number of anchor points required by the IAE is the dimension of the underlying manifold plus one (so to be able to perform interpolations). Three anchor points are selected for the thermal and synchrotron spectra as we generated them with two free parameters. The Gaussian spectra depend on one parameter only, thus two anchor points are needed. The choice of the anchor points is in practice not particularly critical; as long as they are not colinear, the reconstruction performances are satisfying. The anchor points are selected by hand, so that they tend to maximize their contrasts (see Fig.~\ref{fig:spec}).\\
	The three sets of spectra are decomposed into training, validation and test sets. The reconstruction performances are found to stabilize as of four layers, which is therefore the number of layers chosen for all three models. The values of the hyperparameters of the IAE models are summarized in Table~\ref{tab:param_learning}.
	\begin{table}
		\centering
		\begin{tabular}{lccc}
			\toprule
			Parameter & \multicolumn{3}{c}{Value} \\
			\midrule
			& Sync. & Therm. & Gauss. \\
			\cmidrule{2-4}
			Anchor points & 3 & 3 & 2 \\
			Training samples  & 541 & 601 & 600  \\
			Validation and test samples & 178 & 198 & 200 \\
			Number of epochs & 10\,000 & 10\,000 & 100\,000 \\
			\cmidrule{2-4}
			Number of layers & \multicolumn{3}{c}{4} \\
			Type of layers & \multicolumn{3}{c}{Fully connected} \\
			Activation functions &  \multicolumn{3}{c}{$\mathrm{tanh}$} \\
			Skip connection parameter & \multicolumn{3}{c}{0.1}  \\
			Learning rate & \multicolumn{3}{c}{\num{1e-4}} \\
			Solver & \multicolumn{3}{c}{Adam} \\
			Batch size & \multicolumn{3}{c}{25}\\
			\bottomrule
		\end{tabular}
		\caption{Parameters of the IAE networks and the learning stage.}
		\label{tab:param_learning}
	\end{table}
	
\begin{figure}
		\hfill%
		\subfloat[\label{fig:proj_spec_sync}Synchrotron]{\includegraphics[width=0.25\textwidth]{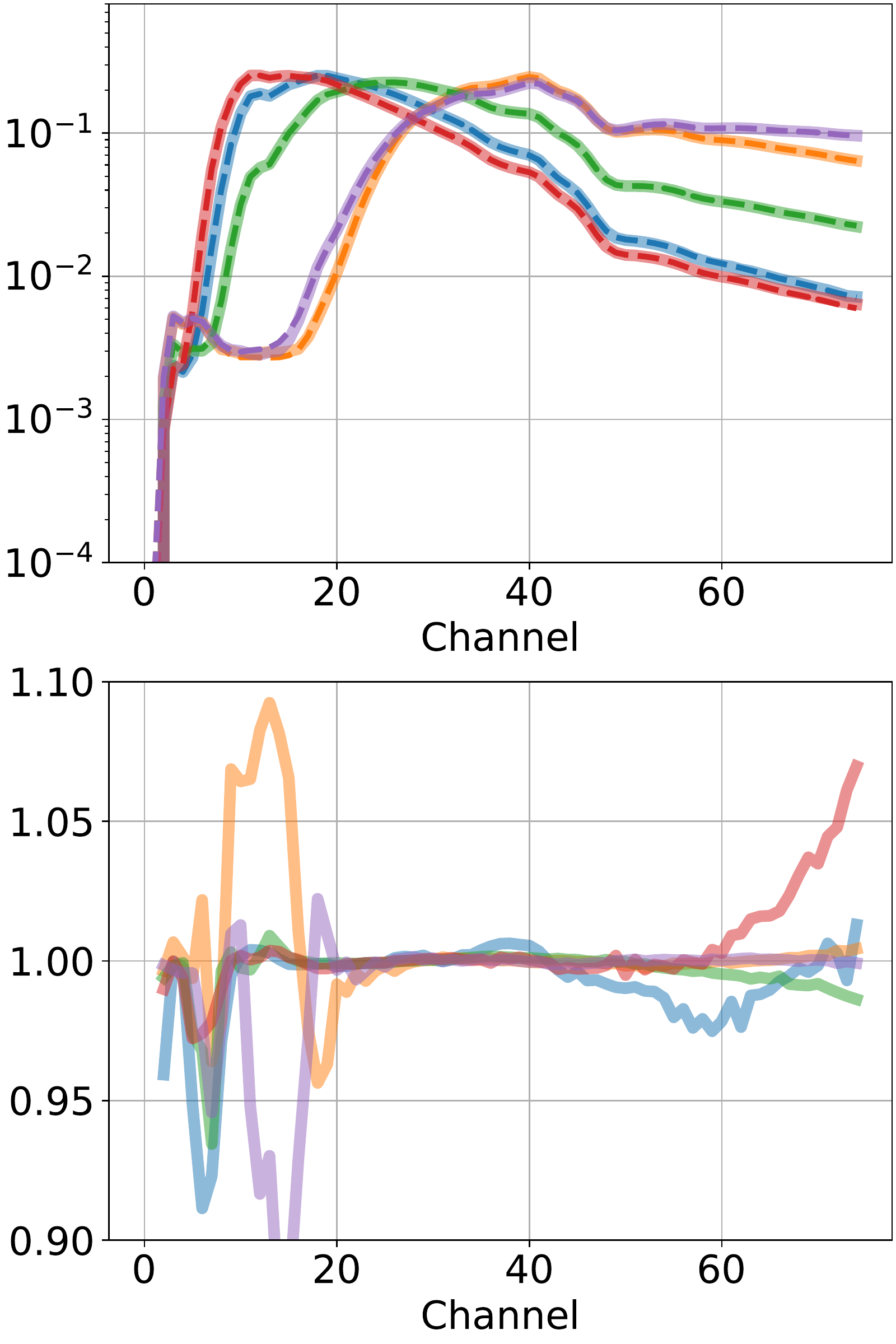}}
		\hfill%
		\subfloat[\label{fig:proj_spec_thermal}Thermal]{\includegraphics[width=0.25\textwidth]{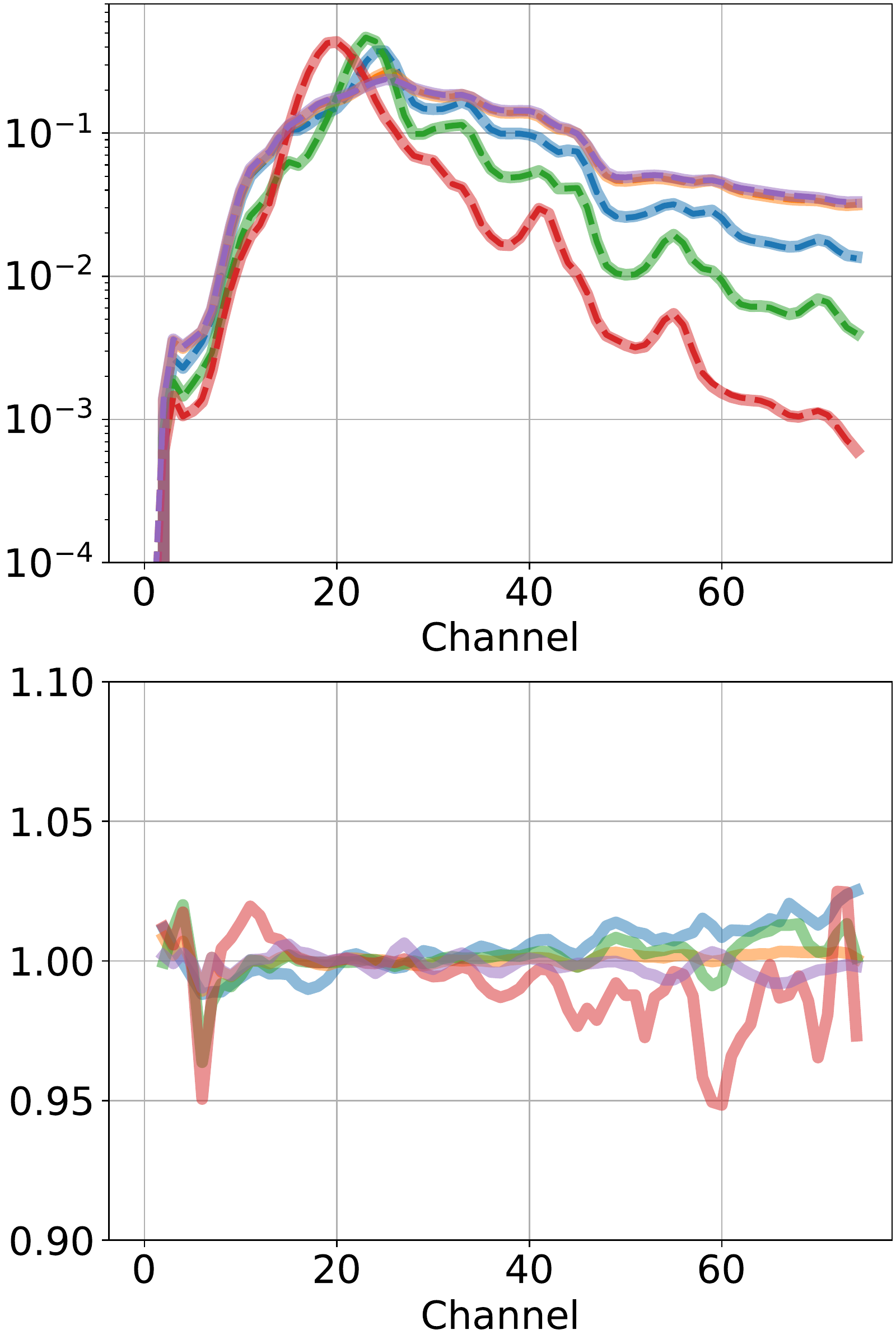}}
		\hfill%
		\subfloat[\label{fig:proj_spec_gauss}Gaussian line emission]{\includegraphics[width=0.25\textwidth]{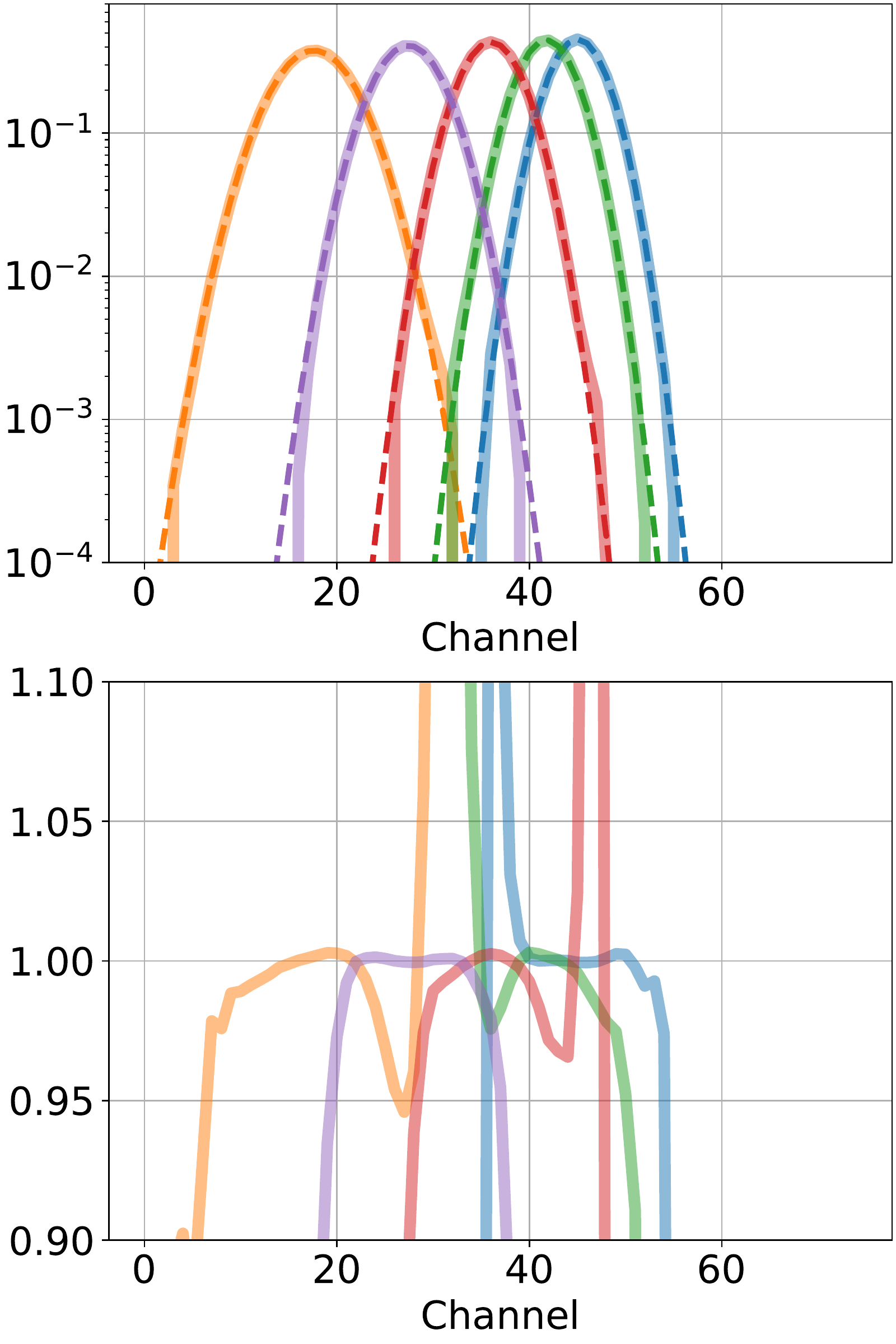}}
		\hfill%
		\hfill%
		\caption{Examples of projections of spectra from the test sets on the manifolds modeled by the IAE. Top row: spectra (solid lines: results of projection, dashed lines: ground truth). Bottom row: projected spectra over ground-truth spectra (plotted if ground truth greater than $\num{1e-4}$). The figures on a same row share the same ordinate range.}
		\label{fig:proj_spec}
	\end{figure}

	Figure \ref{fig:proj_spec} shows examples of projections of spectra from the test sets on the manifolds modeled by the IAE (using Eq.~\ref{eq:proj}), alongside with the projection errors. The spectra are well reproduced on a reasonably wide dynamic, except for the low-amplitude samples. To correct this, the spectra could be learned in a logarithmic scale, but it is out of the scope of the present work. 
	The median SAD of each type of spectra and in overall are reported in Table \ref{tab:proj_spec}. These results give the IAE modeling error bounds, that sGMCA cannot exceed.
	
	\begin{table}
		\centering
		\begin{tabular}{lcccc}
			\toprule
			Spectra & Sync. & Therm. & Gauss. & Overall \\
			\midrule
			SAD (dB) &25.26 & 25.92 & 24.81 & 25.18 \\
			\bottomrule
		\end{tabular}
		\caption{Median reconstruction SAD over the test sets by the IAE.}
		\label{tab:proj_spec}
	\end{table}
	
\begin{figure}
		\begin{center}
			\subfloat{\includegraphics[width=0.3\columnwidth]{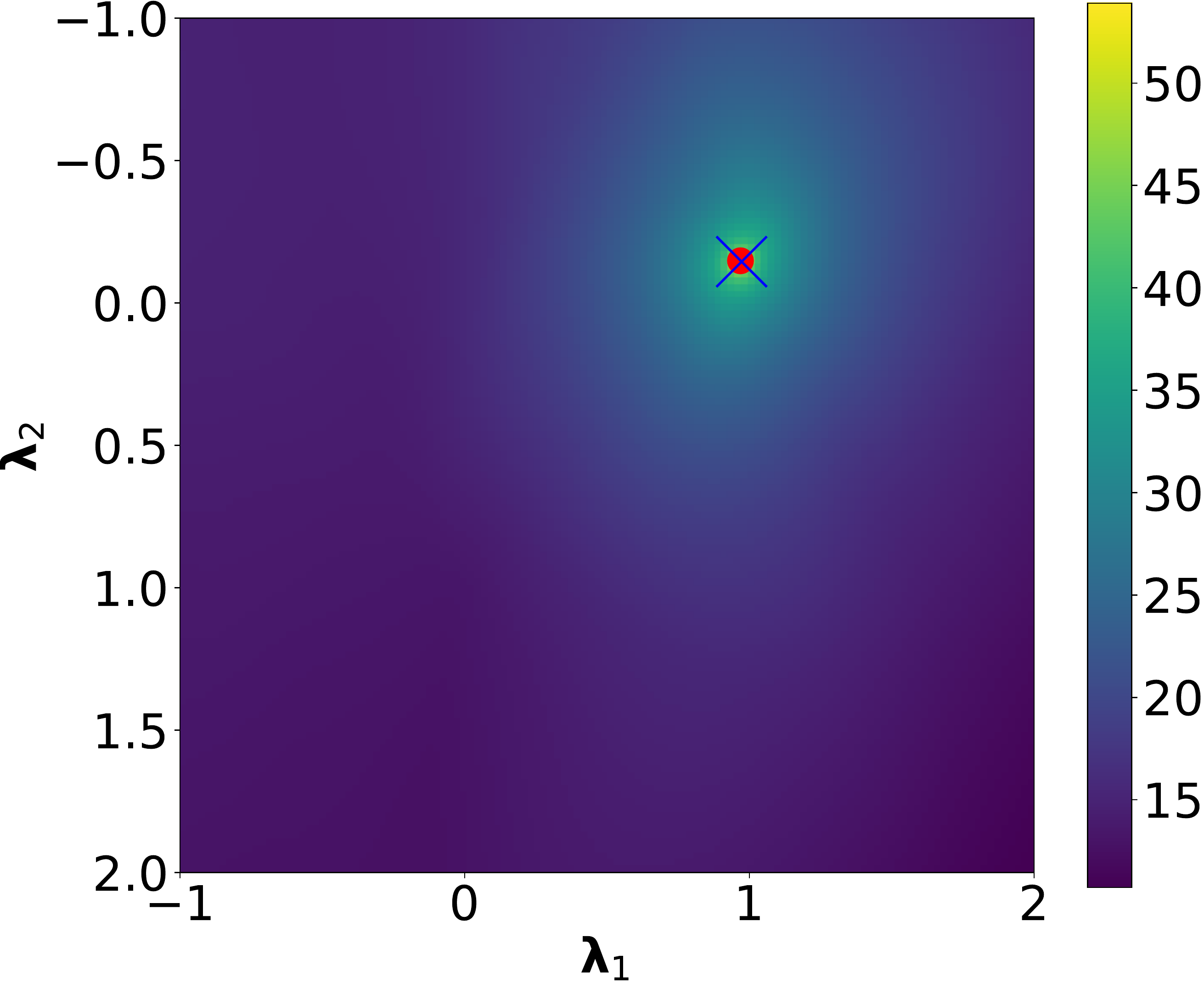}} 
			\subfloat{\includegraphics[width=0.3\columnwidth]{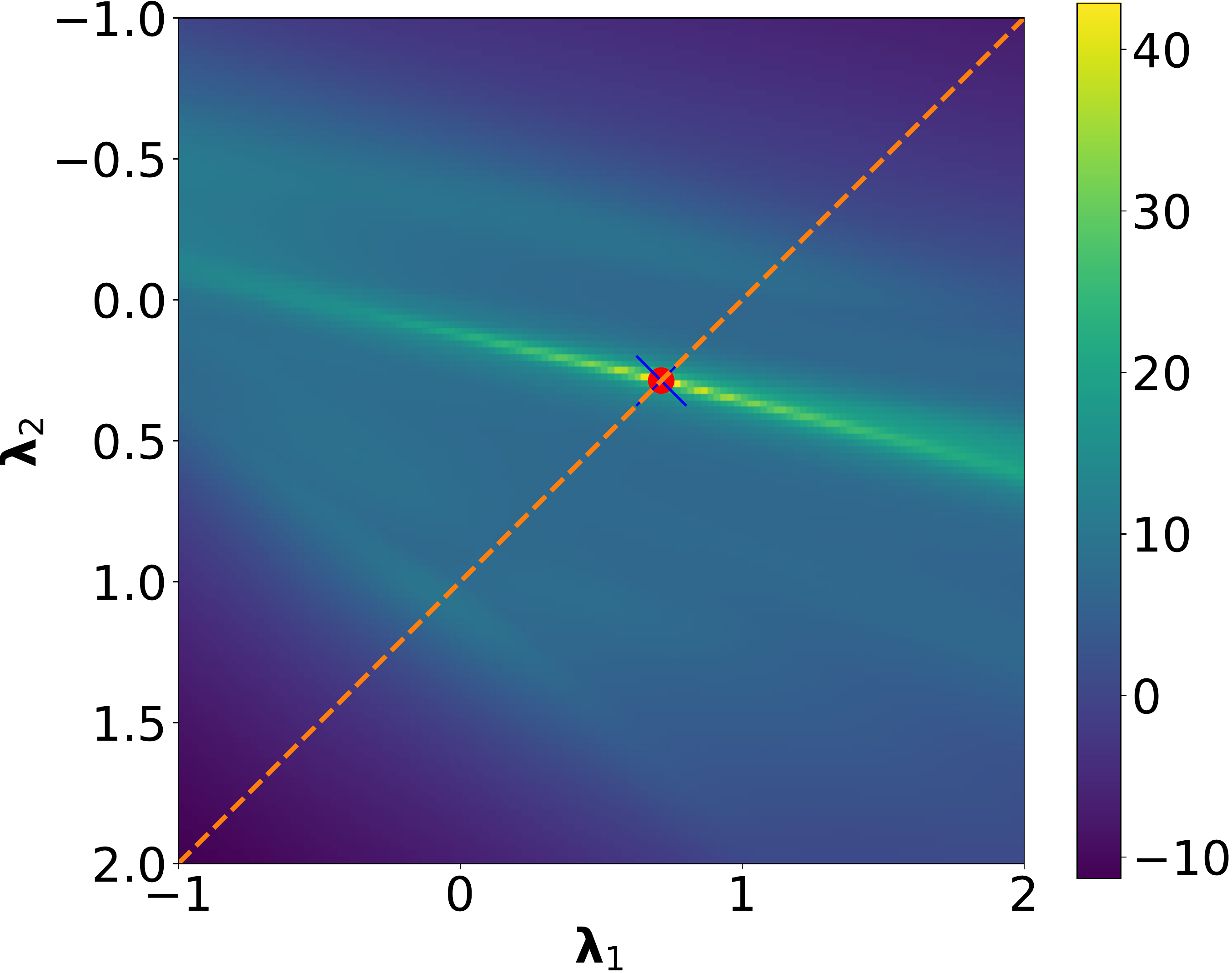}} \\[-0.3cm]
			\addtocounter{subfigure}{-2}
			\subfloat[\label{fig:proj_optim_landscape_sync}]{\includegraphics[width=0.3\columnwidth]{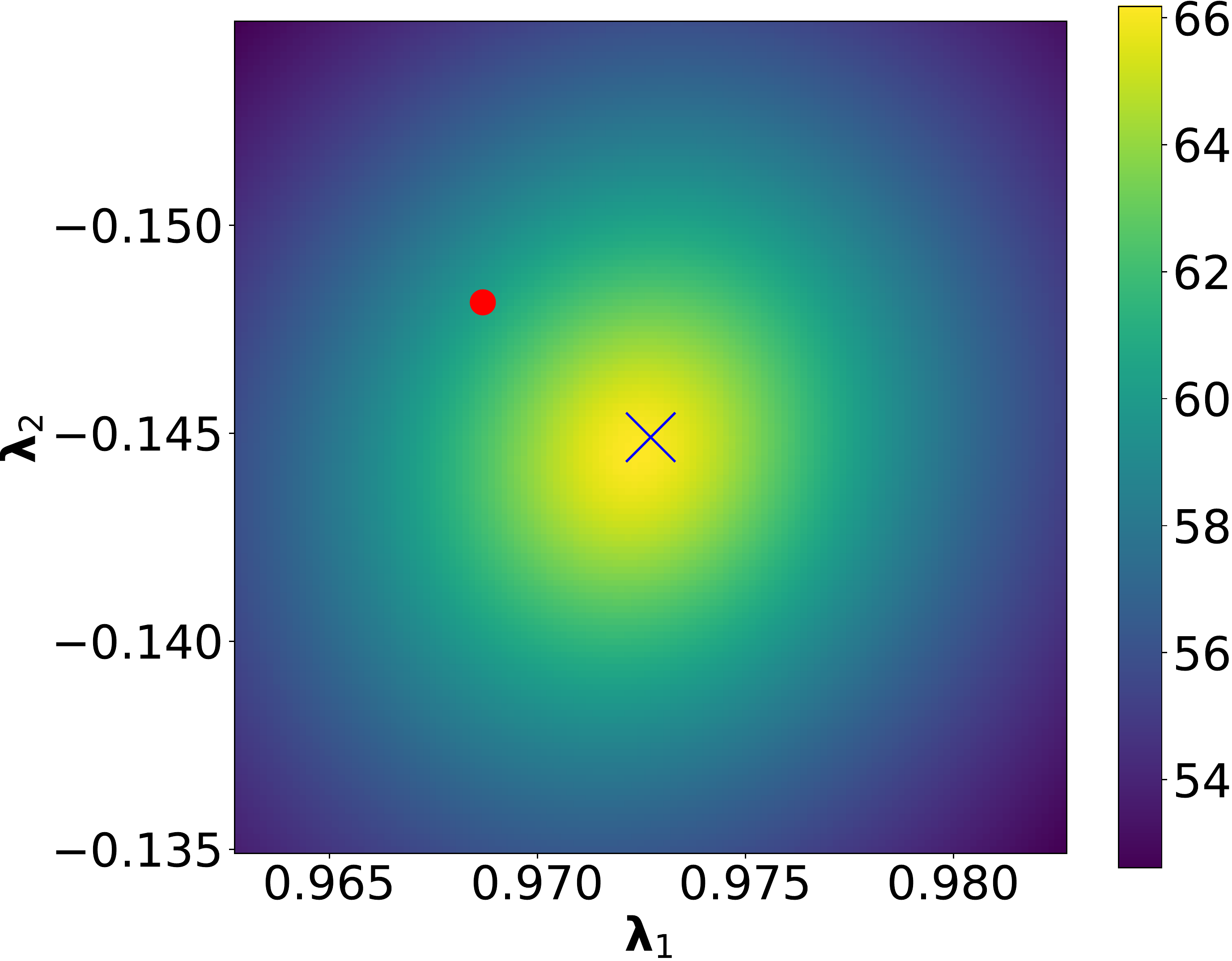}} 
			\subfloat[\label{fig:proj_optim_landscape_gauss}]{\includegraphics[width=0.3\columnwidth]{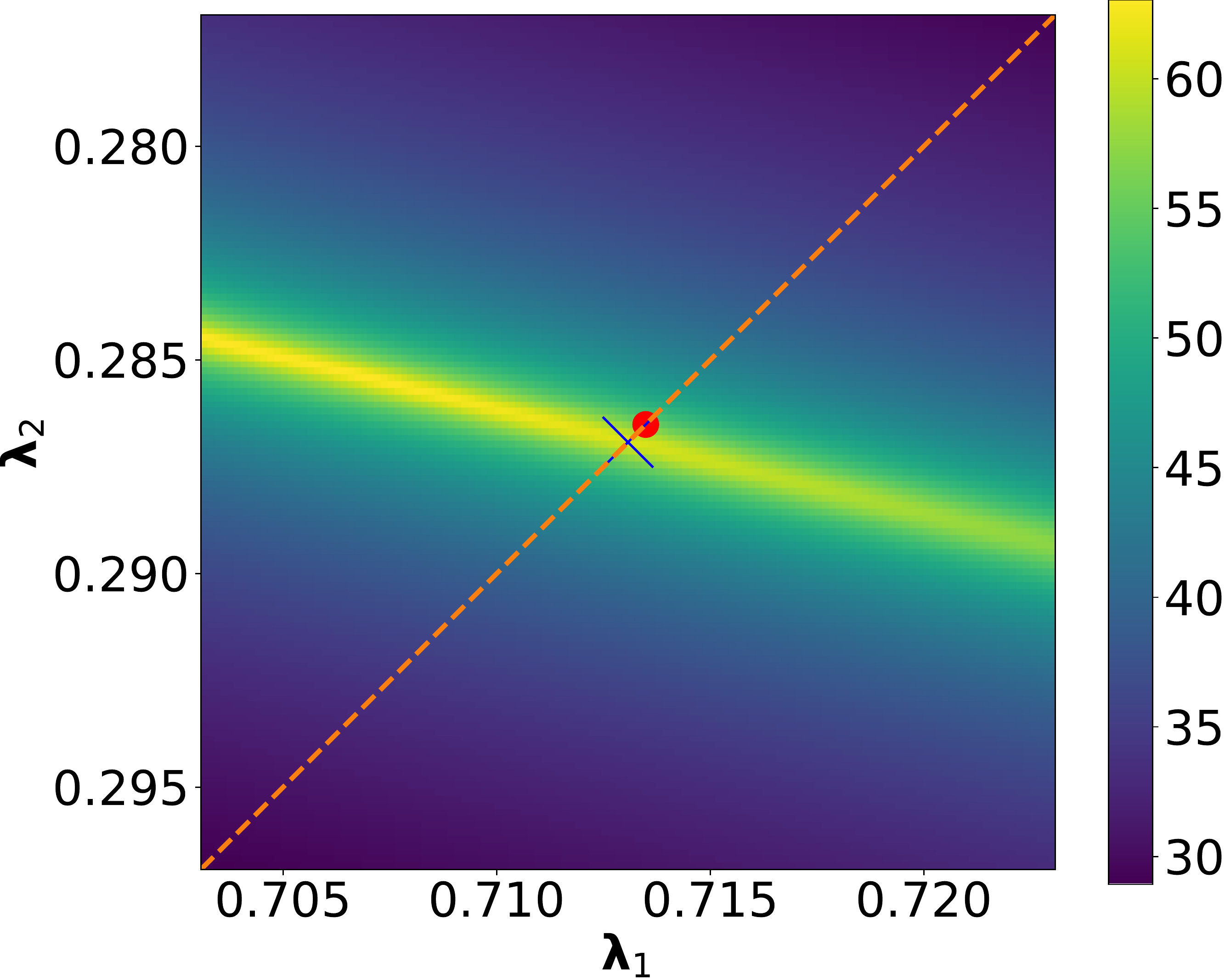}} \\
			\caption{Optimization landscapes of the manifold projections, that is $\left\lVert\mathbf{a} - \rho^*~ g\left(\bm{\lambda}\right)\right\rVert_2^2$ as a function of $\bm{\lambda}$ with the ground truth $\rho^*$, in an inverse logarithmic scale. Blue cross: result of the manifold projection, red circle: initialization (using the IAE's encoder). Top row: overview. Bottom row: zoom on the solution. \protect\subref{fig:proj_optim_landscape_sync} Synchrotron model, with $\bm{\lambda}_3 = 1 - \bm{\lambda}_1 - \bm{\lambda}_2$ to fulfill the sum-to-one constraint of the latent parameter inherent to the IAE. \protect\subref{fig:proj_optim_landscape_gauss} Gaussian model, the dashed line is the sum-to-one constraint $\bm{\lambda}_1 + \bm{\lambda}_2 = 1$.}
			\label{fig:proj_optim_landscape}
		\end{center}
	\end{figure}

	Figure \ref{fig:proj_optim_landscape} shows two optimization landscapes of projections on manifold, that is Eq.~\eqref{eq:proj_lamda} as a function of the latent parameter $\bm{\lambda}$. As stated earlier, the optimization landscape is quite convex near the solution, which allows using proximal minimization schemes as long as the initialization is decent. 
	
	\subsection{Results}
	
	\subsubsection{Overall results}
\begin{figure}
		\begin{center}
			\subfloat[\label{fig:res_sgmca}sGMCA]{\includegraphics[width=0.245\textwidth]{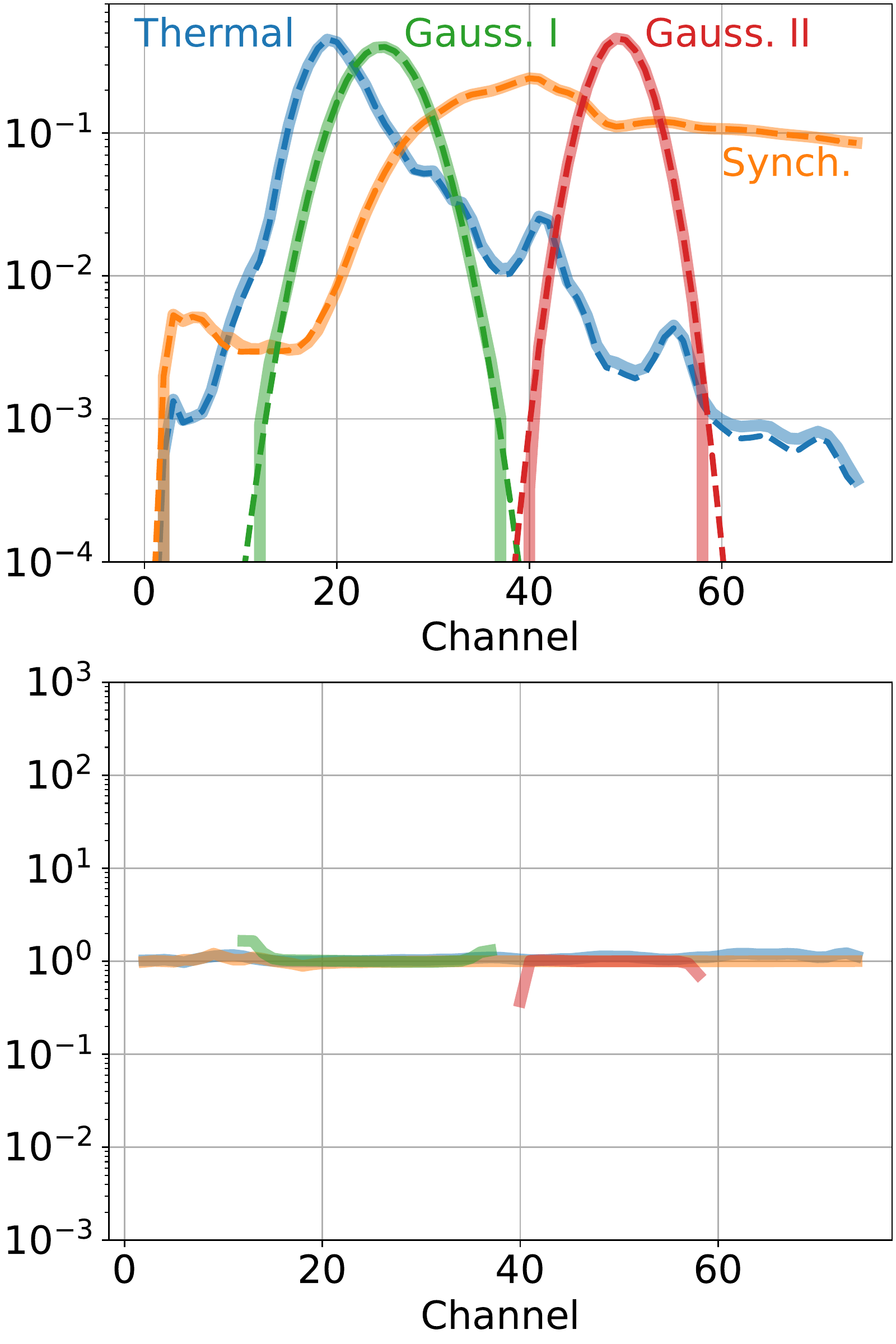}} \hfill
			\subfloat[\label{fig:res_gmca}GMCA]{\includegraphics[width=0.245\textwidth]{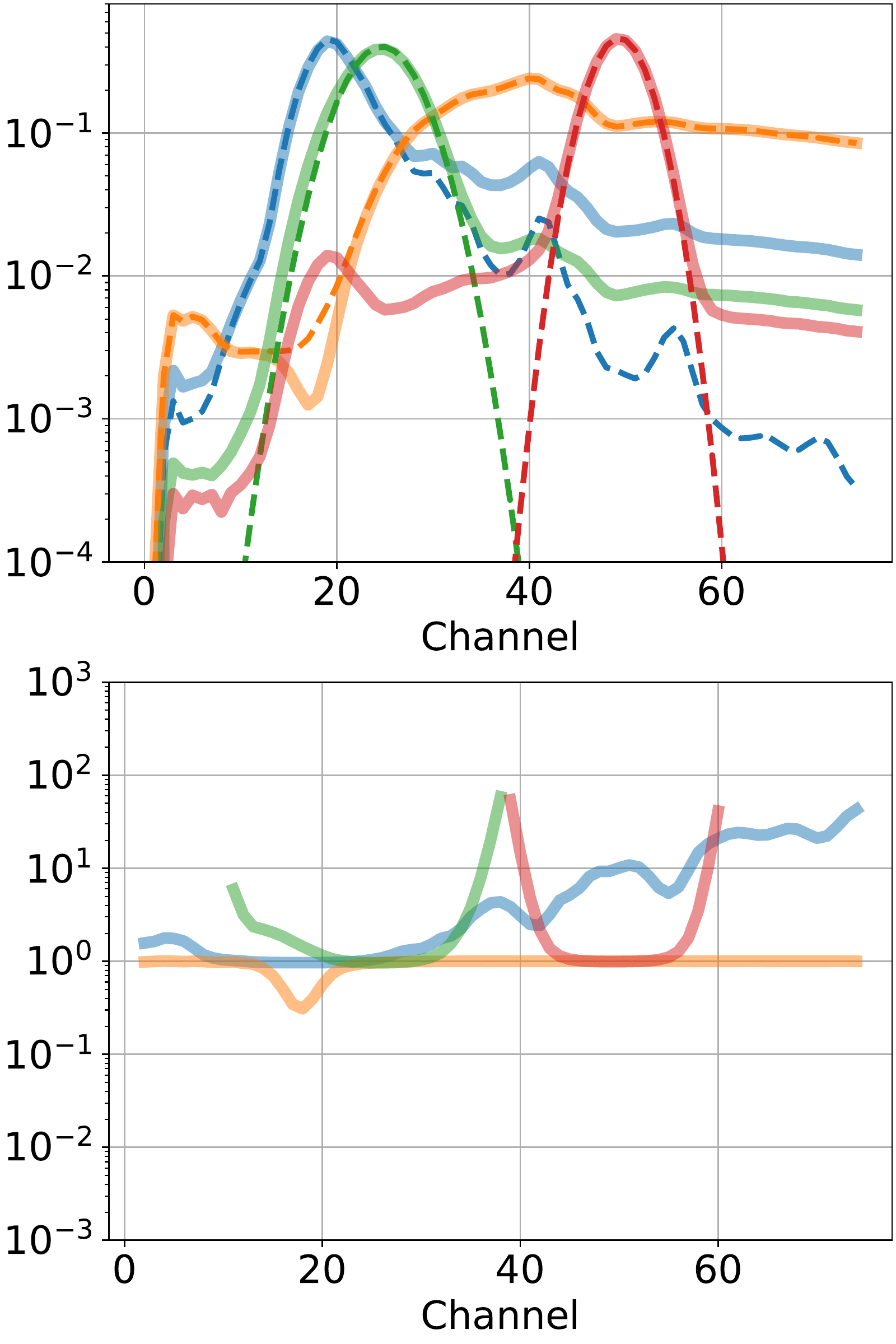}} \hfill
			\subfloat[\label{fig:res_hals}HALS]{\includegraphics[width=0.245\textwidth]{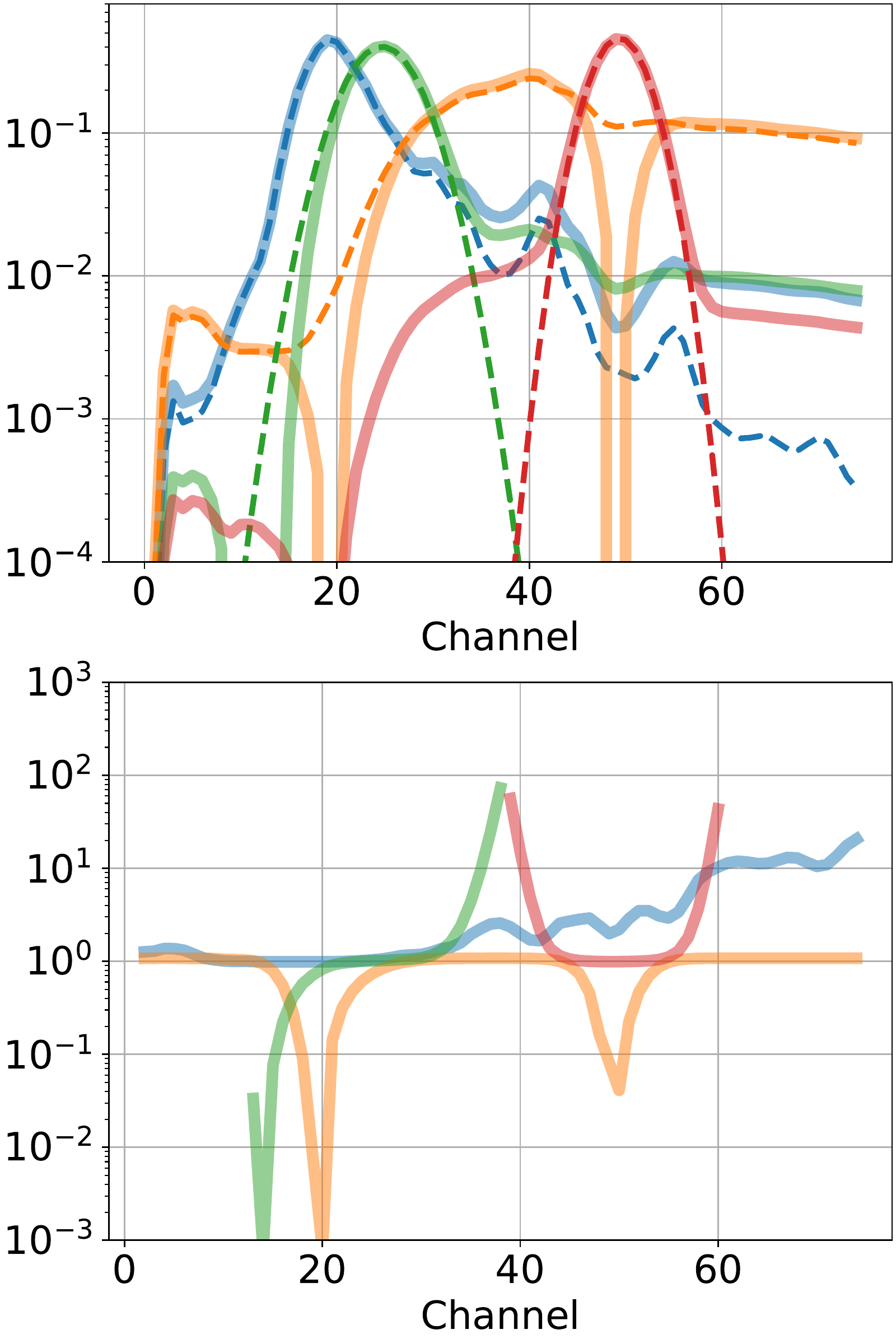}} \hfill
			\subfloat[\label{fig:res_snmf}SNMF]{\includegraphics[width=0.245\textwidth]{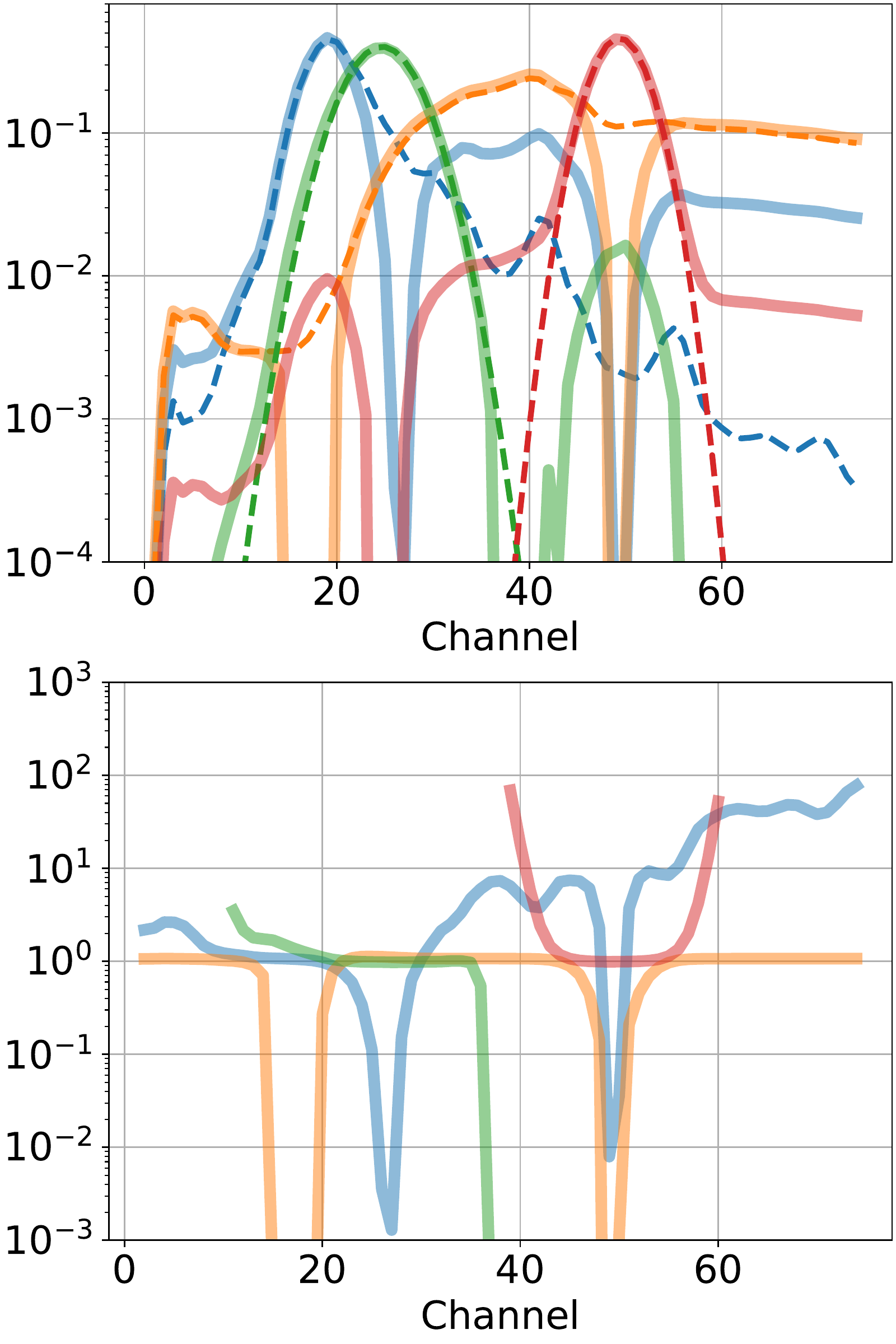}}
			\caption{Example of estimated spectra, with $\textit{SNR}=40~\text{dB}$, $\delta=20$ and $k=1$. Top row: spectra (solid lines: estimation, dashed lines: ground truth). Bottom row: estimated spectra over ground truth spectra (plotted if ground truth greater than $\num{1e-4}$). The figures on a same row share the same ordinate range.}
			\label{fig:res_sp}
		\end{center}
	\end{figure}

\begin{figure}
		\begin{center}
			\subfloat{\includegraphics[width=0.23\textwidth]{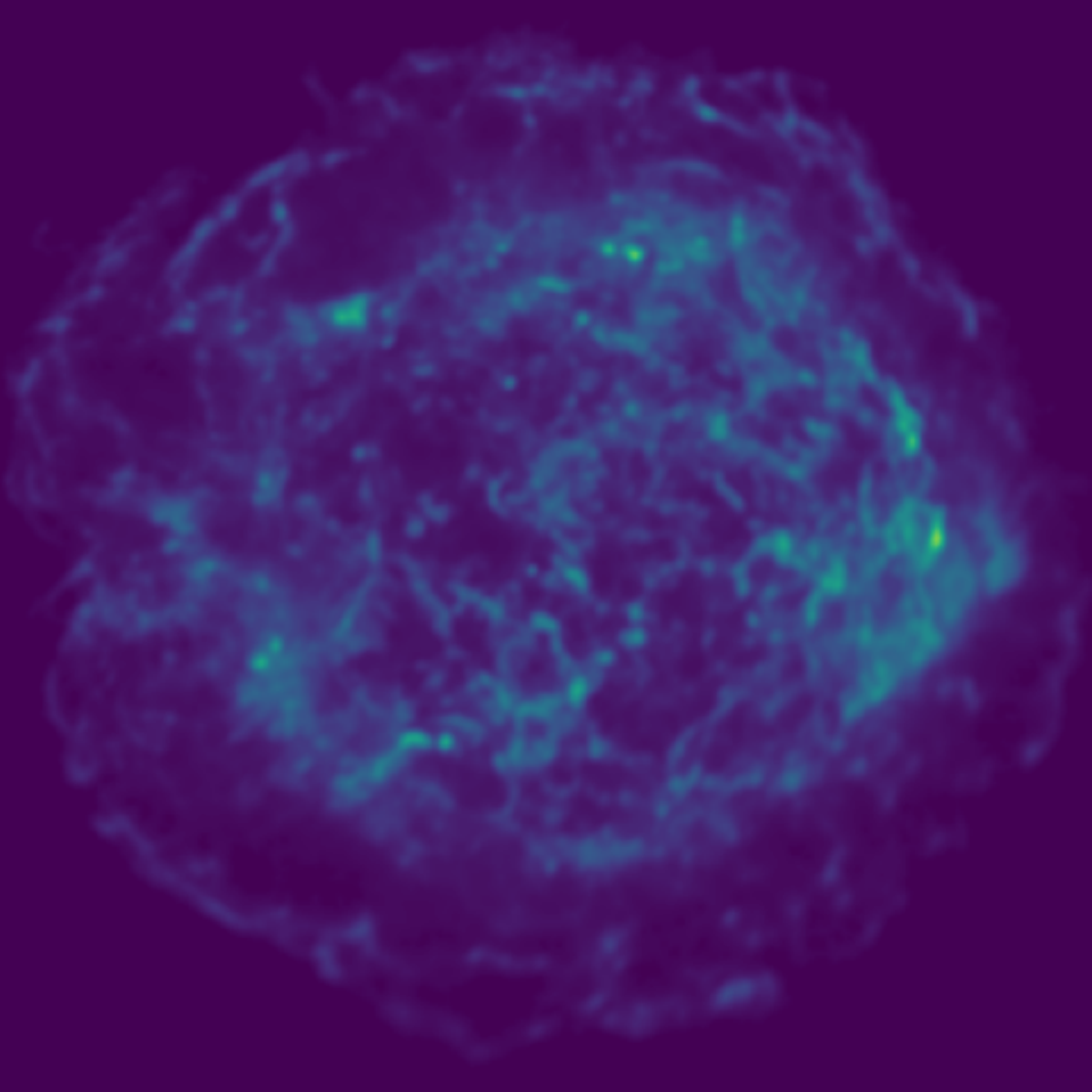}} \hfill
			\subfloat{\includegraphics[width=0.23\textwidth]{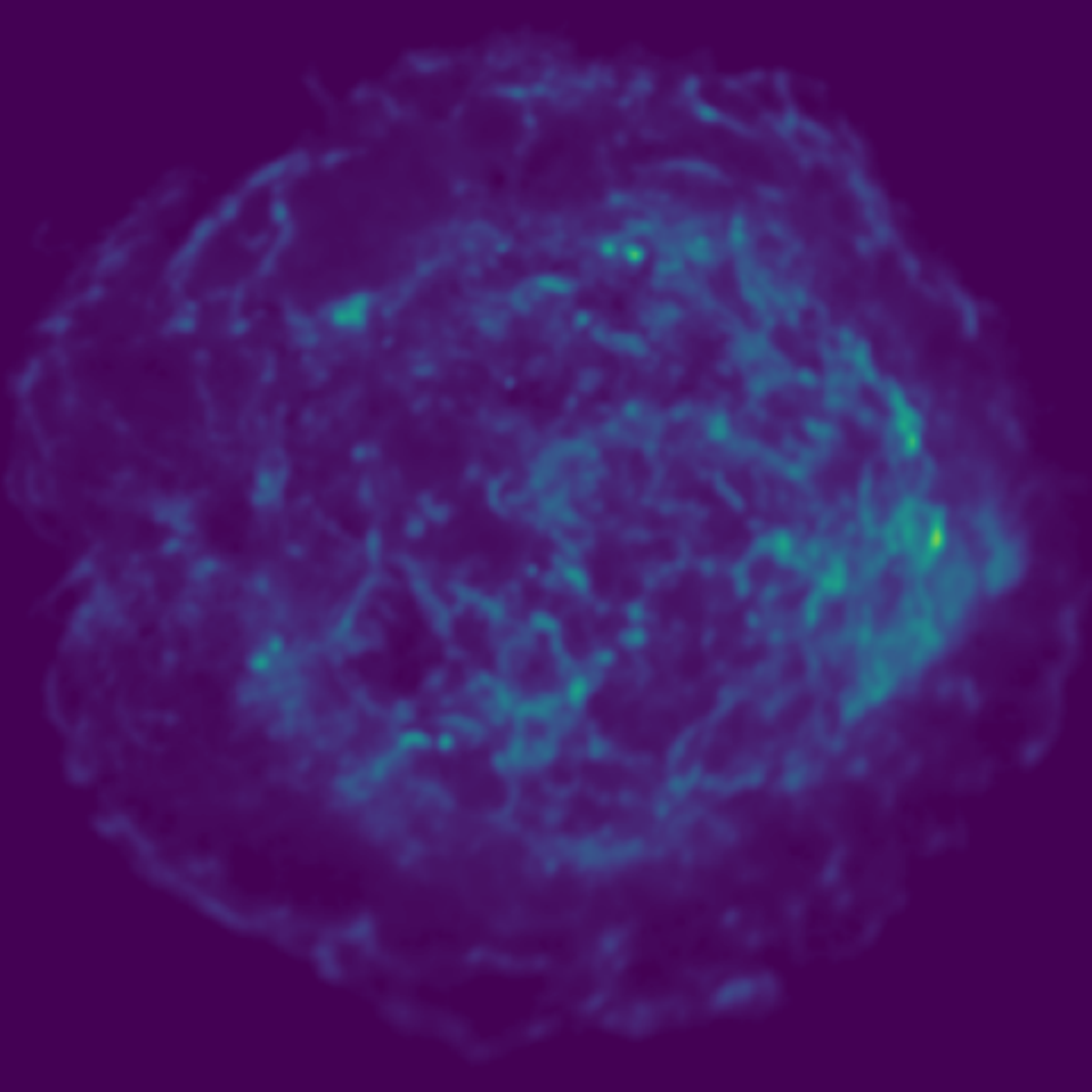}} \hfill
			\subfloat{\includegraphics[width=0.23\textwidth]{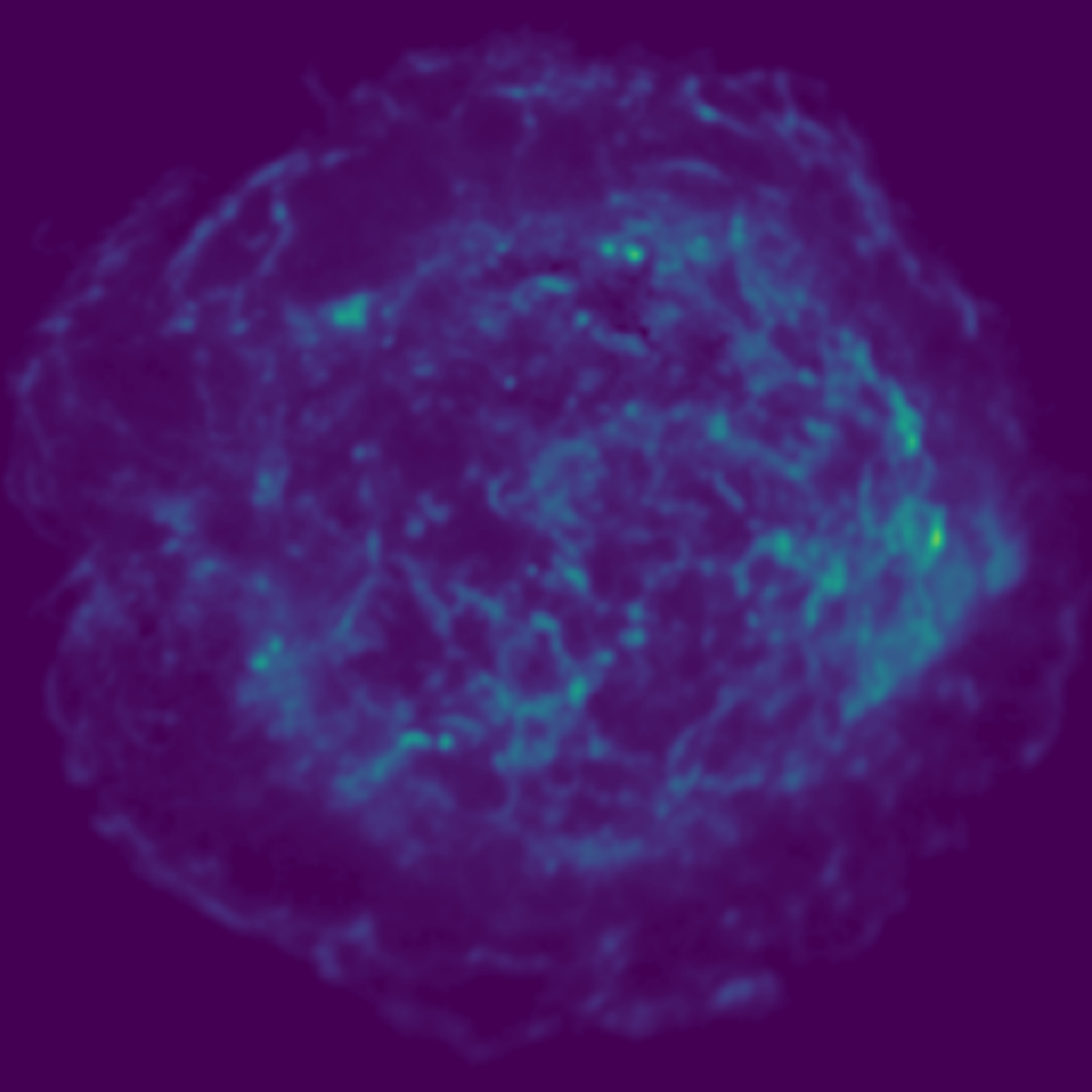}} \hfill
			\subfloat{\includegraphics[width=0.23\textwidth]{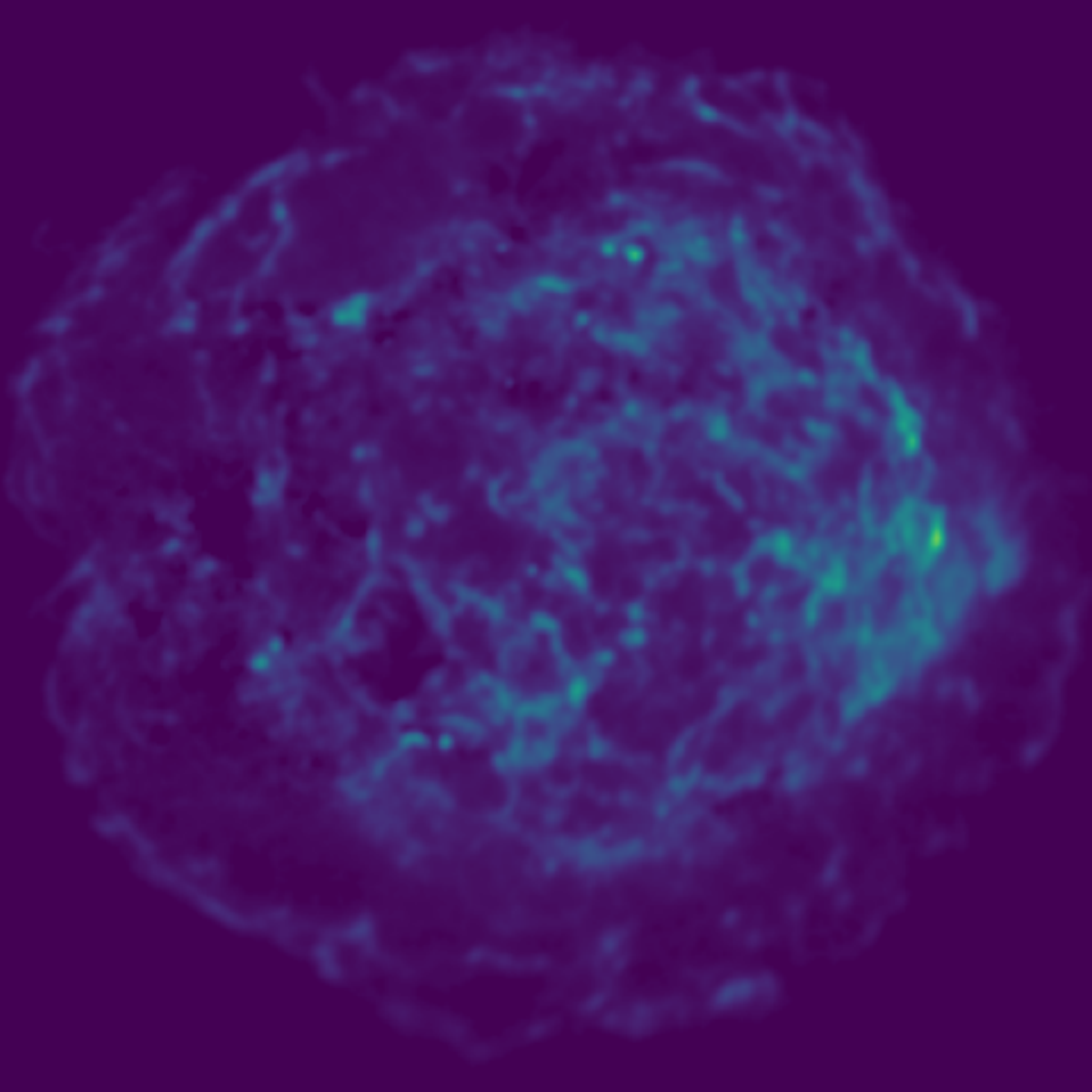}} \hfill 
			\subfloat{\includegraphics[width=0.055\textwidth]{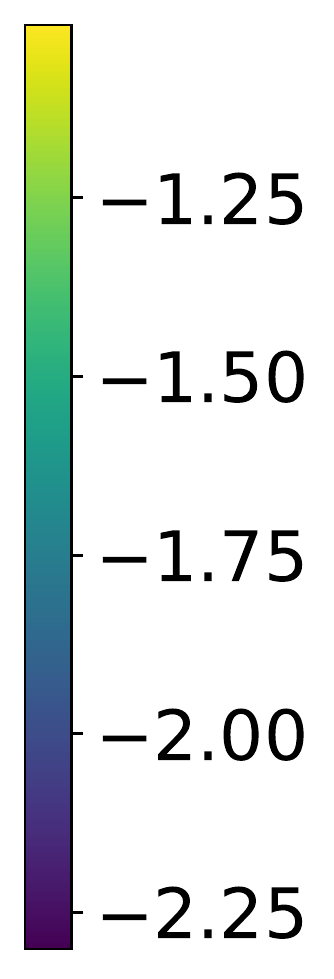}}
			\newline
			\addtocounter{subfigure}{-5}
			\subfloat[\label{fig:source_sgmca_0}sGMCA]{\includegraphics[width=0.23\textwidth]{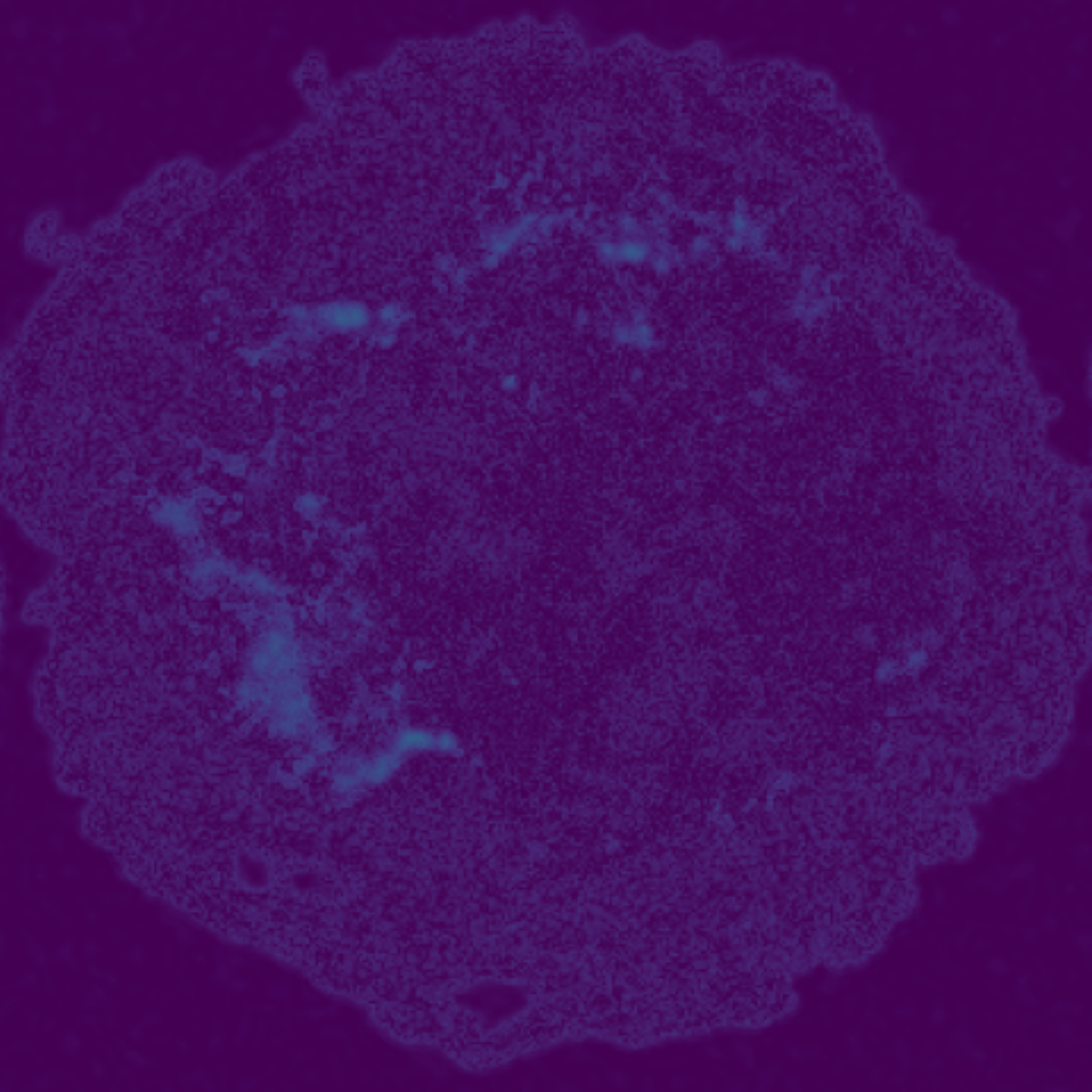}} \hfill
			\subfloat[\label{fig:source_gmca_0}GMCA]{\includegraphics[width=0.23\textwidth]{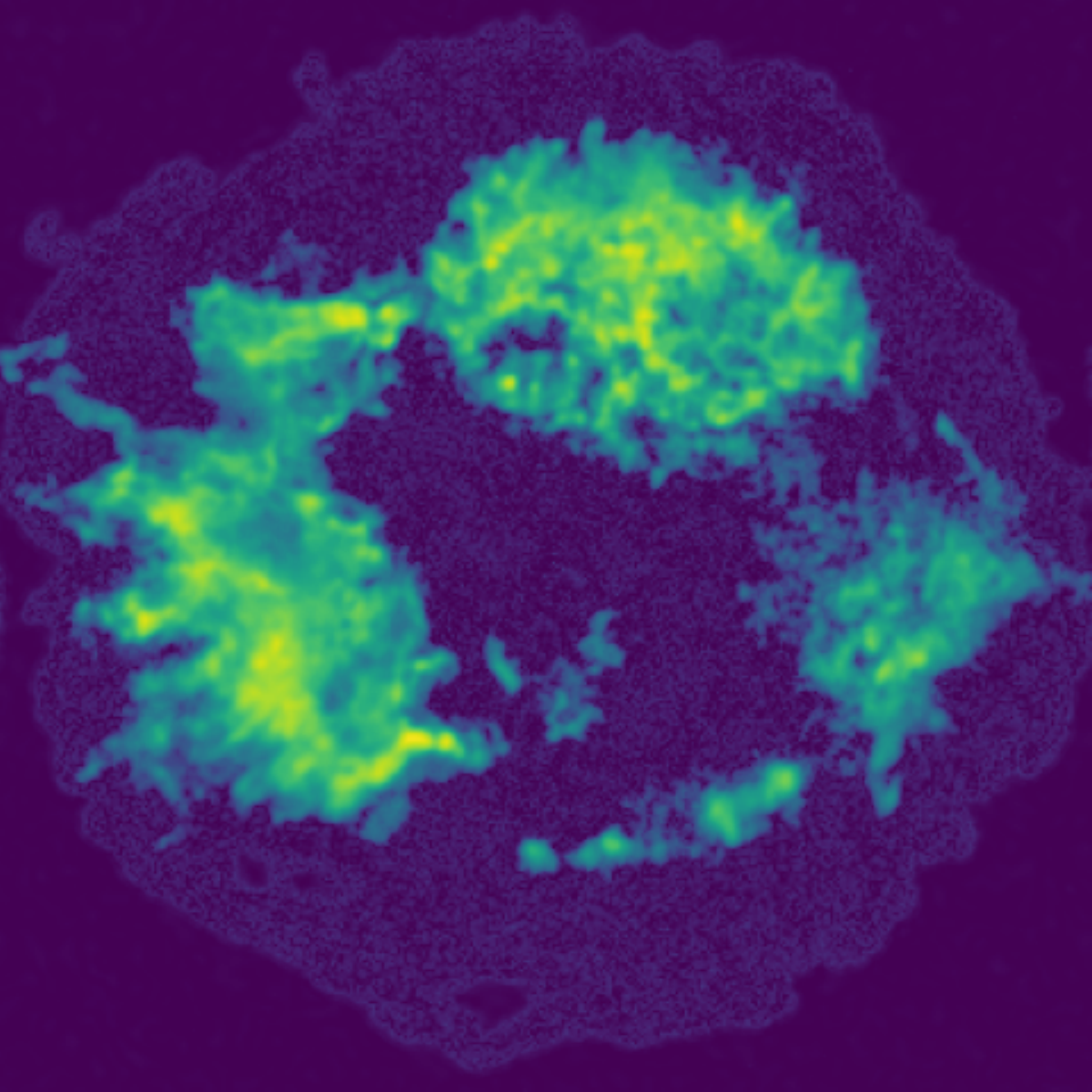}} \hfill
			\subfloat[\label{fig:source_hals_0}HALS]{\includegraphics[width=0.23\textwidth]{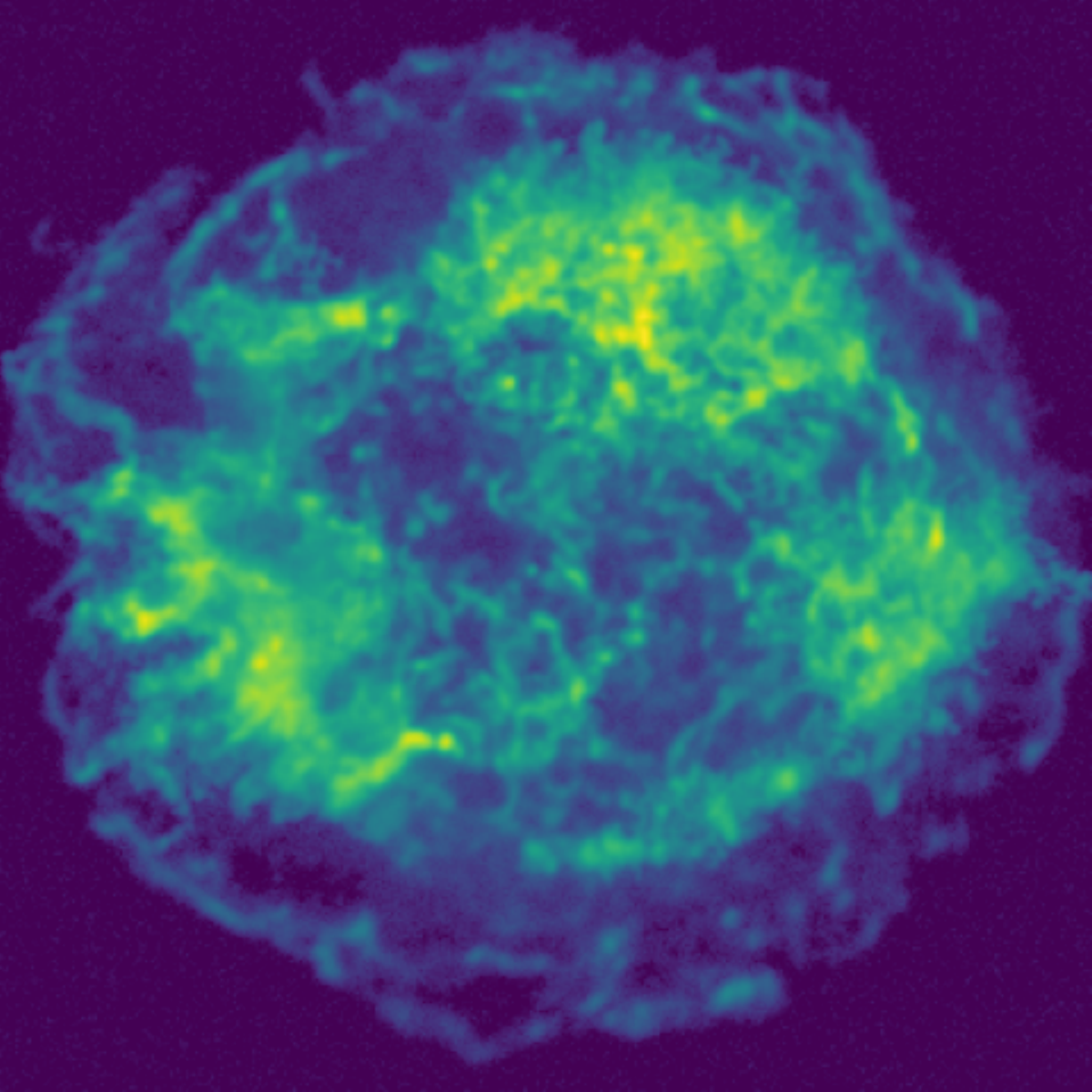}} \hfill
			\subfloat[\label{fig:source_snmf_0}SNMF]{\includegraphics[width=0.23\textwidth]{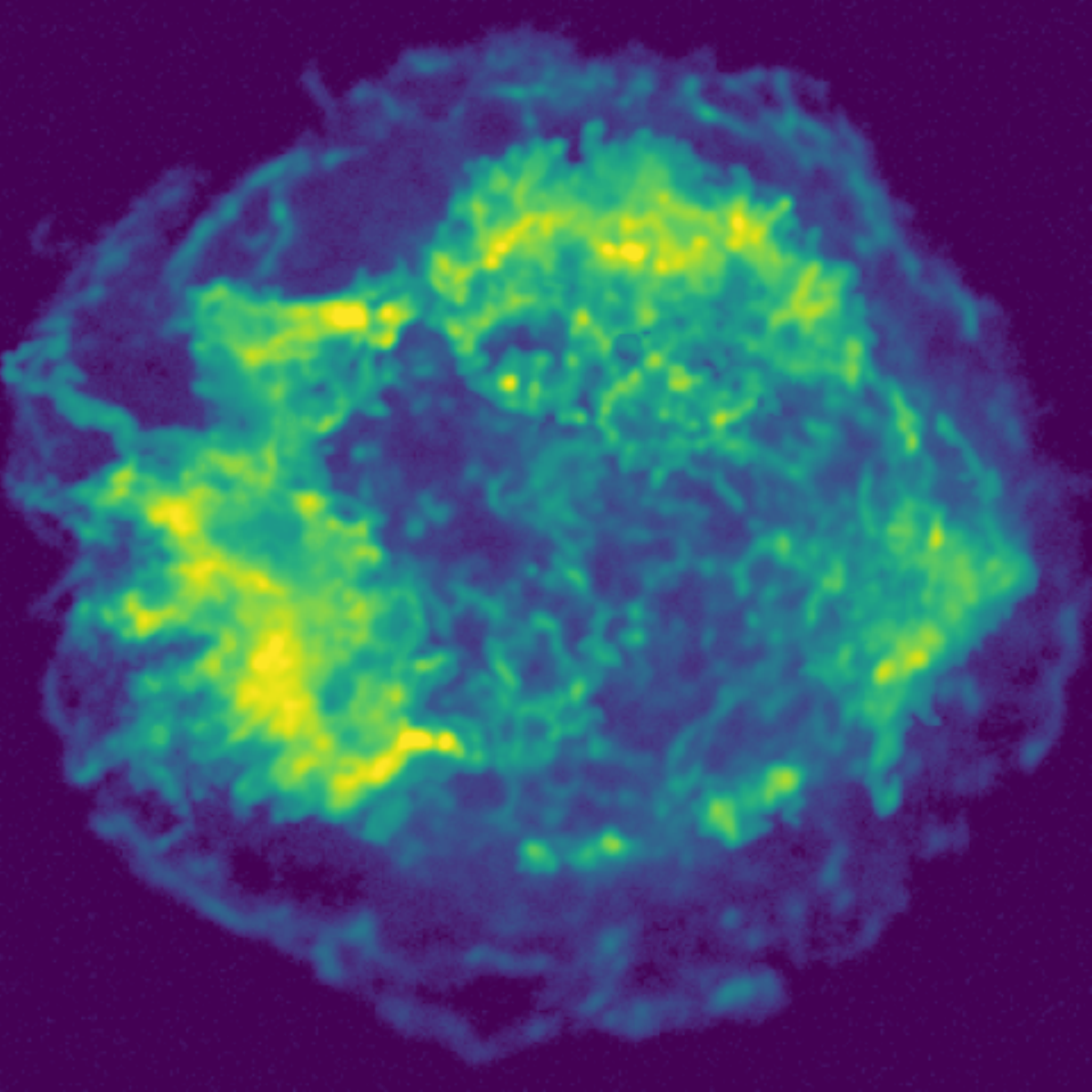}} \hfill		\subfloat{\includegraphics[width=0.055\textwidth]{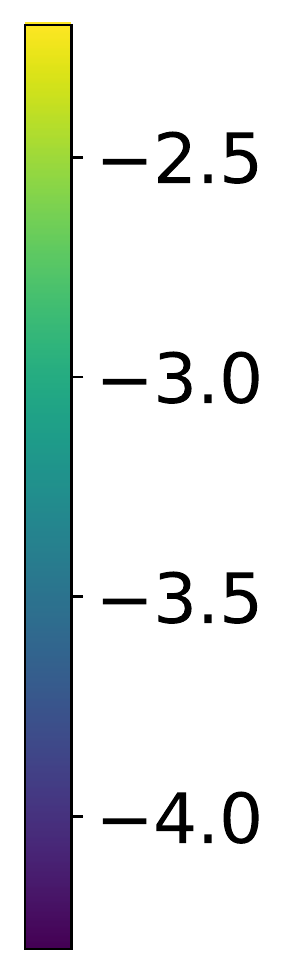}}
			\caption{Example of the estimated synchrotron source, with $\textit{SNR}=40~\text{dB}$, $\delta=20$ and $k=1$. Top: estimations (logarithmic scale), bottom: absolute error (logarithmic scale). The figures on a same row share the same color scale.}
			\label{fig:res_src_0}
		\end{center}
	\end{figure}

\begin{figure}
		\begin{center}
			\subfloat{\includegraphics[width=0.23\textwidth]{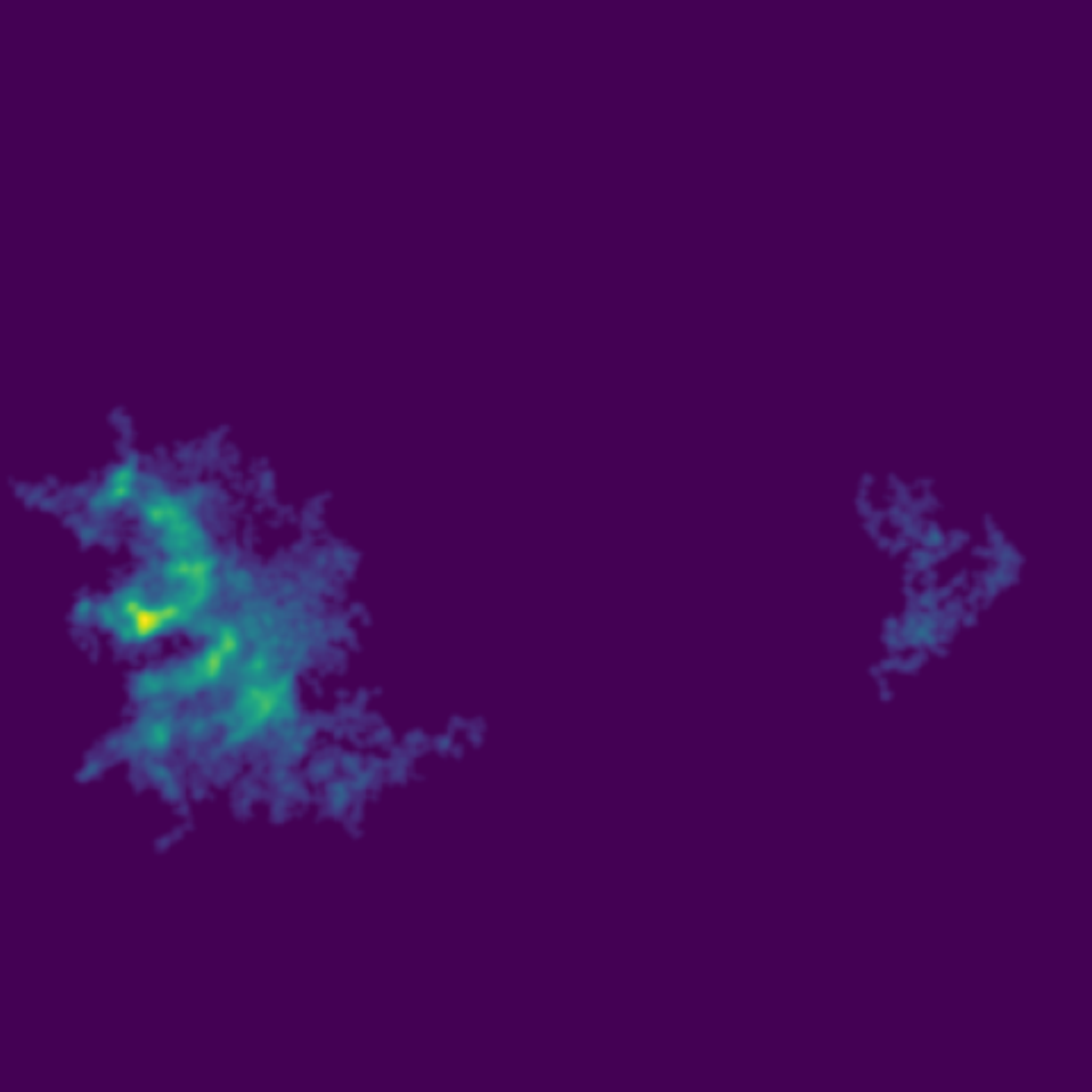}} \hfill
			\subfloat{\includegraphics[width=0.23\textwidth]{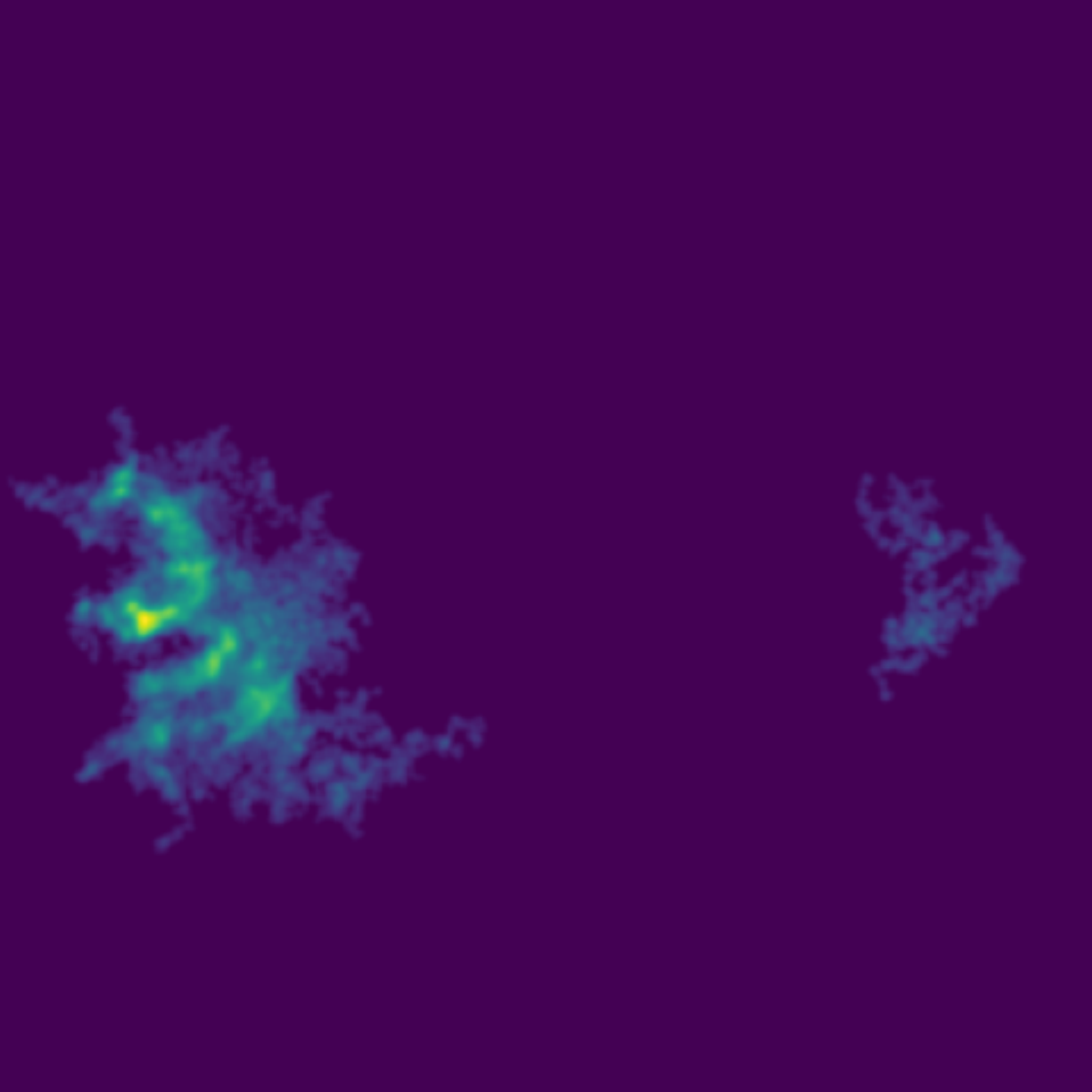}} \hfill
			\subfloat{\includegraphics[width=0.23\textwidth]{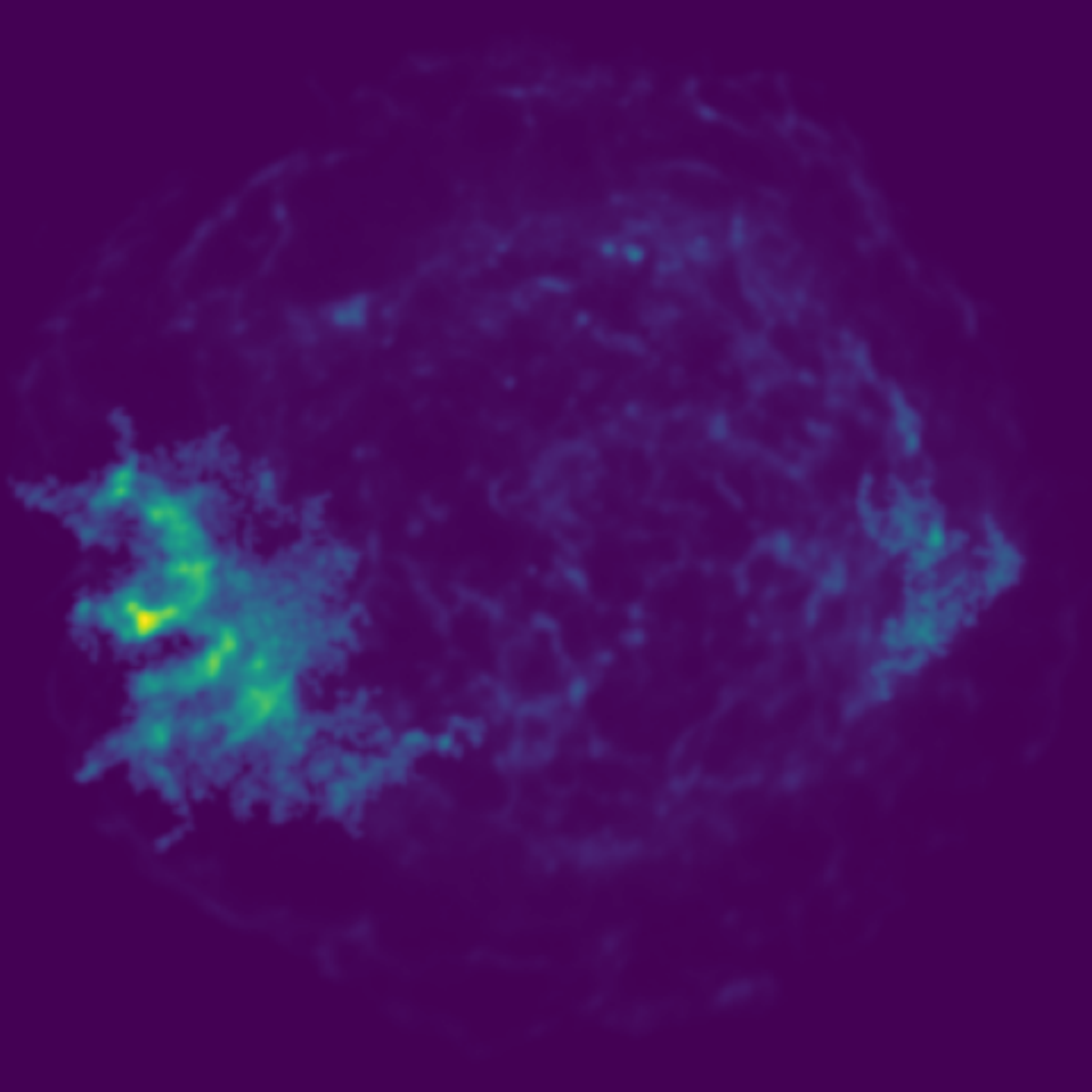}} \hfill
			\subfloat{\includegraphics[width=0.23\textwidth]{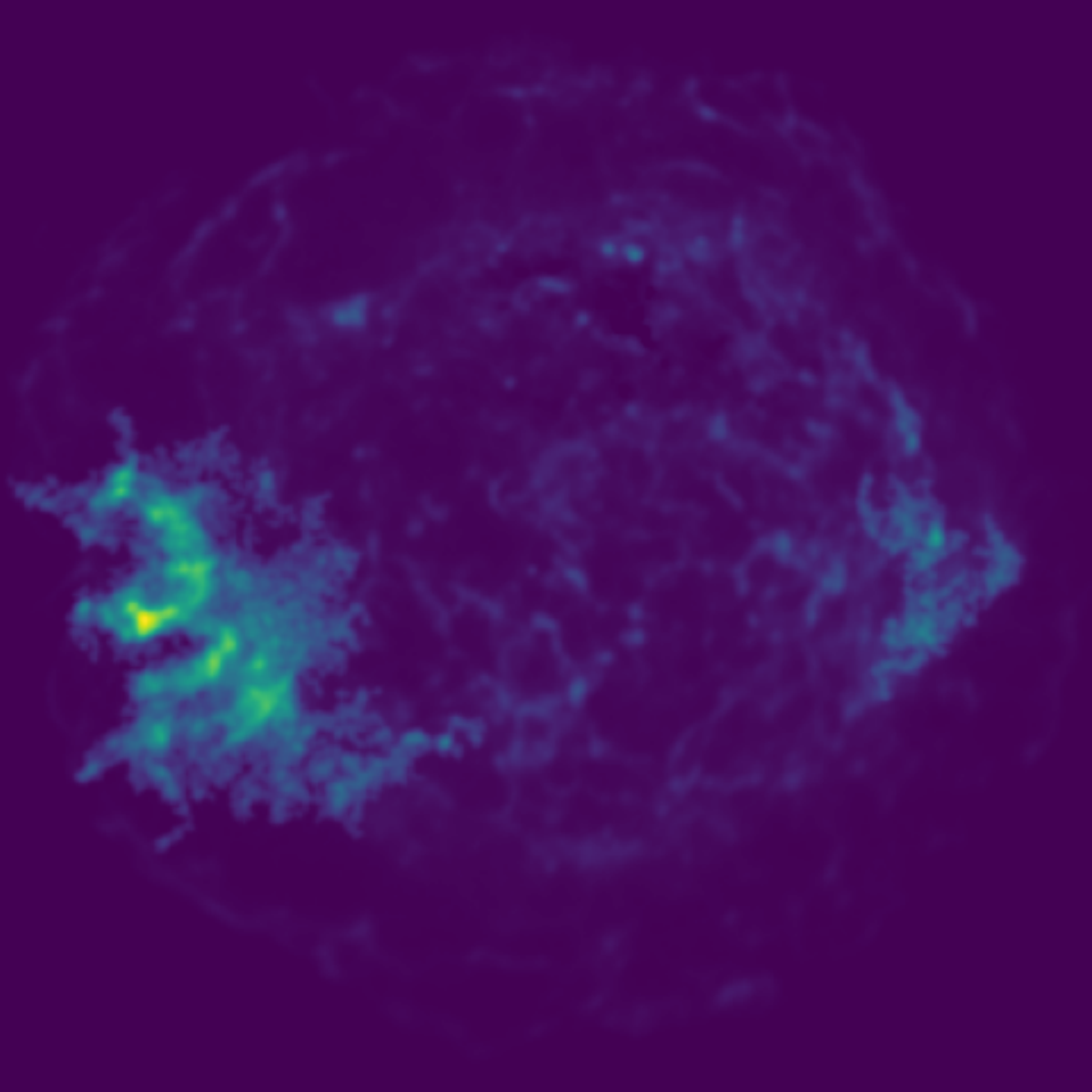}} \hfill 
			\subfloat{\includegraphics[width=0.055\textwidth]{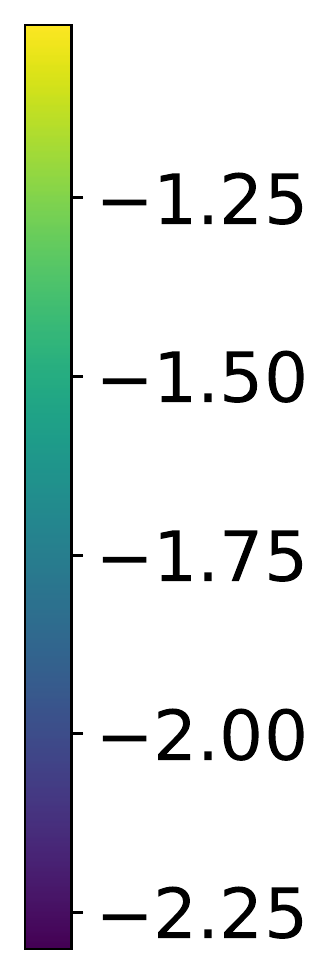}}
			\newline
			\addtocounter{subfigure}{-5}
			\subfloat[\label{fig:source_sgmca_3}sGMCA]{\includegraphics[width=0.23\textwidth]{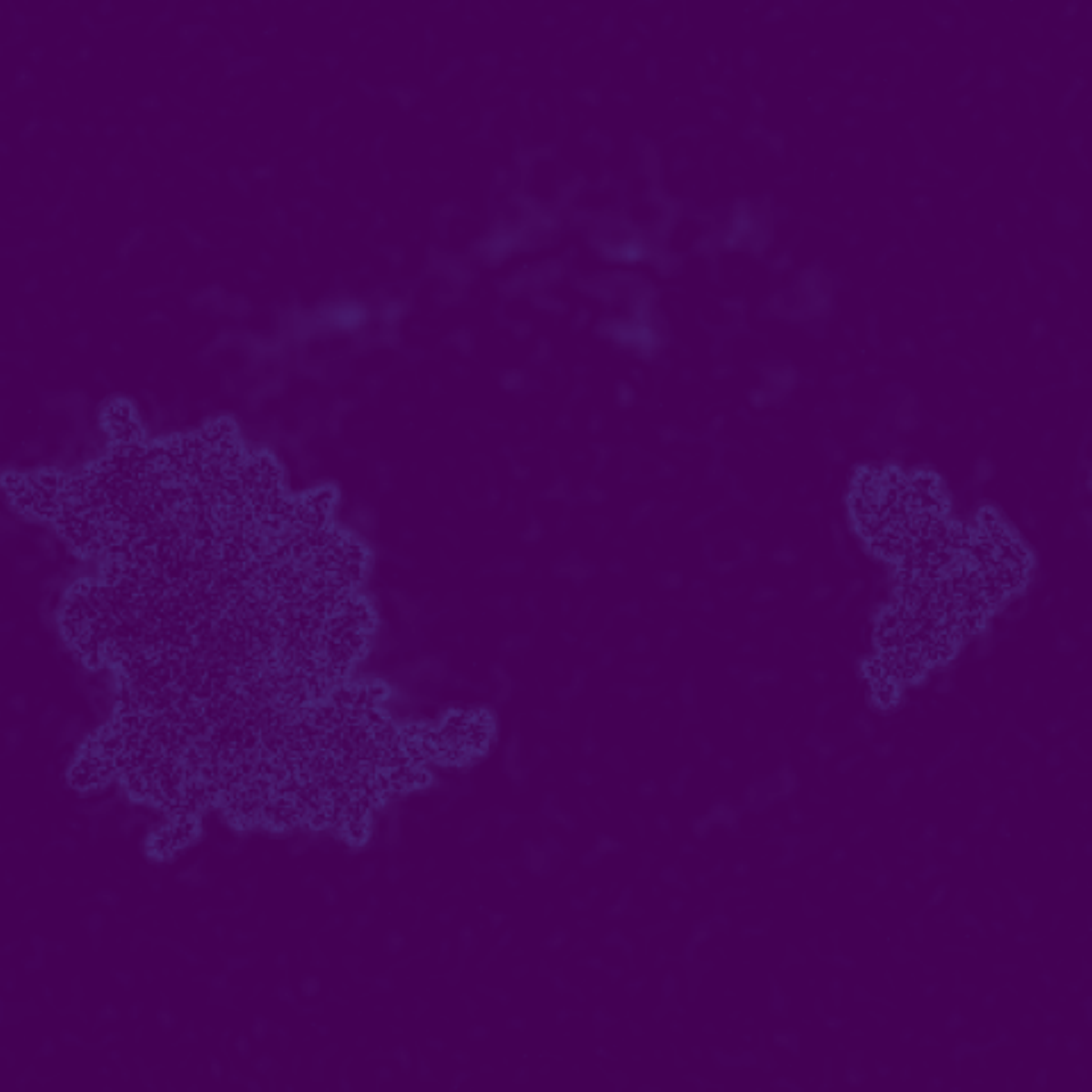}} \hfill
			\subfloat[\label{fig:source_gmca_3}GMCA]{\includegraphics[width=0.23\textwidth]{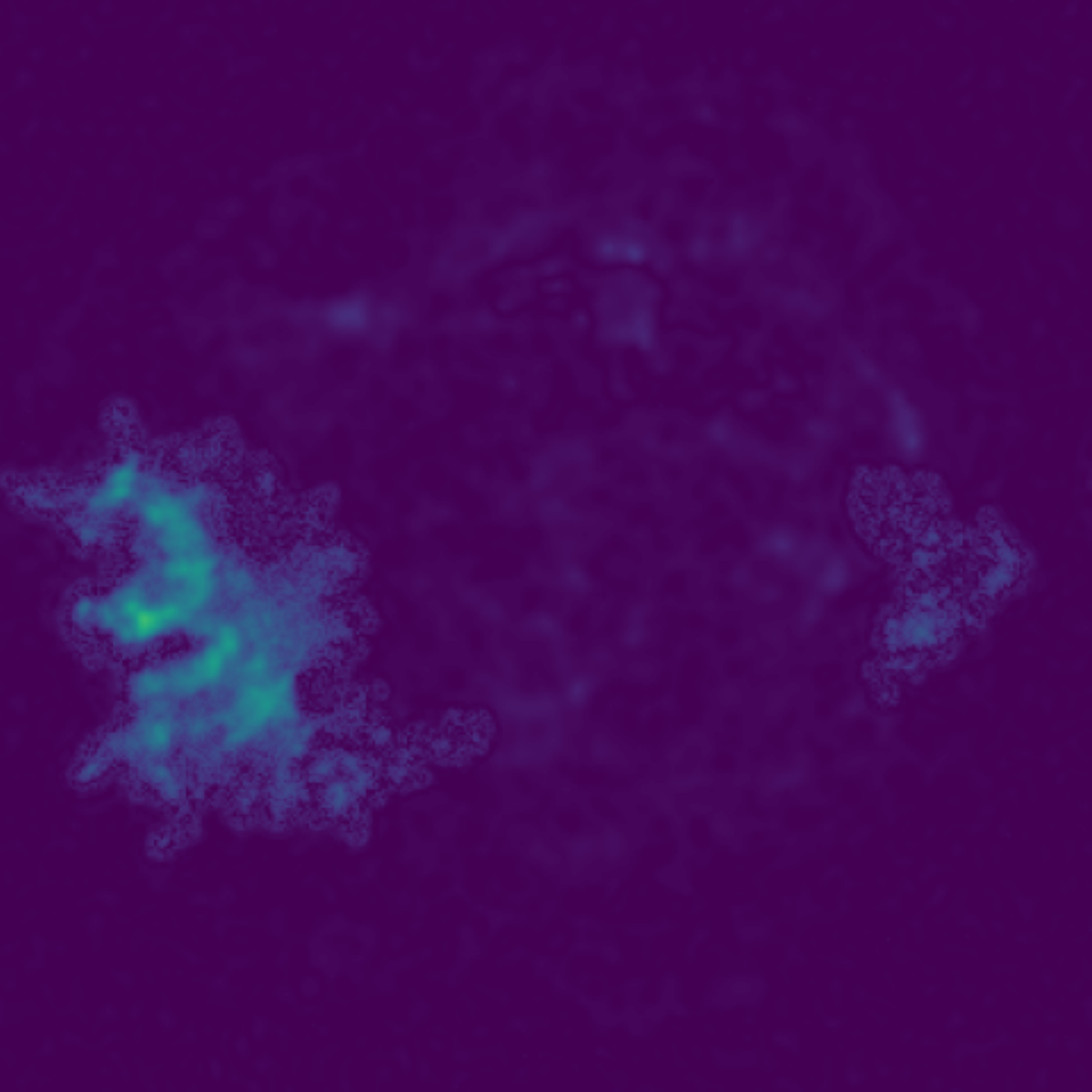}} \hfill
			\subfloat[\label{fig:source_hals_3}HALS]{\includegraphics[width=0.23\textwidth]{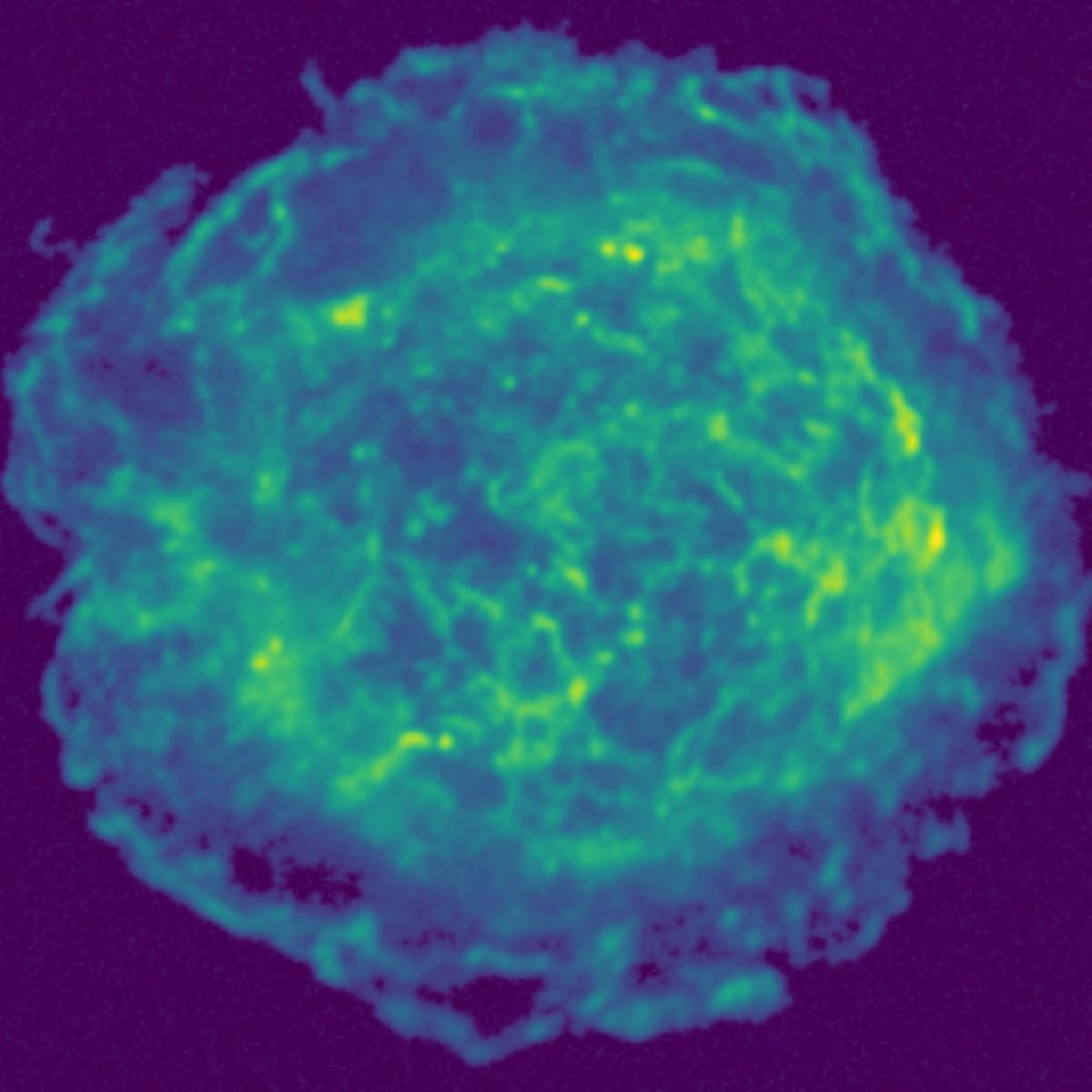}} \hfill
			\subfloat[\label{fig:source_snmf_3}SNMF]{\includegraphics[width=0.23\textwidth]{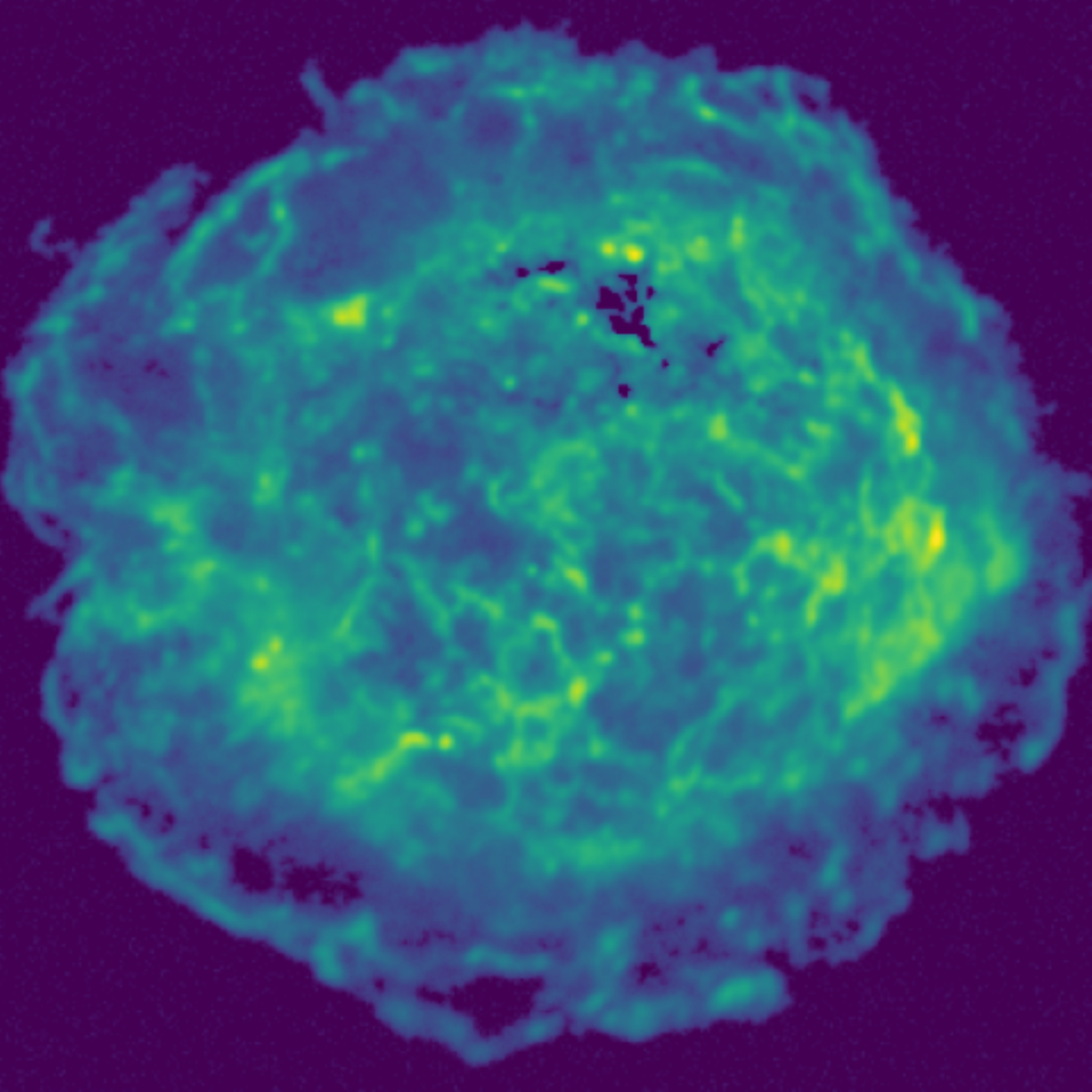}} \hfill		\subfloat{\includegraphics[width=0.055\textwidth]{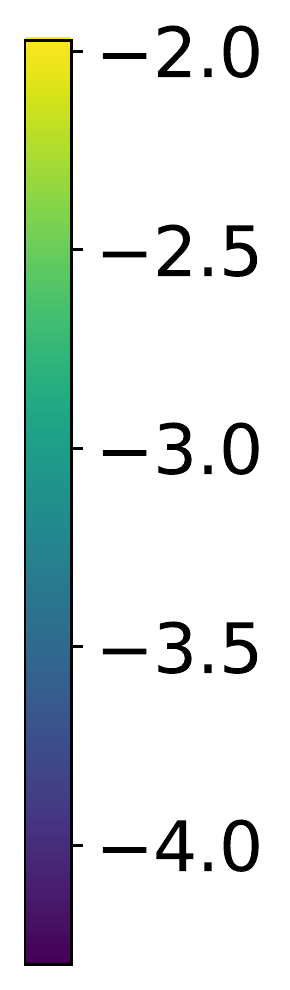}}
			\caption{Example of the estimated Gaussian II source, with $\textit{SNR}=40~\text{dB}$, $\delta=20$ and $k=1$. Top: estimations (logarithmic scale), bottom: absolute error (logarithmic scale). The figures on a same row share the same color scale.}
			\label{fig:res_src_3}
		\end{center}
	\end{figure}

	Firstly, let us compare qualitatively sGMCA to the BSS algorithms on a typical run. Figure \ref{fig:res_sp} shows the estimated spectra along with the estimation errors. The three BSS methods are particularly prone to interferences, which result in either portions of spectra misestimated at zero or with leakages of other components, which is problematic for astrophysical interpretations. On the contrary, sGMCA manages to remove most interferences and recovers satisfactory spectra. 
	Figures \ref{fig:res_src_0} and \ref{fig:res_src_3} show the estimates of respectively the synchrotron source and the Gaussian II source, as well as the associated estimation errors. Residuals of other sources and/or reconstruction artifacts are clearly visible in the estimation errors of the three blind methods. The sGMCA algorithm provides a more accurate source, whose error equally originates from interference, noise contamination and artifacts.\\
	Concerning computation times, sGMCA is undeniably slower than the blind methods due to the manifold projections. For example, in the tests performed, the projection of the four spectra takes approximately $3~\si{s}$ and 50 iterations are typically necessary. In contrast, the blind methods run in a few seconds.

	
	\subsubsection{Impact of the experimental parameters}
	
	In order to quantitatively confirm these results, three experiments are performed, in each of which an experimental parameter is varied. They are constituted of Monte-Carlo trials with varying spectra (taken from the test sets) and noise realizations. The overall results are reported in Fig.~\ref{fig:exp} and are commented on in the paragraphs below.\\
	The sparsity hyperparameter of SNMF is found to be very sensitive to the experimental parameters; it is set to a globally "good" value $\lambda = \num{1e-7}$.
	
	\paragraph{Noise level}
	The overall tendencies are consistent; the higher the \textit{SNR}, the better the estimations. As we had observed previously on the example, sGMCA estimates particularly well the spectra; the gain in SAD is from a few to 10 dB compared to GMCA, depending on the noise level. Since the sources are better disentangled, this results in a significant gain of the SIR and thus the SDR.\\
	The additional prior information to which the nearest-neighbor benchmark algorithm has access leads to improved performances compared to GMCA, but not as much as sGMCA. 
	This emphasizes the advantage of the regularization provided by the IAE; by modeling the manifolds on which the spectra evolve, the latter are reconstructed much more precisely.\\
	It is noted that the performance metrics of sGMCA and GMCA can reach a plateau at a high \textit{SNR}. According to the hyperparameter tuning strategy of GMCA and sGMCA, the thresholds applied to the sources are low at a low noise level, inducing an underregularization. Interestingly, the sGMCA is less sensitive to this effect, most likely because the components are better separated.

	\paragraph{Collinearity of the spectra} Similarly to the previous experiment, sGMCA outperforms the BSS algorithms, with appreciable gains in SAD by 10 to 20 dB and in SDR by 5 to 10 dB compared to GMCA. When the two Gaussian spectra tend to coincide ($\delta\rightarrow0$) the source estimates of the GMCA-based algorithms decline because the least-square update of the sources in the minimization scheme becomes ill-conditioned. Moreover, the GMCA metrics are relatively insensitive to $\delta$; a more detailed analysis shows that the GMCA errors are dominated by the synchrotron component, on which $\delta$ has indeed a negligible effect. 
	
\begin{figure}
		\begin{center}
			\subfloat{\includegraphics[width=0.23\textwidth]{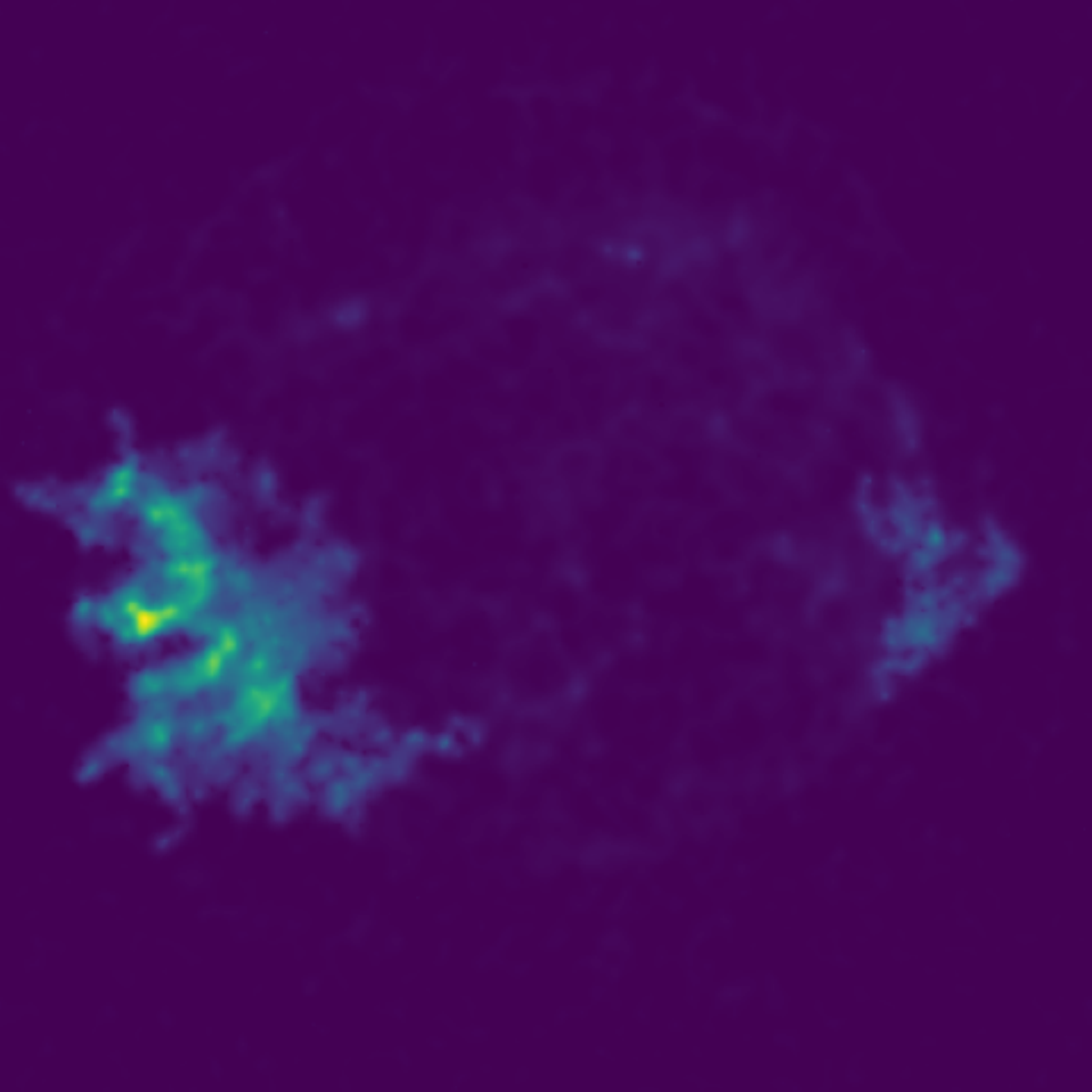}} \hfill
			\subfloat{\includegraphics[width=0.23\textwidth]{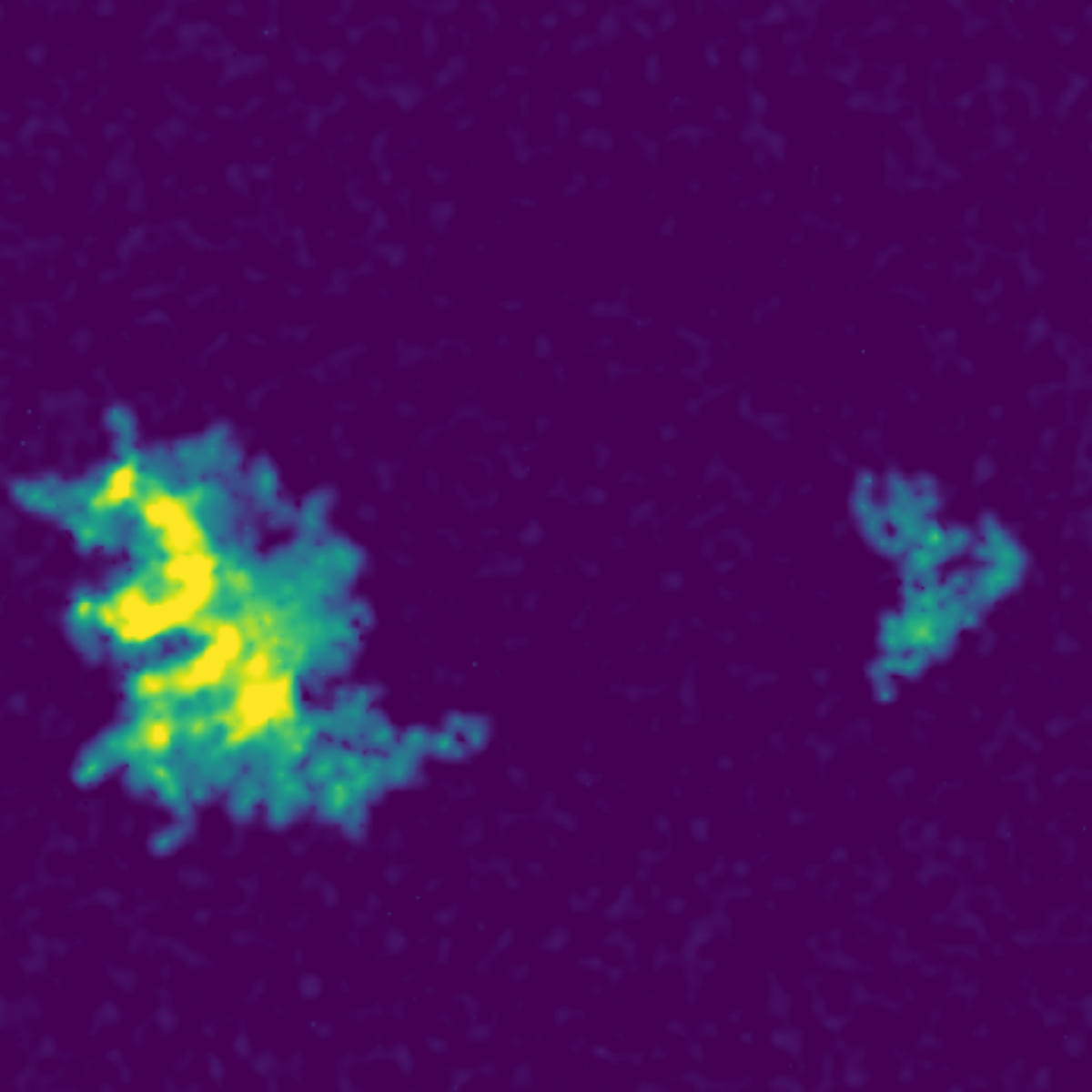}} \hfill
			\subfloat{\includegraphics[width=0.23\textwidth]{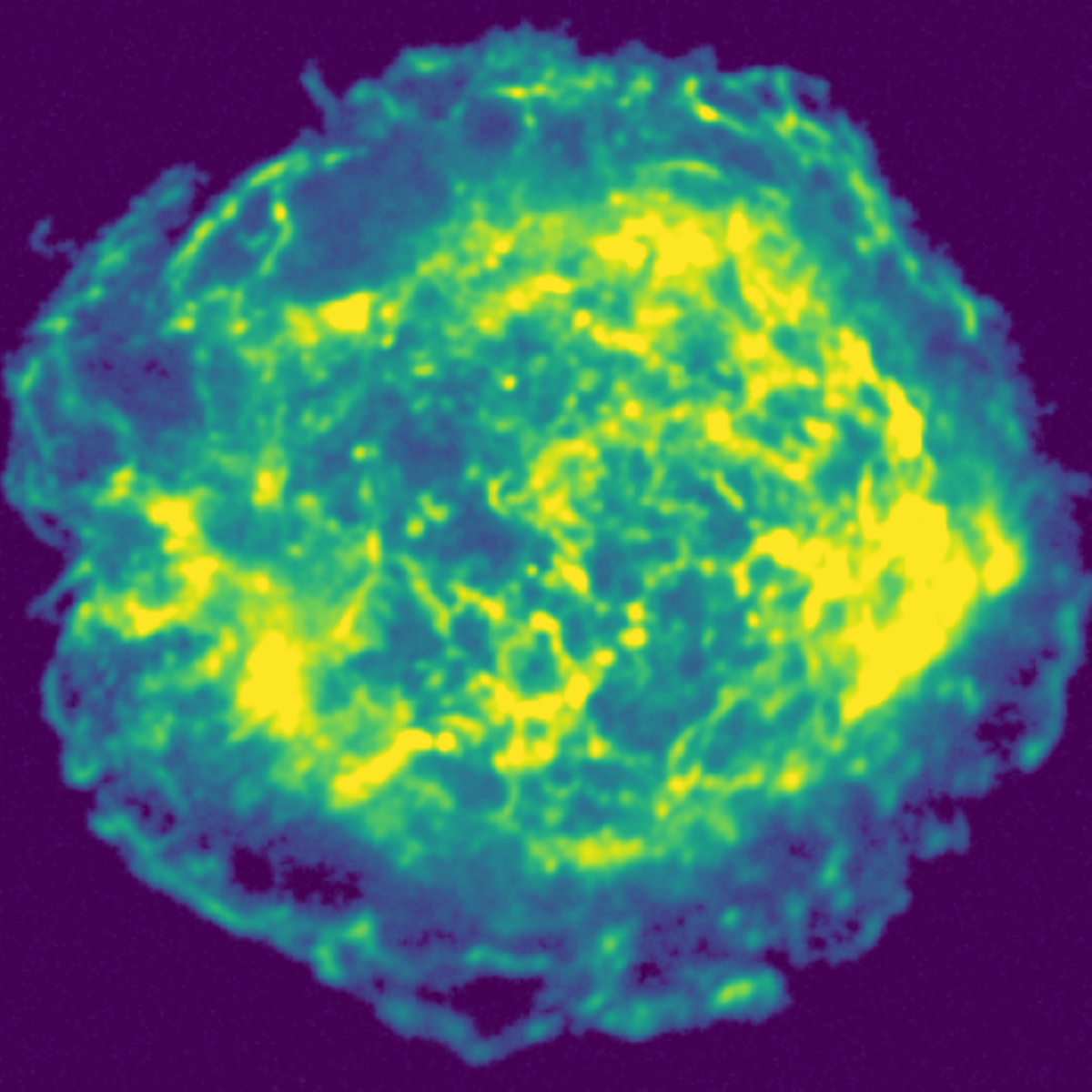}} \hfill
			\subfloat{\includegraphics[width=0.23\textwidth]{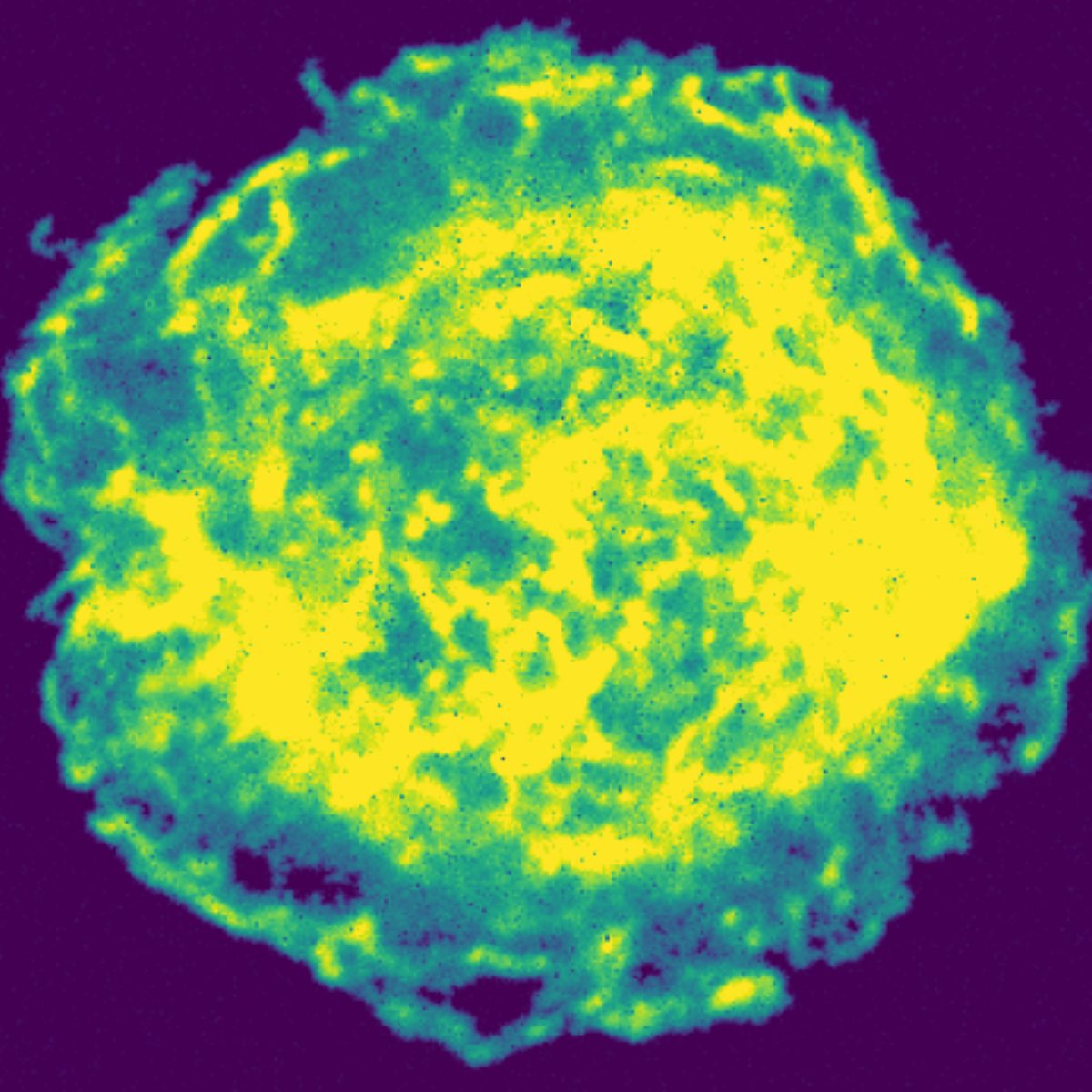}} \hfill 
			\subfloat{\includegraphics[width=0.055\textwidth]{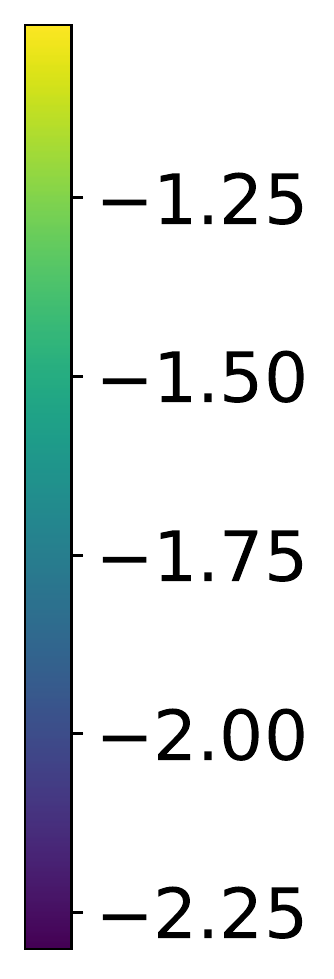}}
			\newline
			\addtocounter{subfigure}{-5}
			\subfloat[\label{fig:source_sgmca_3_unbal}sGMCA]{\includegraphics[width=0.23\textwidth]{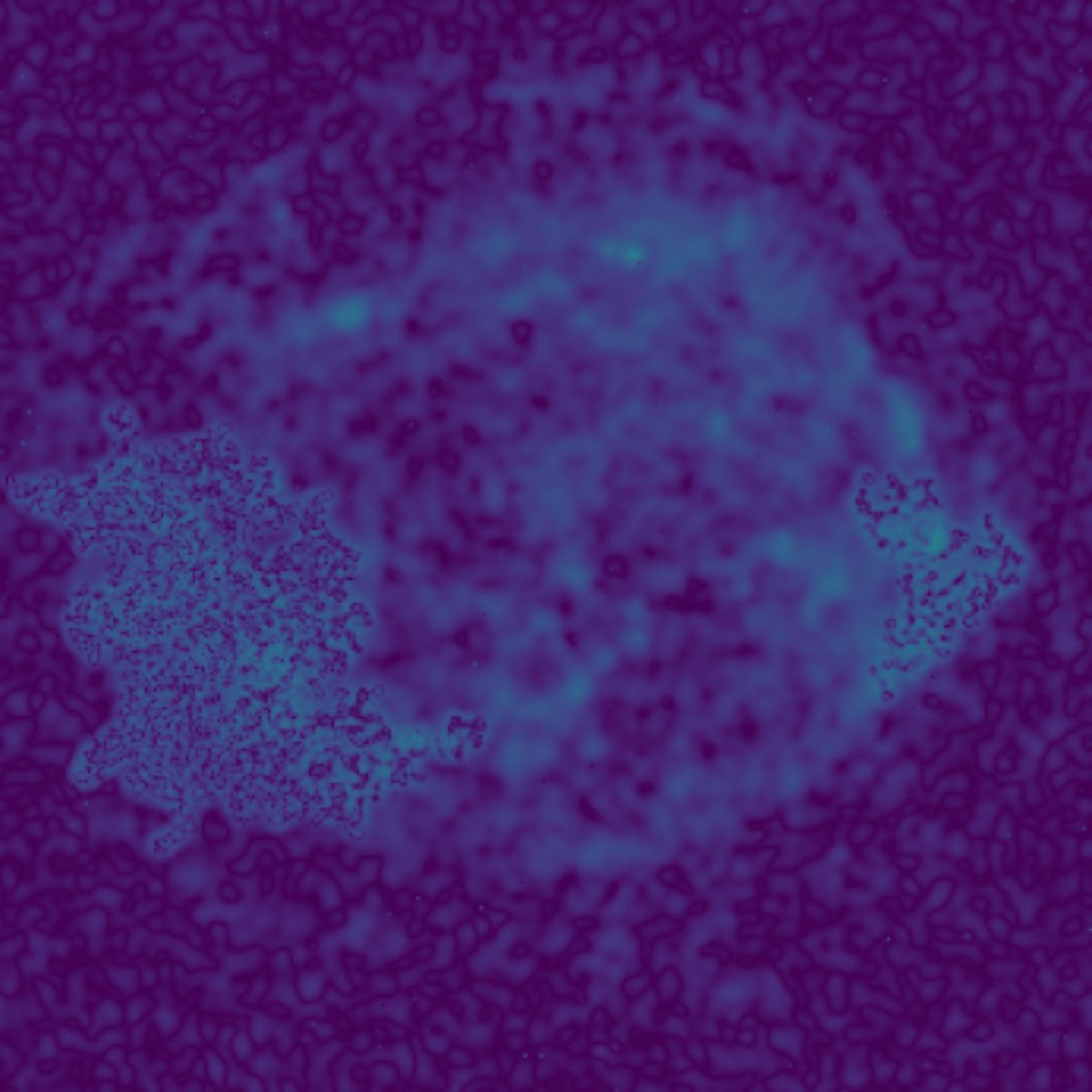}} \hfill
			\subfloat[\label{fig:source_gmca_3_unbal}GMCA]{\includegraphics[width=0.23\textwidth]{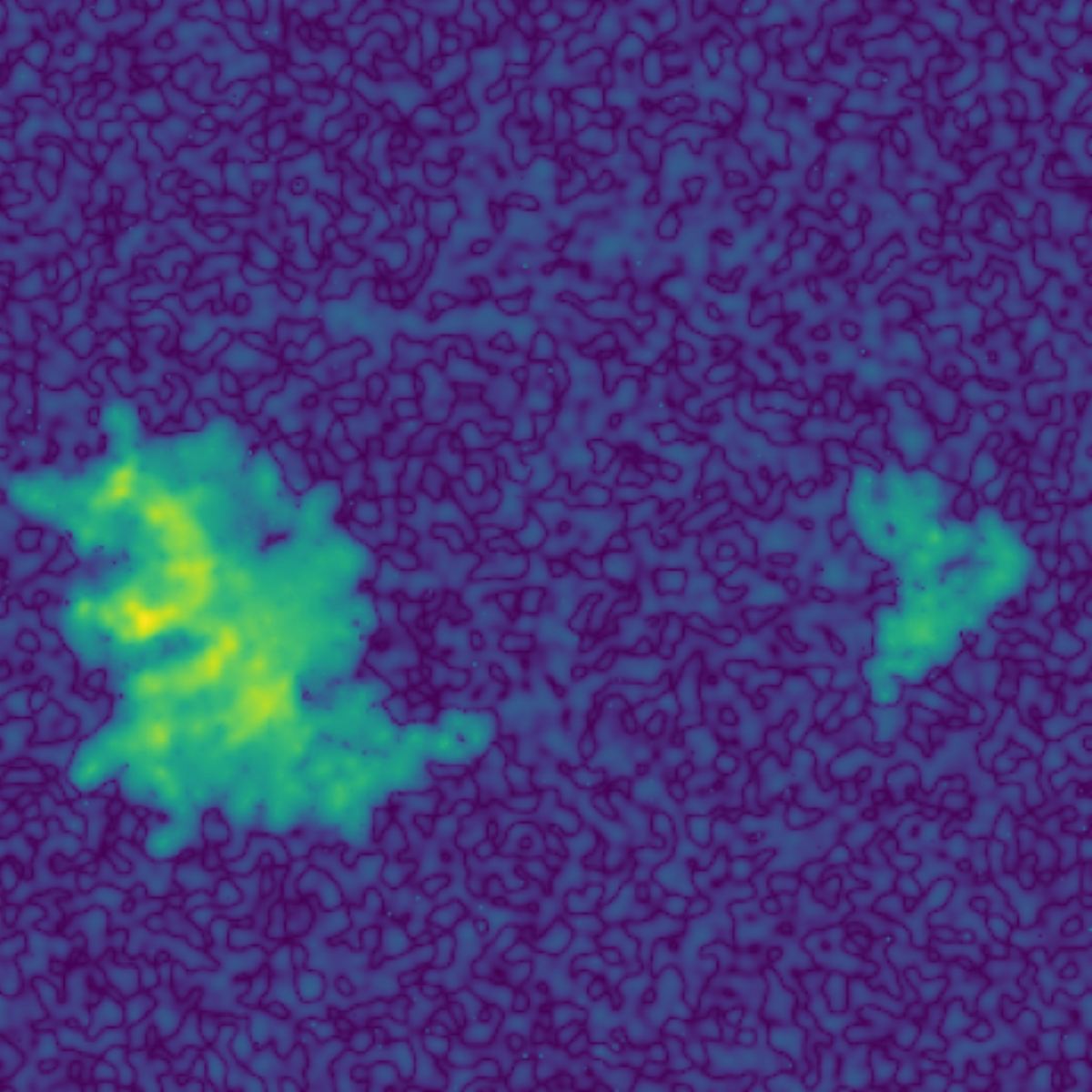}} \hfill
			\subfloat[\label{fig:source_hals_3_unbal}HALS]{\includegraphics[width=0.23\textwidth]{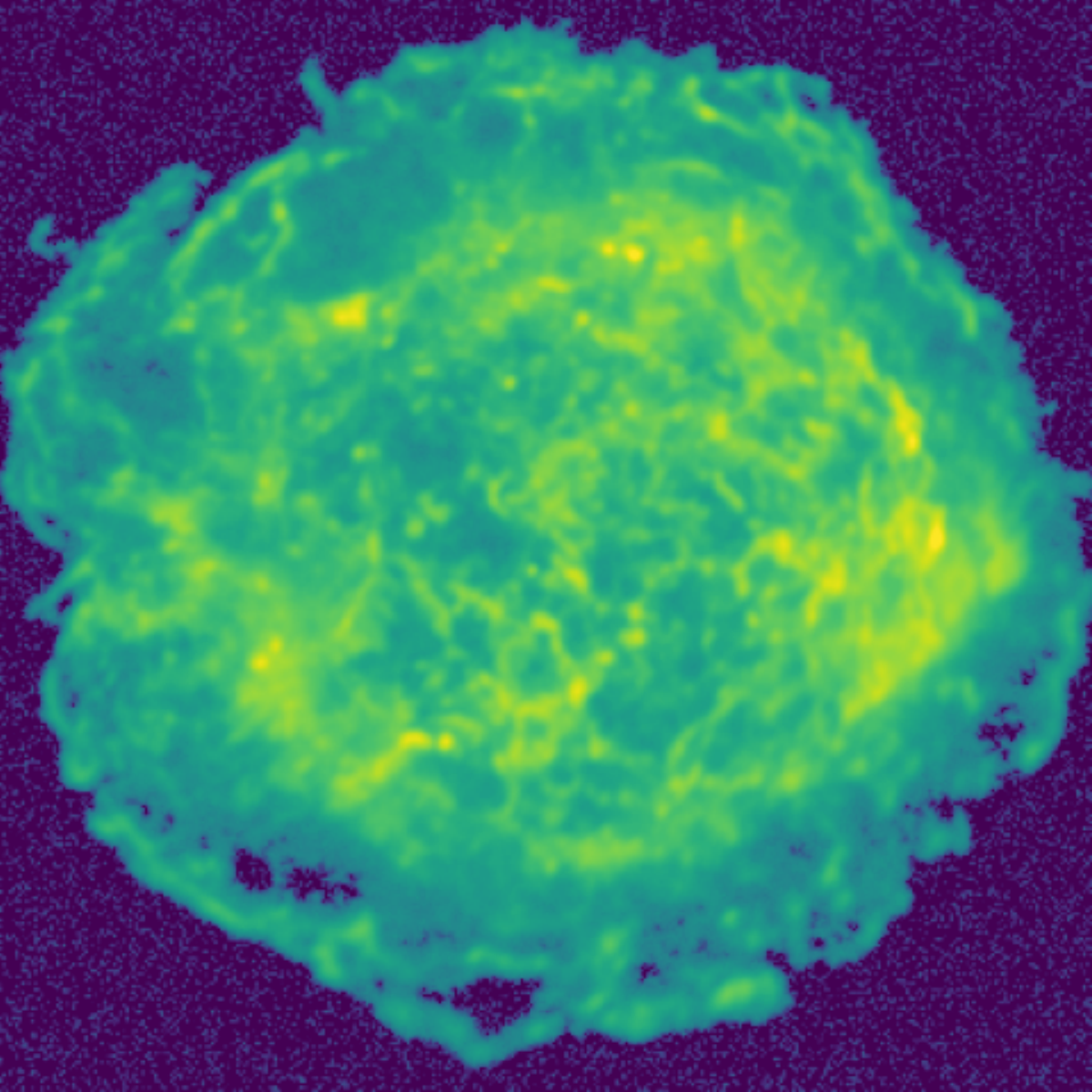}} \hfill
			\subfloat[\label{fig:source_snmf_3_unbal}SNMF]{\includegraphics[width=0.23\textwidth]{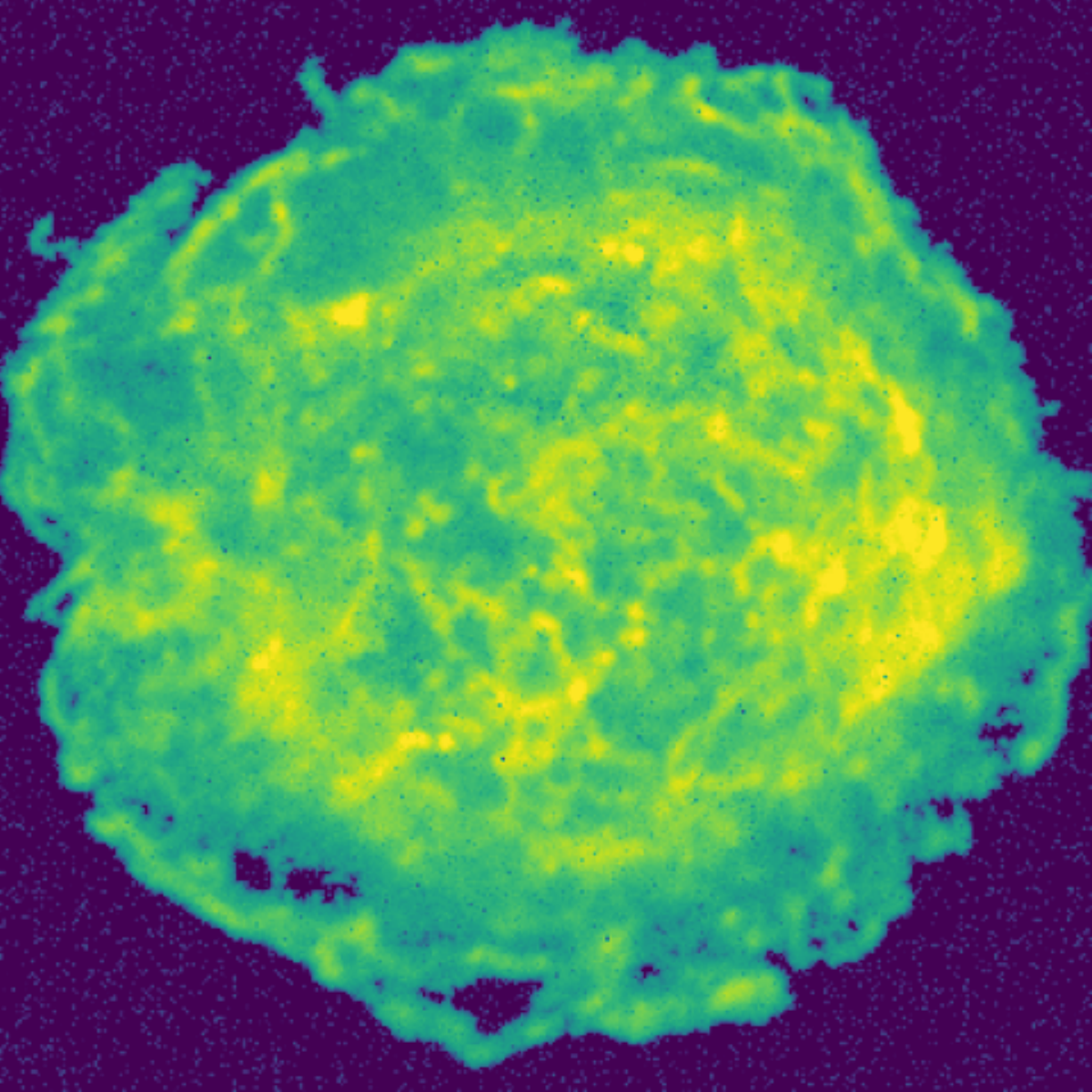}} \hfill	\vspace{0.01\textwidth}	\subfloat{\includegraphics[width=0.045\textwidth]{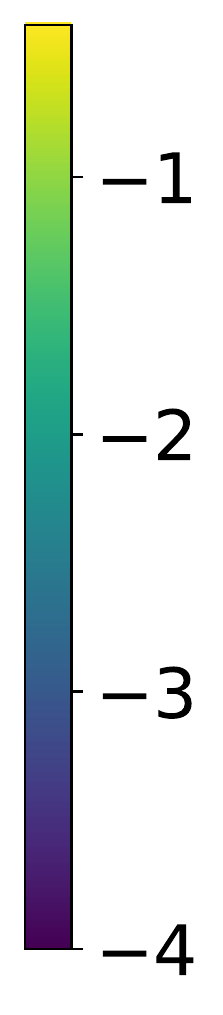}}
			\caption{Example of the estimated Gaussian II source, with $\textit{SNR}=40~\text{dB}$, $\delta=20$ and $k=0.01$ (\textit{i.e.}, the synchrotron source is a hundred times brighter). Top: estimations (logarithmic scale), bottom: absolute error (logarithmic scale). The figures on a same row share the same color scale.}
			\label{fig:res_src_3_unbal}
		\end{center}
	\end{figure}

	\paragraph{Unbalance of the sources} The reported source metrics concern only the thermal and Gaussian components, since we want to assess the impact on the hidden sources.
	Again, sGMCA allows to recover more precisely the spectra, with a considerable gain in SAD ranging from 8 dB to 12 dB compared to GMCA, and up to 5 dB for the SDR.\\
	When the sources are unbalanced, GMCA estimates precisely the synchrotron spectrum but very poorly the three other spectra (they are contaminated by the synchrotron component). The accurate estimation of the synchrotron spectrum makes it possible to separate the synchrotron source from the data in the least-square update of the sources, which allows for the retrieval of the hidden sources, hence the acceptable SDR (see examples of estimated Gaussian II source when the synchrotron source is a hundred times brighter in Fig.~\ref{fig:res_src_3_unbal}).\\
	On the contrary, the nearest-neighbor benchmark algorithm performs particularly poorly; it generally fails to identify the spectra of the three hidden sources, because at the first iteration (which we recall to be the output of GMCA) the nearest neighbors of all four spectra are synchrotron spectra. 
	
	\newcommand{\sizeres}{0.315}
	\begin{figure}
		\begin{center}
			\subfloat{\includegraphics[width=\sizeres\textwidth]{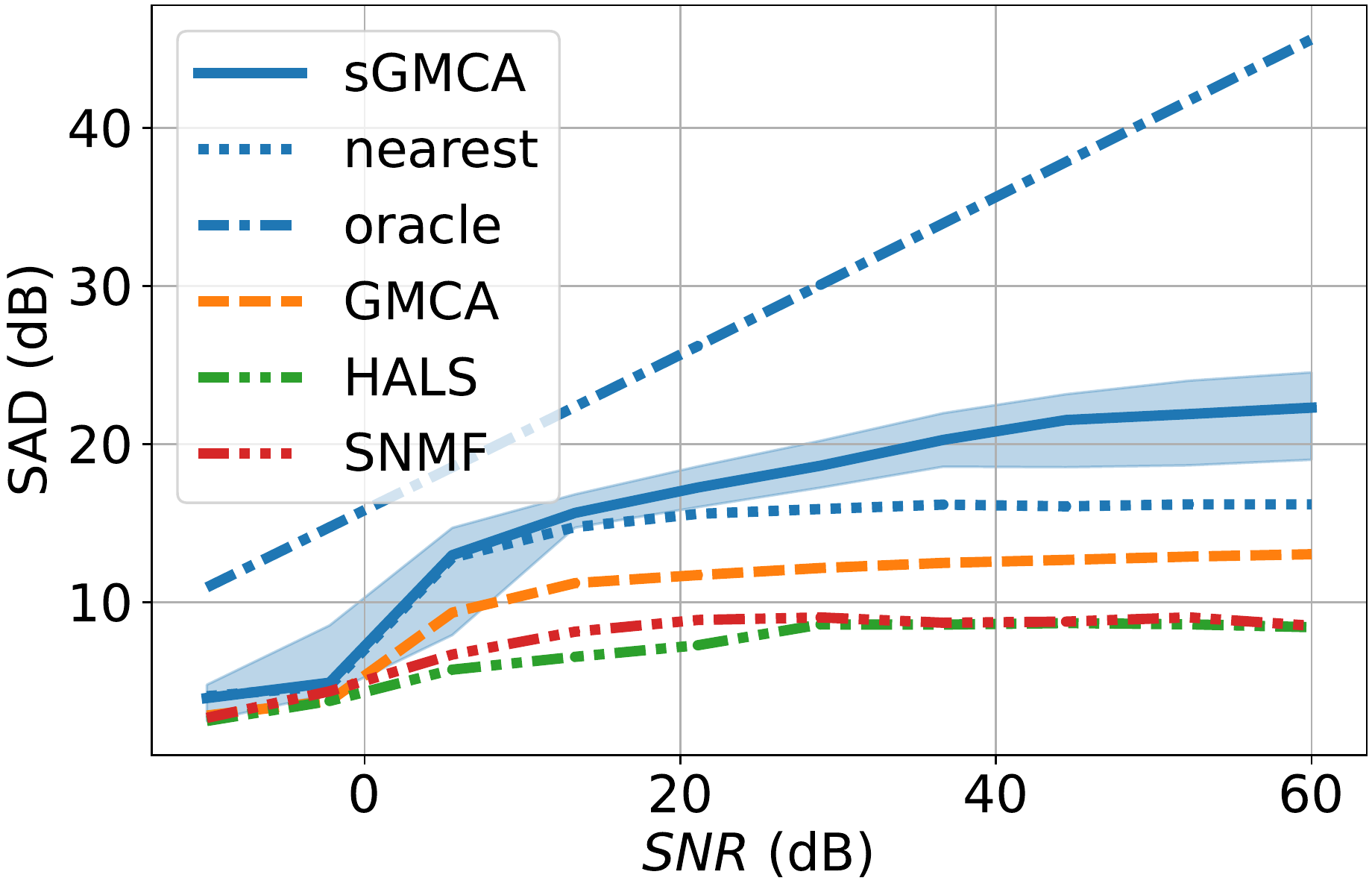}} \hfill
			\subfloat{\includegraphics[width=\sizeres\textwidth]{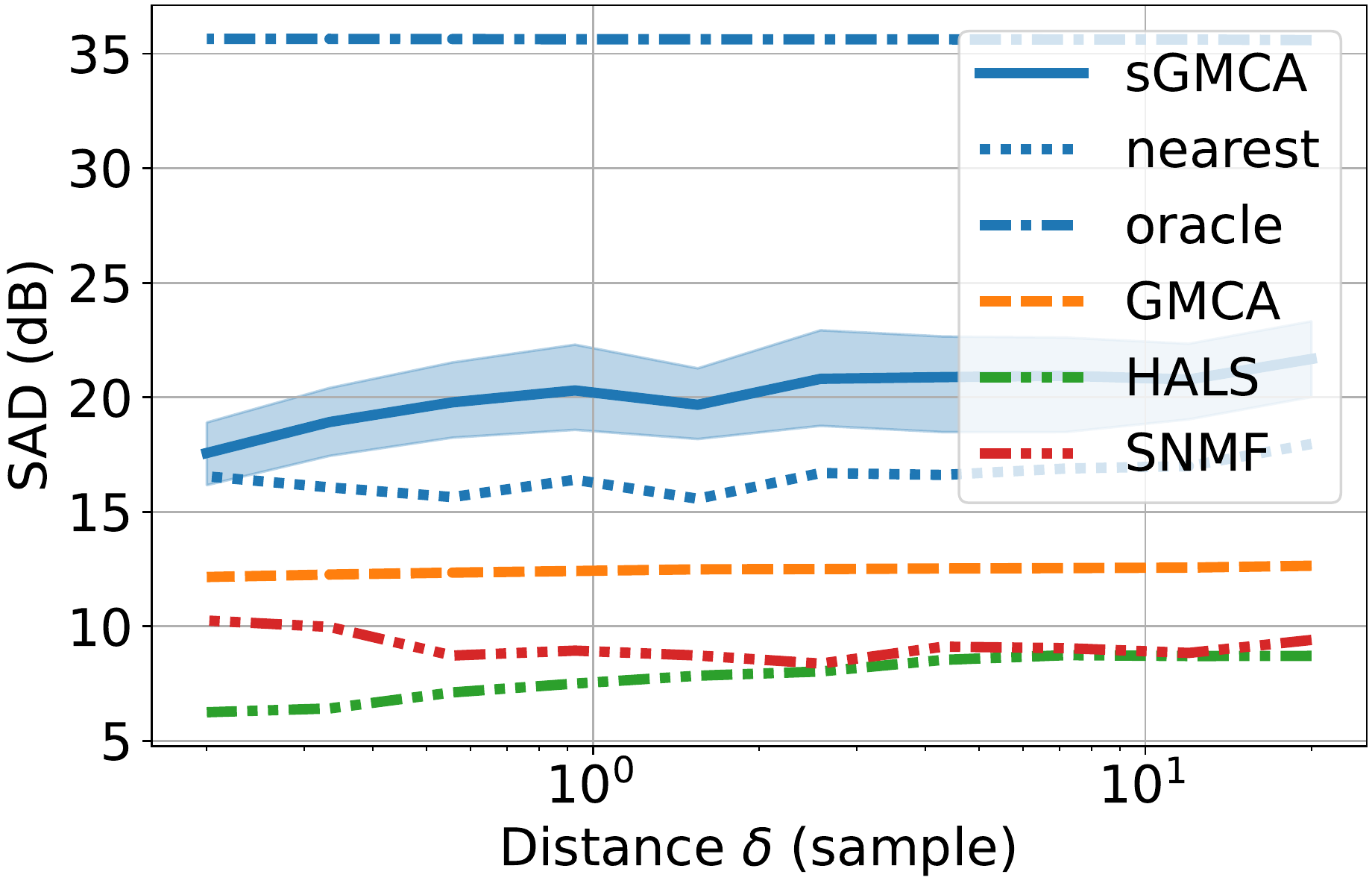}} \hfill
			\subfloat{\includegraphics[width=\sizeres\textwidth]{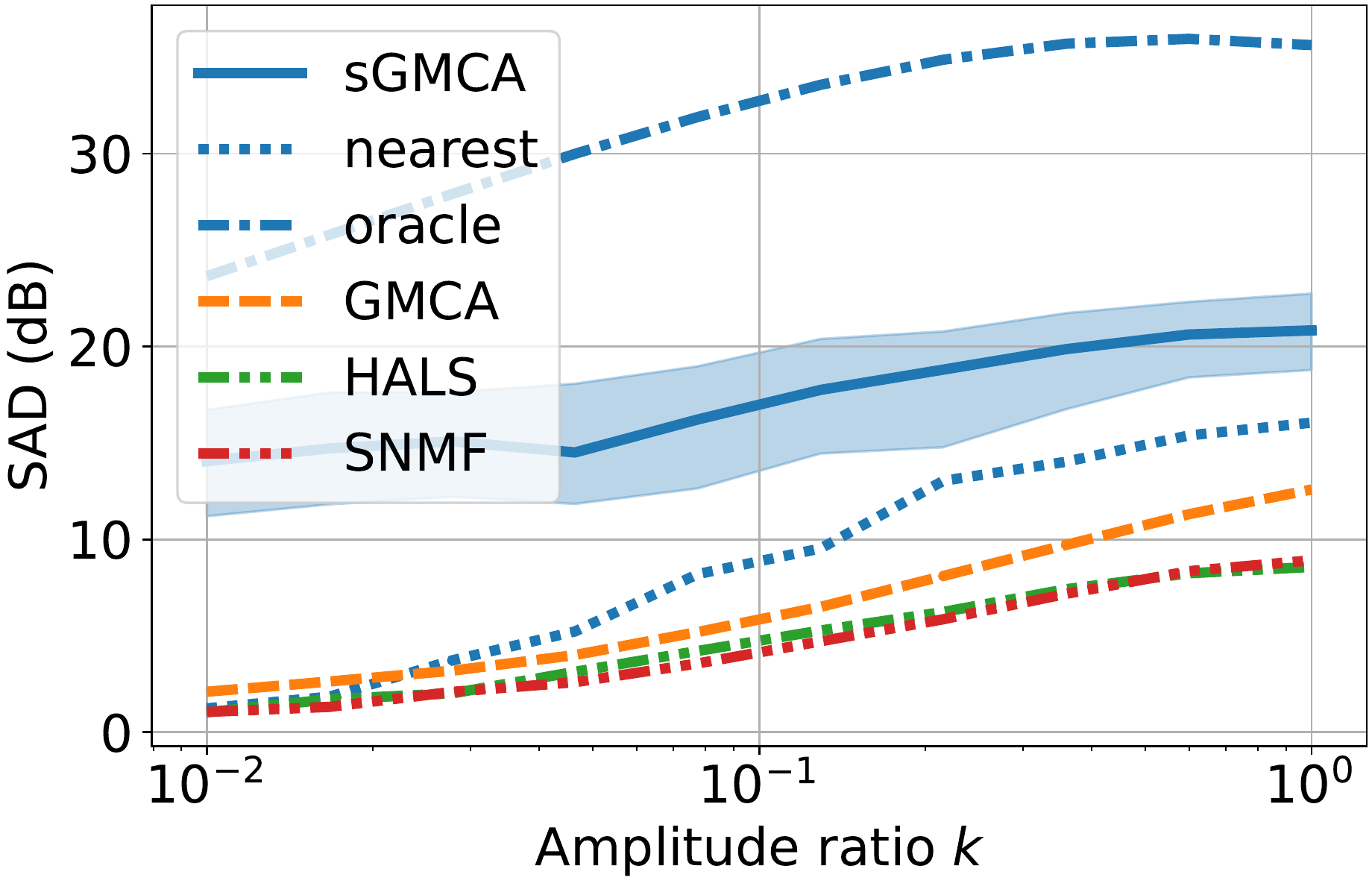}}
			\vspace{-1em}\newline
			\subfloat{\includegraphics[width=\sizeres\textwidth]{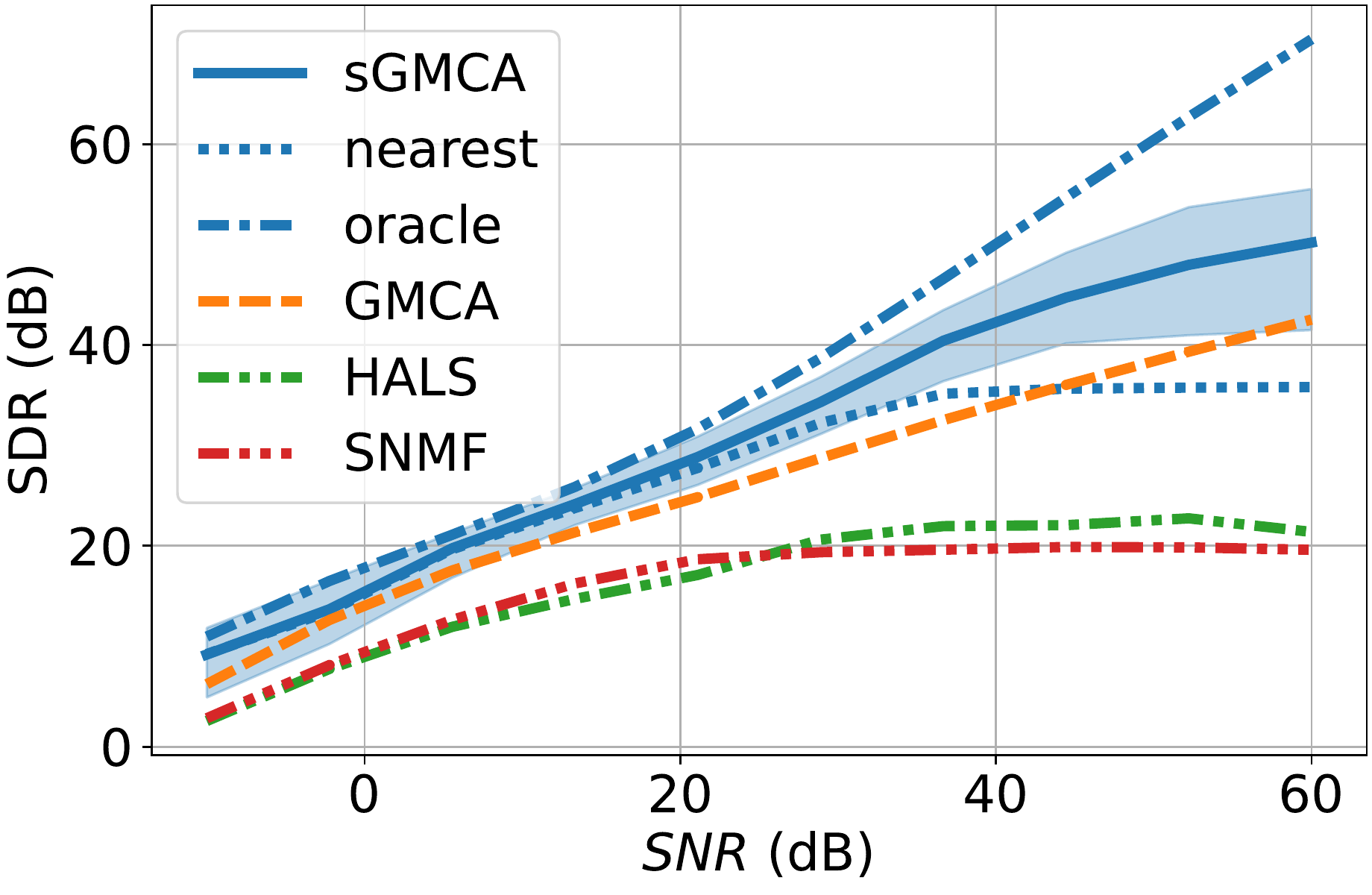}} \hfill
			\subfloat{\includegraphics[width=\sizeres\textwidth]{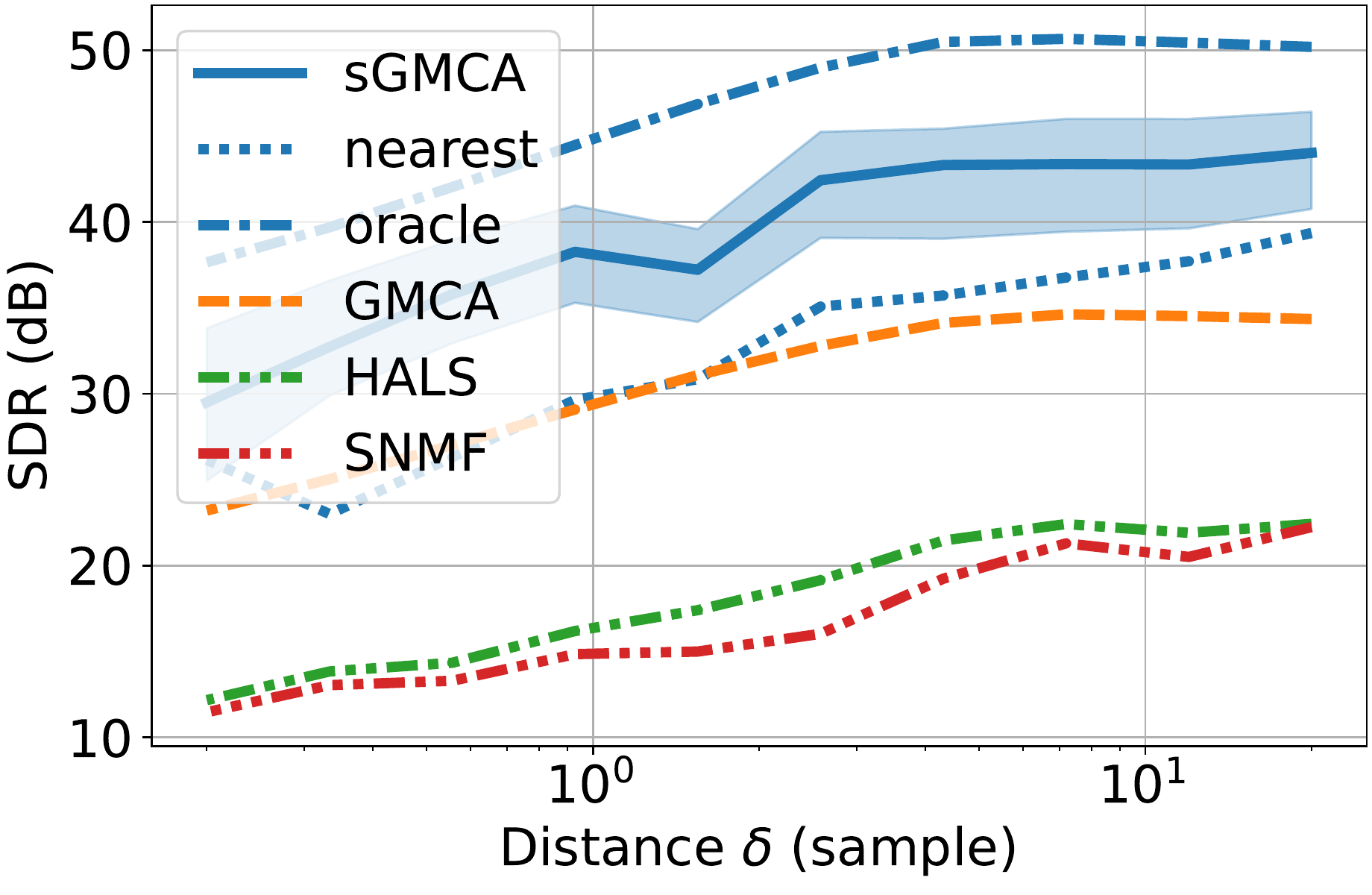}} \hfill
			\subfloat{\includegraphics[width=\sizeres\textwidth]{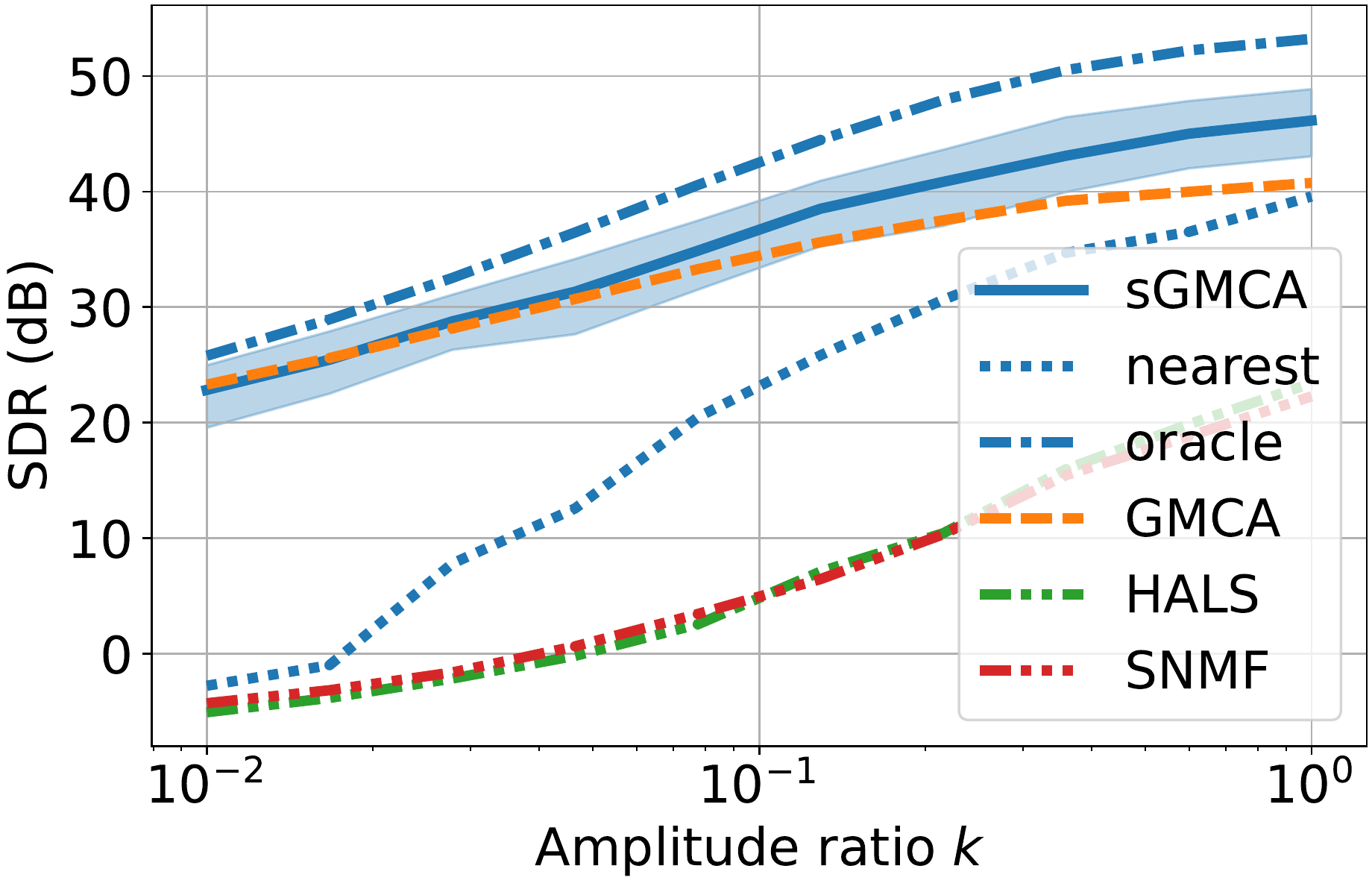}}
			\vspace{-1em}\newline
			\subfloat{\includegraphics[width=\sizeres\textwidth]{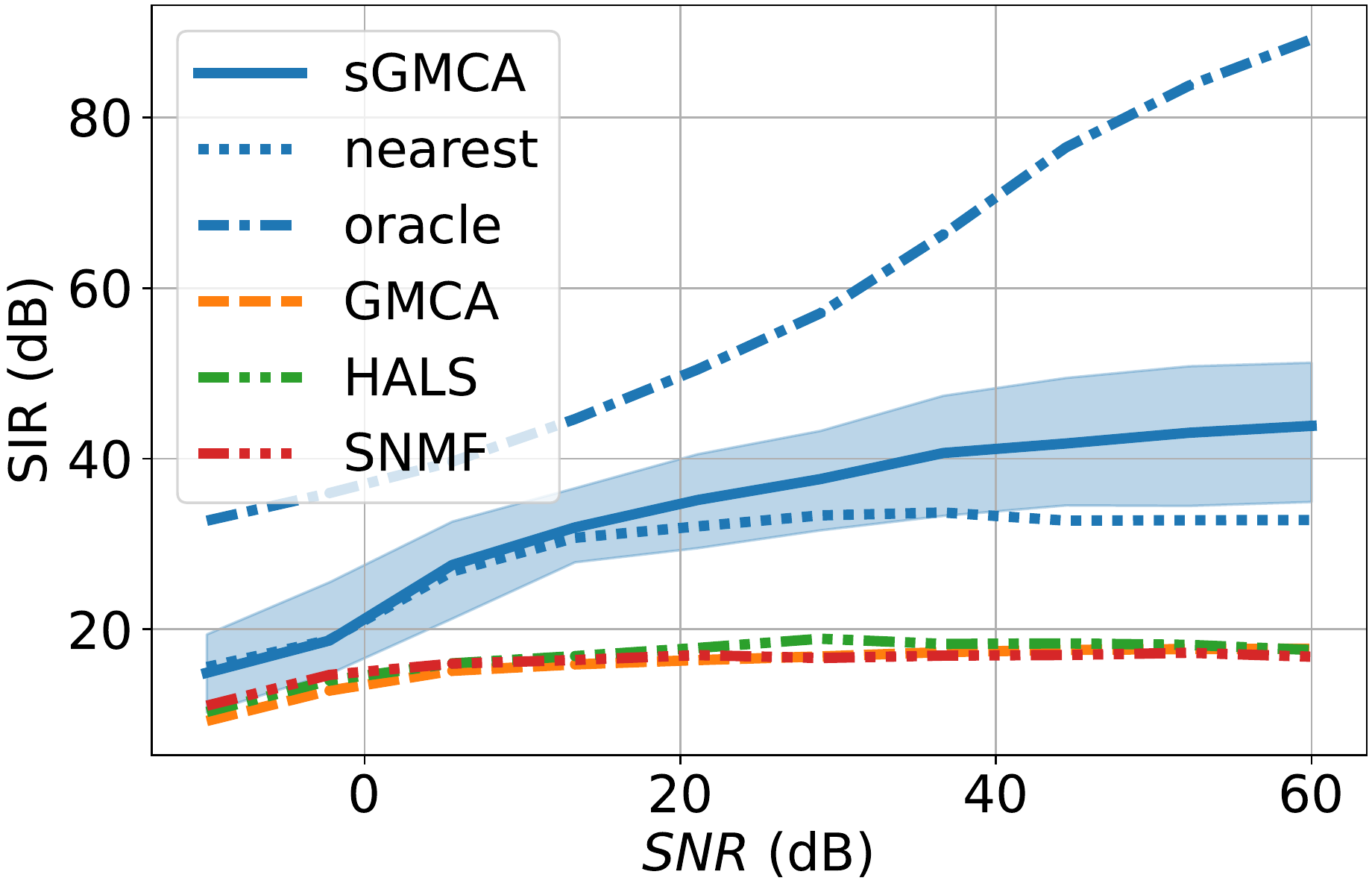}} \hfill
			\subfloat{\includegraphics[width=\sizeres\textwidth]{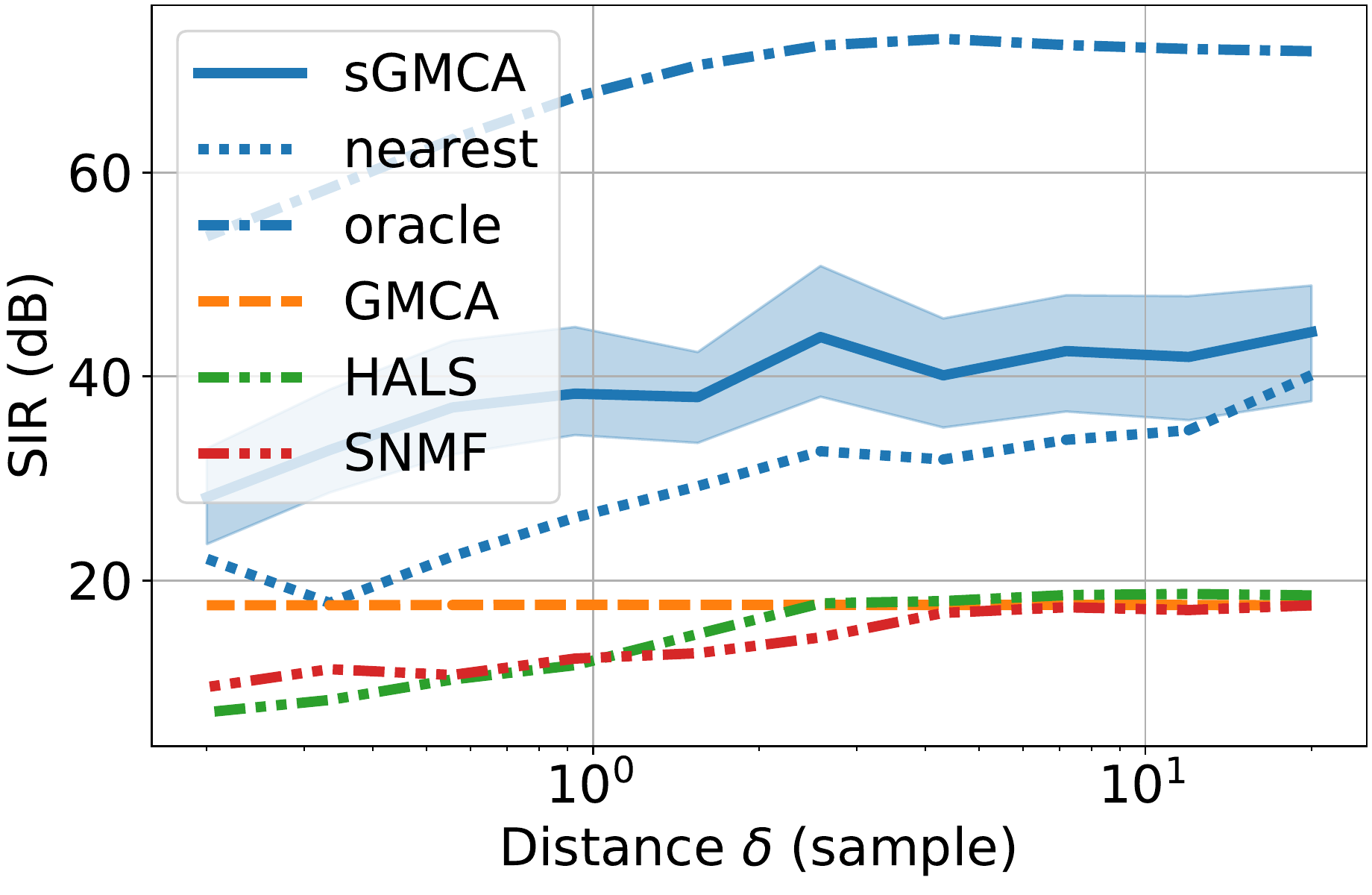}} \hfill
			\subfloat{\includegraphics[width=\sizeres\textwidth]{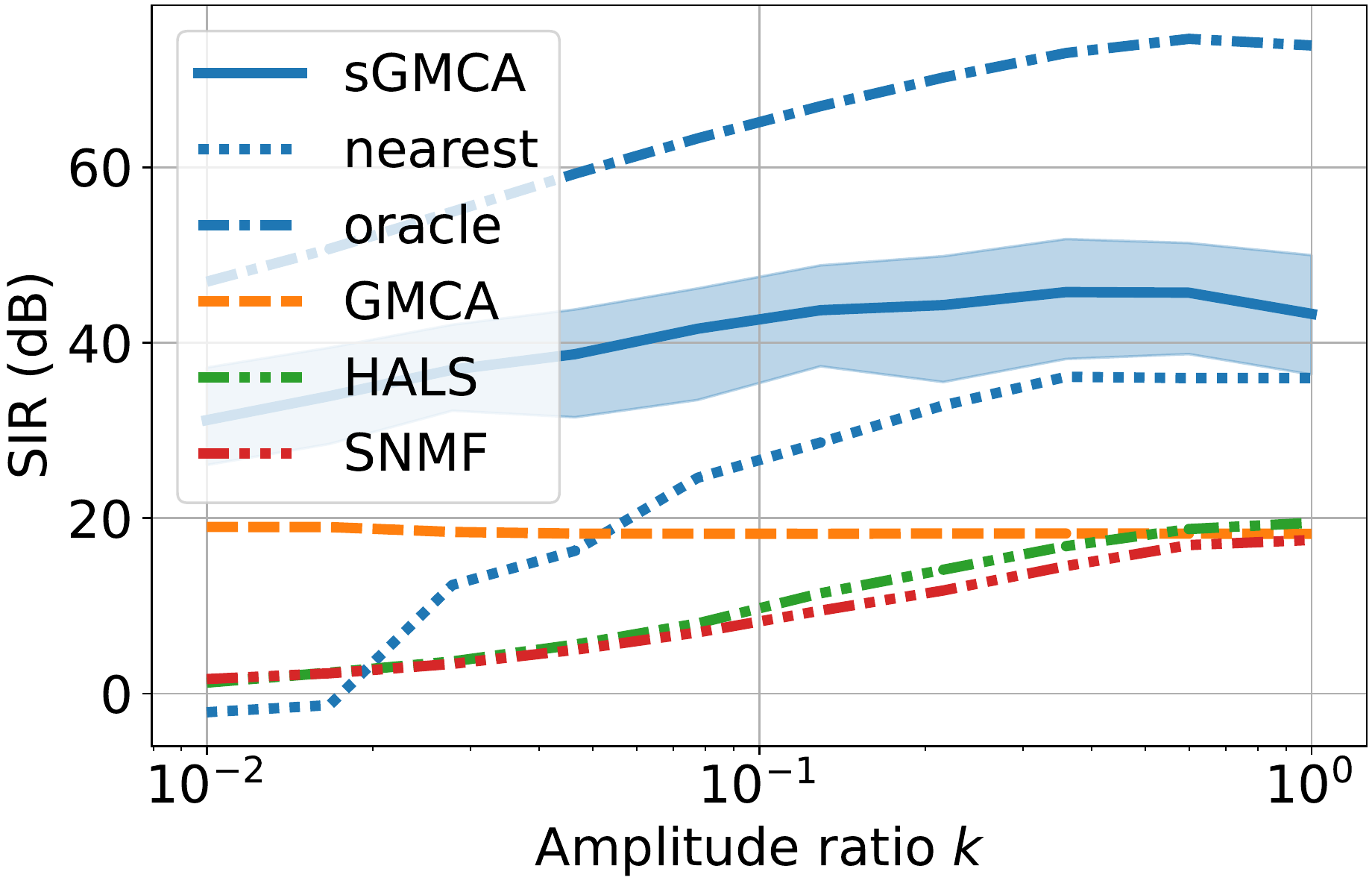}}
			\vspace{-1em}\newline
			\subfloat{\includegraphics[width=\sizeres\textwidth]{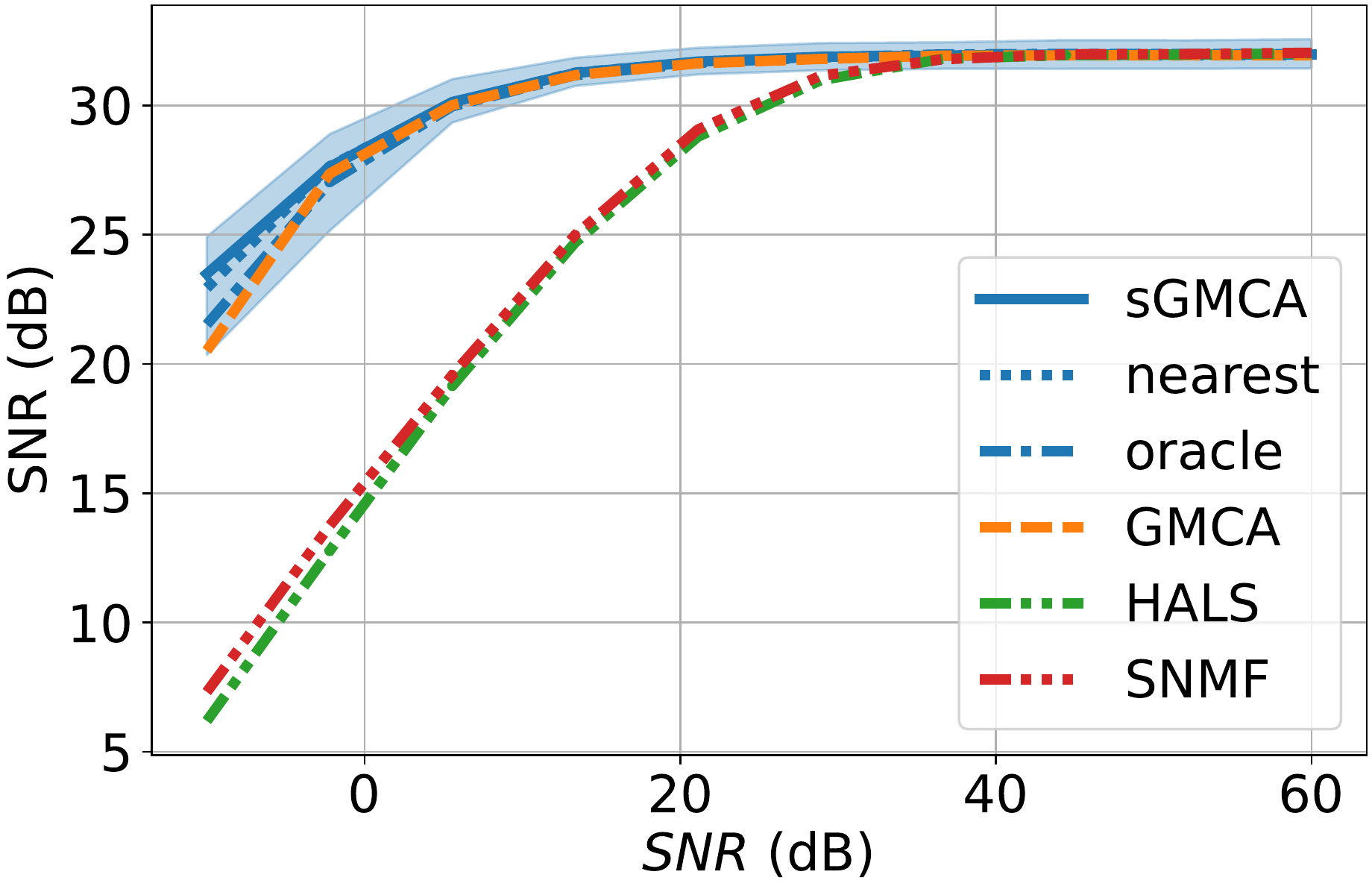}} \hfill
			\subfloat{\includegraphics[width=\sizeres\textwidth]{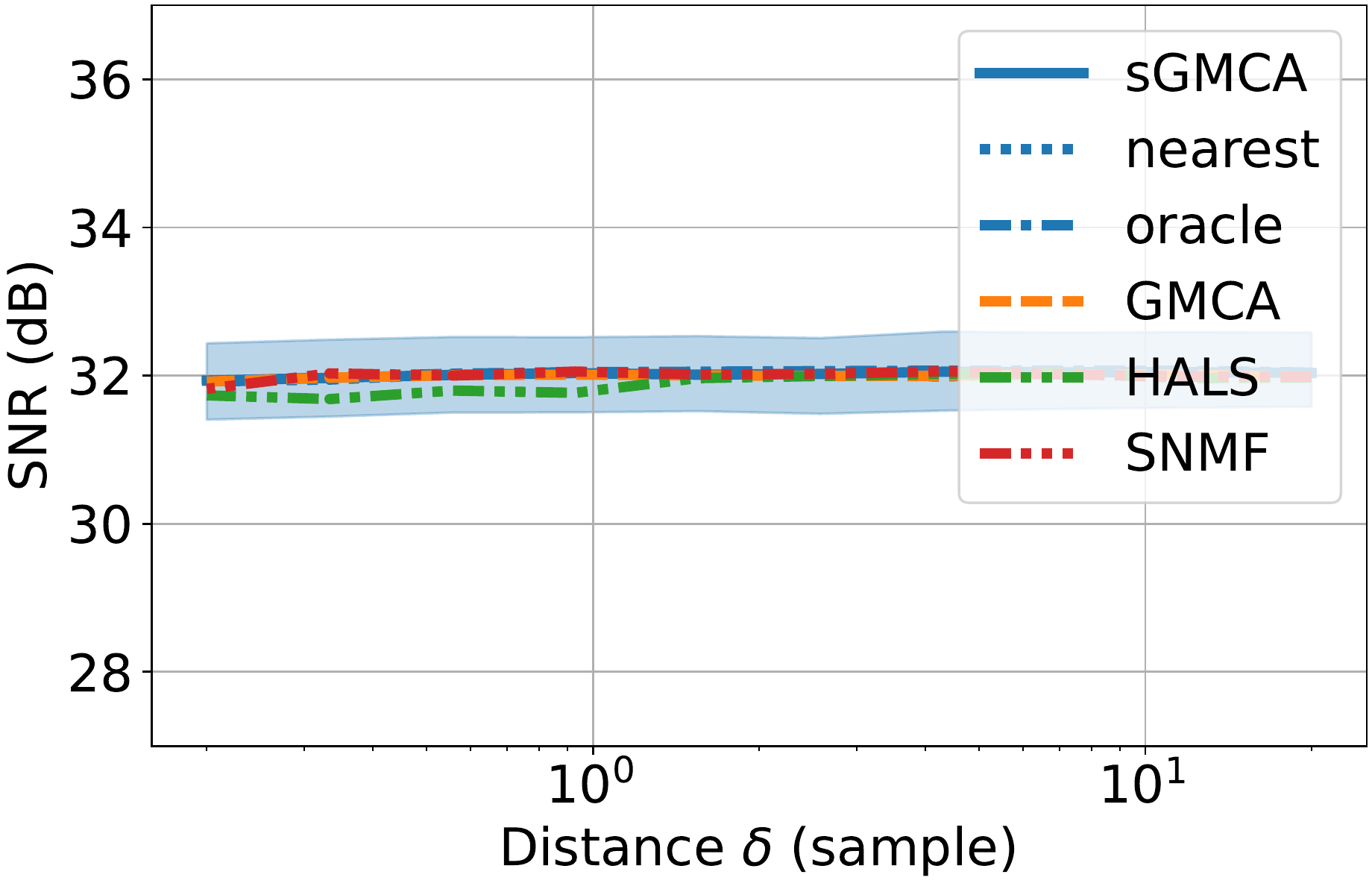}} \hfill
			\subfloat{\includegraphics[width=\sizeres\textwidth]{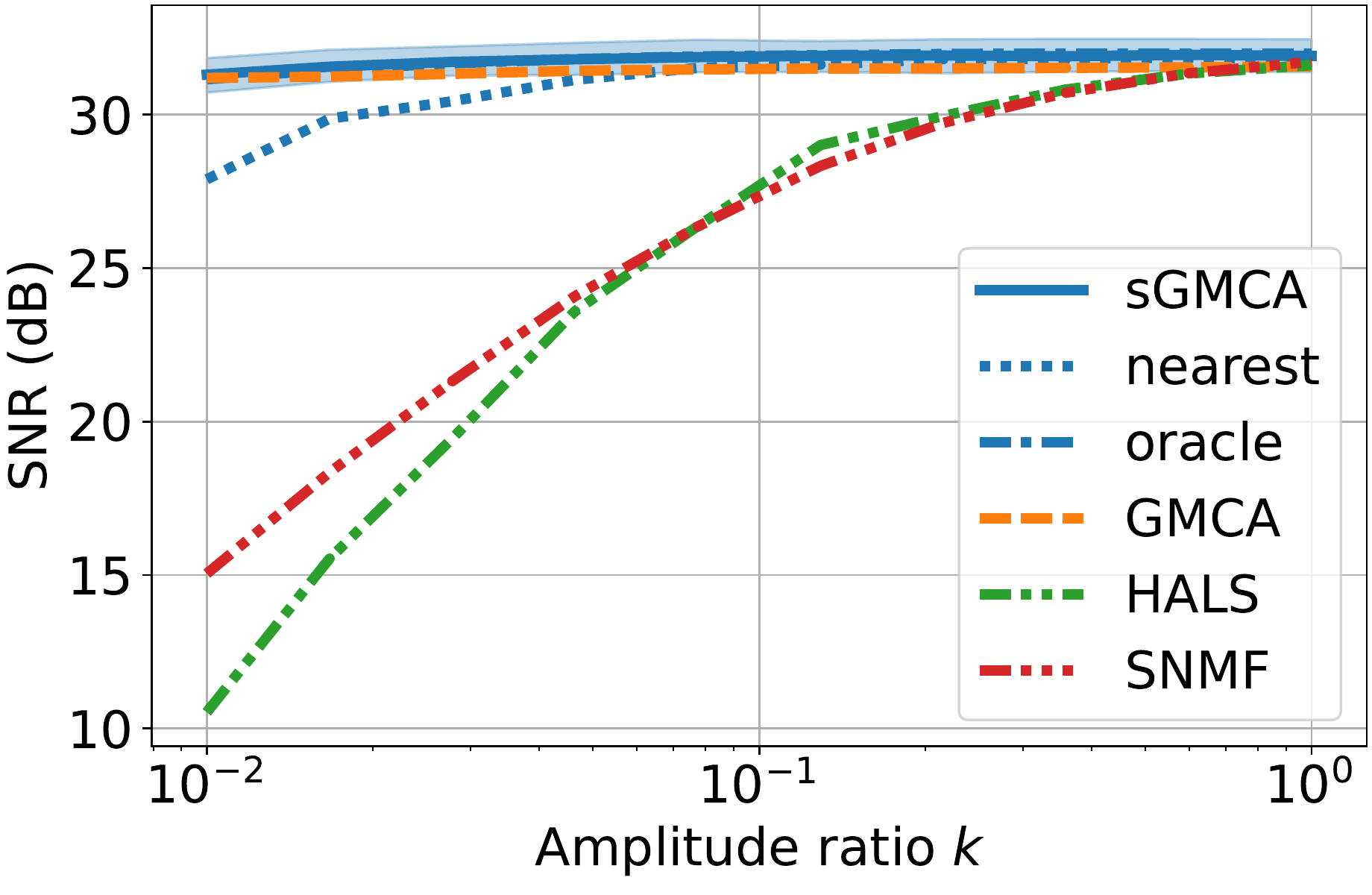}}
			\vspace{-1em}\newline
			\addtocounter{subfigure}{-12}
			\subfloat[\label{fig:snr}]{\includegraphics[width=\sizeres\textwidth]{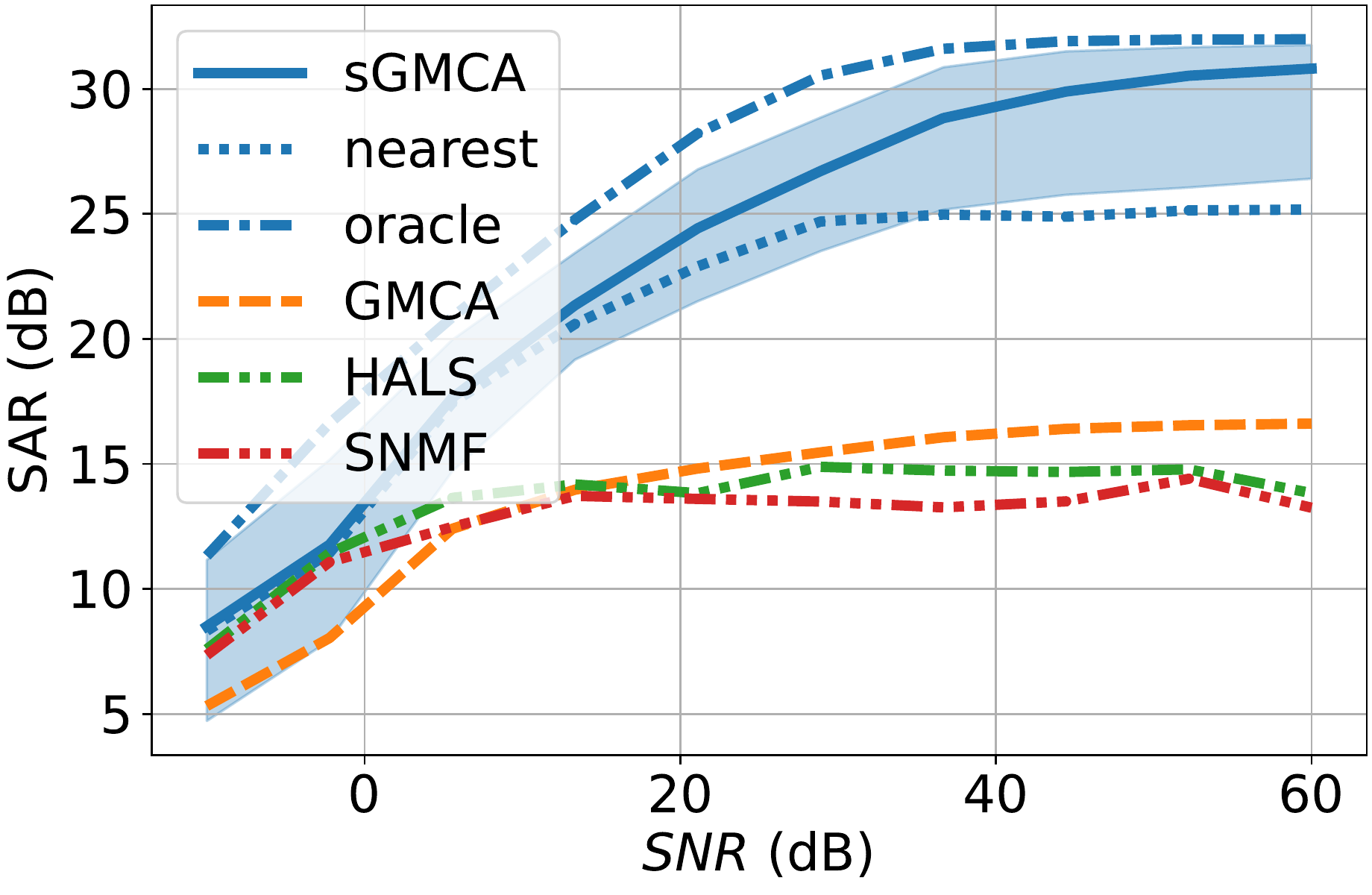}} \hfill
			\subfloat[\label{fig:dist}]{\includegraphics[width=\sizeres\textwidth]{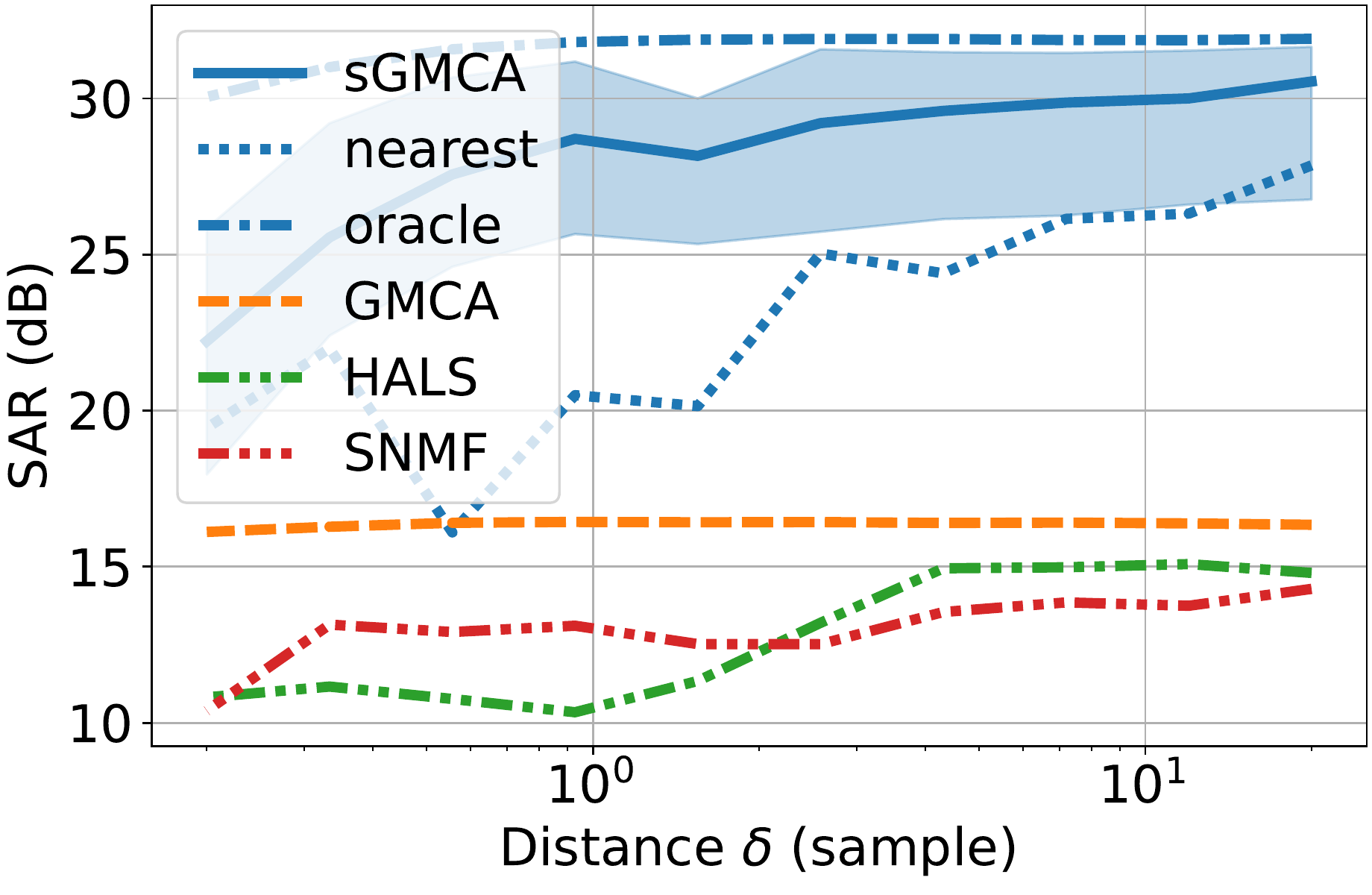}} \hfill
			\subfloat[\label{fig:amp}]{\includegraphics[width=\sizeres\textwidth]{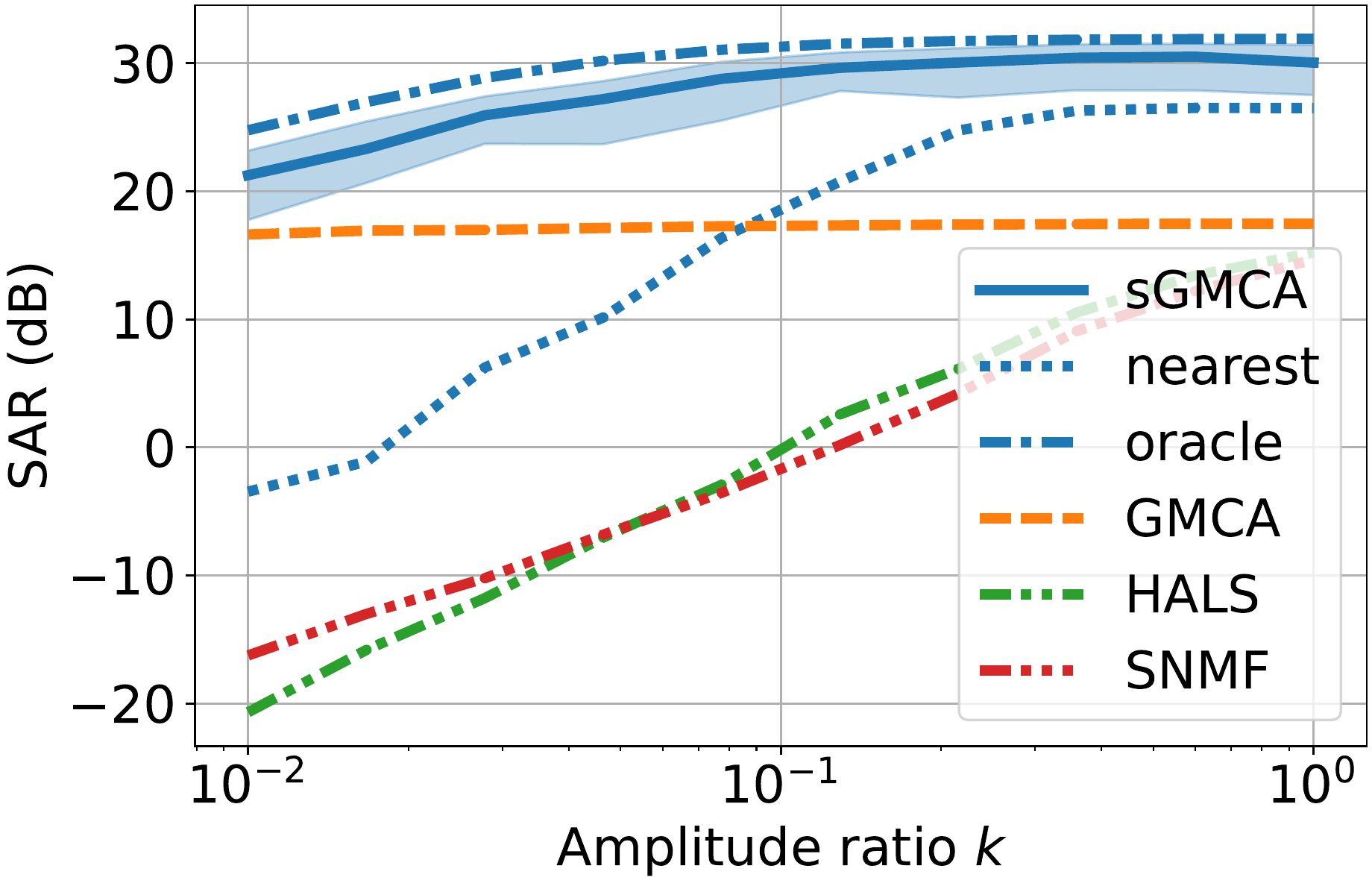}}
			\caption{Median metrics over 100 realizations, and first and third quartile for sGMCA, when varying the \protect\subref{fig:snr} noise level, \protect\subref{fig:dist} collinearity of the Gaussian line spectra, \protect\subref{fig:amp} unbalance of the sources. The top row (SAD) concerns only the spectra, the other rows (SDR, SIR, SNR and SAR) concern only the sources. \protect\subref{fig:amp} The reported SDR, SIR, SNR and SAR are calculated over the thermal and Gaussian sources only.}
			\label{fig:exp}
		\end{center}
	\end{figure}
	
	\subsubsection{Impact of the generative modeling}
	In the previous experiments, all four components were constrained by generative models. In this subsection, we focus on the impact of having one or more components fully unknown. To that end, Monte Carlo experiments are performed where the IAE models of the synchrotron, thermal and/or Gaussian spectra are removed, both in the balanced and an unbalanced case. The results are reported in Table \ref{tab:semiblind}.
	
	In the balanced case, the performance metrics are particularly sensitive to the presence or not of the thermal model. The thermal source strongly correlates with the two Gaussian line sources, it is therefore not surprising that constraining the thermal spectrum notably improves the unmixing. 
	
	In the unbalanced case, where the synchrotron source has a hundred times larger norm than the other sources, it is particularly advantageous to include the synchrotron model in addition to the thermal model. This reduces the leakage of the synchrotron component into the estimates of the other hidden components. 
	
	\newcommand\widthtable{5em}
	\begin{table}
		\begin{small}
			\begin{center}
				\begin{tabular}{@{}rp{3.25em}p{\widthtable}p{\widthtable}p{\widthtable}p{\widthtable}p{\widthtable}@{}}
					\toprule
					& & \multicolumn{5}{c}{Components modeled with a generative model}        \\ \cmidrule(l){3-7} 
					$k$ & Metric (dB) & All & Therm.~\& Gauss. & Sync.~\& Gauss. & Gauss. & None \\ \midrule
					
					& SAD & $21.67\pm1.33$  &  $20.71\pm1.71$   &  $15.59\pm1.16$   &  $15.32\pm0.77$&  $12.54\pm0.23$  \\
					&SDR &  $43.93\pm2.40$ &  $42.41\pm2.58$   &  $39.06\pm1.72$   &  $37.74\pm1.56$&  $34.24\pm0.54$   \\
					1 & SIR &  $44.16\pm5.63$ &  $42.16\pm5.10$   &  $33.35\pm3.17$   &  $33.33\pm2.74$&  $17.49\pm0.20$ \\
					& SNR &  $31.98\pm0.46$ &  $31.97\pm0.49$   &  $31.96\pm0.50$   &  $31.91\pm0.48$&  $31.91\pm0.45$ \\
					& SAR &  $30.48\pm0.91$ &  $29.42\pm1.36$   &  $22.81\pm0.49$   &  $22.57\pm0.47$&  $16.30\pm0.20$ \\ \midrule
					
					& SAD &  $17.73\pm2.32$ & $12.27\pm5.04$ & $8.83\pm1.26$ & $8.27\pm0.91$ & $5.78\pm0.63$ \\
					& SDR & $36.64\pm2.19$ &  $34.39\pm3.87$   &  $36.96\pm2.42$   &  $37.02\pm1.84$ &  $34.56\pm0.80$ \\
					0.01 & SIR &$42.54\pm5.06$ &  $35.97\pm8.89$   &  $36.70\pm4.38$   &  $37.05\pm5.03$ &  $18.22\pm0.22$ \\
					& SNR &  $31.87\pm0.49$ &  $31.68\pm0.61$   &  $31.67\pm0.51$   &  $31.62\pm0.56$ &  $31.47\pm0.48$ \\
					& SAR &  $29.24\pm1.07$ &  $27.00\pm2.49$   &  $25.05\pm0.99$   &  $25.00\pm0.47$ &  $17.31\pm0.18$ \\
					\bottomrule
				\end{tabular}
			\end{center}
		\end{small}
		\caption{Median metrics over 100 realizations. The spread is the maximum difference between the median and the first or third quartile. $k$ is recalled to be the ratio between the norm of the thermal or a Gaussian source (they have the same norm) and the norm of the synchrotron source. For $k=0.01$, the source metrics are calculated over the thermal and Gaussian sources.}
		\label{tab:semiblind}
	\end{table}

	\section{Conclusion}
	
	We introduce a novel source separation approach to tackle physical hyperspectral data. Compared to standard blind source separation methods, the objective is twofold: to better discriminate between sources and to ensure the provision of physically relevant information. For this purpose, we make use of learned priors, which are based on generative models, on the spectra of the sought-after components in a standard variational framework. Extensive numerical experiments on realistic astrophysical data show that the introduced regularization efficiently rejects inter-source leakages, thus improving significantly the estimations of both the sources and the spectra, including in challenging settings.
	
	We choose to specifically illustrate the application of the proposed sGMCA algorithm in the context of hyperspectral imaging. If remote sensing is a natural application of such kind of methods, it can be applied to a large number of matrix factorization problems with manifold-valued/manifold-constrained factors. One can think of biomedical signal processing ({\it e.g.}~LC-MS data analysis) or EEG and MEG data analysis in neurosciences. In the former, the temporal response of the device (\textit{e.g.}~retention time) naturally belongs to an unknown manifold. In the latter, the sGMCA algorithm could be applied to extract certain brain activity patterns from MEG data.
	
	\section*{Acknowledgment}
	This work is supported by the European Community through the grant LENA (ERC StG - contract no.~678282).
	
	\appendix
	\section{Open-source code} \label{app:code}
	The code is open source and can be found online at  \href{https://github.com/RCarloniGertosio/sGMCA}{\texttt{github.com/RCarloniGertosio/sGMCA}} on version 3 of the LGPL.
	
	\section{The interpolatory autoencoder} \label{app:iae}
	\begin{wrapfigure}{r}{0.33\textwidth}
		\vspace{-3cm}
		\begin{flushright}
			\footnotesize
			\begin{tikzpicture}[node distance=.5cm and .5cm]  
				
				\def\innersep{.5em}
				
				\tikzstyle{mainblock}=[rectangle,draw,fill=blue!10,inner sep=\innersep,align=center]
				\tikzstyle{secondaryblock}=[rectangle,draw,fill=blue!10,inner sep=\innersep,align=center]
				\tikzstyle{mainarrow}=[->,>=stealth,thick,rounded corners=4pt,align=center]
				\tikzstyle{secondaryarrow}=[->,>=stealth,thick,rounded corners=4pt,dashed]
				\tikzstyle{mainline}=[thick,rounded corners=4pt,align=center]
				\tikzstyle{secondaryline}=[thick,rounded corners=4pt,dashed]
				\tikzstyle{empty} = [inner sep=\innersep,align=center]
				
				
				\node[empty] (ap) at (0, 0) {Anchor points\\ $\left\{\bm{\alpha}^{(n)}\right\}_{n}$};
				\node[mainblock, below=of ap] (encoder) {Encoder $\phi$};
				\node[empty, below=of encoder] (encodedap) {Encoded\\anchor points\\$\left\{\phi\left(\bm{\alpha}^{(n)}\right)\right\}_{n}$};
				\node[mainblock, below=of encodedap] (interpolator) {Interpolator};
				\node[empty, below=of interpolator] (barycenter) {Barycenter\\$\sum_{n} \lambda^{(n)} \phi\left(\bm{\alpha}^{(n)}\right)$};
				\node[mainblock, below=of barycenter] (decoder) {Decoder $\psi$};
				\node[empty, below=of decoder] (output) {Generated sample\\$\hat{\mathbf{x}}\coloneqq\psi\left(\sum_{n} \lambda^{(n)} \phi\left(\bm{\alpha}^{(n)}\right)\right)$};
				
				\node[secondaryblock, left=of encoder] (encoder2) {Encoder $\phi$};
				\node[empty, above=of encoder2] (sample) {Sample \\$\mathbf{x}$};
				\node[empty, below=of encoder2] (encodedsample) {Encoded\\sample\\$\phi\left(\mathbf{x}\right)$};
				
				\draw[mainarrow] (ap) -- (encoder);
				\draw[mainline] (encoder) -- (encodedap);
				\draw[mainarrow] (encodedap) -- (interpolator);
				\draw[mainline] (interpolator) -- (barycenter);
				\draw[mainarrow] (barycenter) -- (decoder);
				\draw[mainarrow] (decoder) -- (output);
				
				\draw[secondaryarrow] (sample) -- (encoder2);
				\draw[secondaryline] (encoder2) -- (encodedsample);
				\draw[secondaryarrow] (encodedsample.south) |- (interpolator.west);
				
				\draw[dashed,thick,color=blue!30] (encoder2) -- (encoder);
				
			\end{tikzpicture}
			\vspace{-.5cm}
			\caption{Diagram of the interpolatory autoencoder}
			\label{fig:iaediagram}
			\vspace{-1cm}
		\end{flushright}
	\end{wrapfigure}
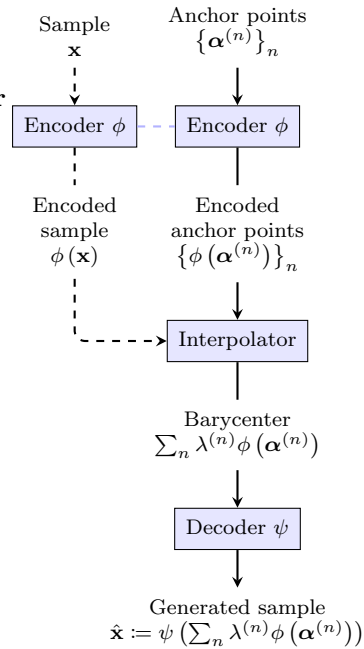
	
	The interpolatory autoencoder (IAE) is a particular neural-network-based AE that we introduced in \cite{BobinIAE2021}. Compared to standard autoencoders, such as VAEs, the IAE is showed to require fewer training samples and the inducted projection on manifold demonstrates a remarkable robustness to contaminants.
	
	\paragraph{Principle} Rather than learning the underlying manifold structure of a dataset, the gist of the IAE is to learn to travel on it by interpolation between so-called "anchor points", which are samples that belong to the manifold. To this effect, a neural-network-based autoencoder that builds maps between the sample space and a code space is trained so that the code of each sample can be expressed as a barycenter or affine combination of the codes of the chosen anchor points; in doing so, the manifold tends to be linearized in the code space.\\
	The architecture of the IAE is represented in Figure \ref{fig:iaediagram}. Let $\{\bm{\alpha}^{(n)}\in\mathcal{M}\}_{n\in[1\dots N]}$ be a set of $N$ anchor points of a manifold $\mathcal{M}$ that we seek to model. Let $\phi: \mathbb{R}^J \rightarrow \mathbb{R}^{J'}$ and $\psi: \mathbb{R}^{J'} \rightarrow \mathbb{R}^J$ be a forward encoder and a backward decoder, respectively, parameterized by a given number of fully connected layers with skip connections. The latter paragraph can be resumed by:
	\begin{equation}
		\forall \mathbf{x} \in \mathcal{M}, \exists \{\lambda^{(n)}\}_{n\in[1\dots N]} \in\mathbb{R}^N, \phi(\mathbf{x}) = \sum_{n=1}^N \lambda^{(n)} \phi(\bm{\alpha}^{(n)}), \sum_{n=1}^N \lambda^{(n)}=1.
	\end{equation}
	The latent parameters $\{\lambda^{(n)}\}_{n\in[1\dots N]}$ are the barycentric weights; they are enforced to sum to one so that the linear combination is indeed a barycenter or affine combination \cite{Grunbaum03} of the encoded anchor points.\\
	The encoder and decoder are optimized so as to minimize the reconstruction error between a training dataset $\{\mathbf{x}^{(t)}\}_{t\in[1\dots T]}$ and the barycenters in the code space decoded in the sample space, \textit{i.e.}: 
	\begin{equation}
		\argmin_{\phi, \psi} \sum_{t=1}^T \left\lVert\mathbf{x}^{(t)} - \psi\left(\sum_{n=1}^N \lambda^{(n,t)} \phi\left(\bm{\alpha}^{(n)}\right)\right)\right\rVert^2_2,
	\end{equation}
	where $\{\lambda^{(n,t)}\}_{n\in [1\dots N]}$ are the barycentric weights associated to $\phi(\mathbf{x}^{(t)})$, which have an analytical expression.\\
	Once the autoencoder is learned, the manifold can be approximated as the affine span of the anchor points in the code domain decoded in the sample domain:
	\begin{equation} \label{eq:manifold_iae}
		\mathcal{M} \approx \left\{\mathbf{x}\in\mathbb{R}^J, \exists \{\lambda^{(n)}\}_{n\in[1 \dots N]} \in \mathbb{R}^N, \mathbf{x} = \psi\left(\sum_{n=1}^N \lambda^{(n)} \phi\left(\bm{\alpha}^{(n)}\right)\right), \sum_{n=1}^N \lambda^{(n)}=1\right\}.
	\end{equation}
	The IAE allows to build an efficient regularization scheme in a variational inference approach by constraining signals to belong to the resulting affine span, that is the modeled manifold. In contrast to the aforementioned data-driven methods, we show that the resulting interpolatory scheme can perform well even when training samples are scarce \cite{BobinIAE2021}. The detailed architecture and a discussion of the hyperparameters of this autoencoder are presented in \cite{BobinIAE2021}.
	
	\paragraph{Notations} In the present article, the barycentric weights are assembled in a vector: $\bm{\lambda}_n \coloneqq \lambda^{(n)}$. 
	Denoting $\bm{\Phi} \in \mathbb{R}^{J'\times N}$ the column-wise application of $\phi$ to the anchor points, \textit{i.e.}~$\bm{\Phi}_{:n}\coloneqq\phi\left(\bm{\alpha}^{(n)}\right)$, the generator is given by $g(\bm{\lambda}) \coloneqq \psi\left(\bm{\Phi\lambda}\right)$.

	\paragraph{Fast projection} The projection on manifold defined in Eq.~\eqref{eq:proj} may be time-consuming as it relies on an iterative scheme. In this regard, it is possible to benefit from the IAE's encoder to define a surrogate fast projection $\widetilde{\Pi_{\mathcal{M}}}(\mathbf{a}) \coloneqq g(\widetilde{\bm{\lambda}})$,
	with $\widetilde{\bm{\lambda}}$ the barycentric weights of the encoding of $\mathbf{a}$ in the IAE's latent domain: 
	\begin{equation}
		\begin{array}{rcl}
			\widetilde{\bm{\lambda}} &\coloneqq& \argmin\limits_{\bm{\lambda}\in\mathcal{T}} \left\lVert\phi\left(\mathbf{a}\right) - \bm{\Phi} \bm{\lambda}\right\rVert_2^2 \\
			&\approx& \Pi_{\mathcal{T}}\left({\bm{\Phi}}^+ ~ \phi\left(\mathbf{a}\right) \right),
		\end{array}
	\end{equation}
	Here the solution is approximated by a projected least squares, in which $\Pi_{\mathcal{T}}$ is further approximated by a rescaling ($\Pi_{\mathcal{T}}(\bm{\lambda}) \approx \bm{\lambda}/\sum_{n=1}^{N} \bm{\lambda}_n$). The coefficient $\mathbf{\rho}$ that accounts for the amplitude of $\mathbf{a}$ can then be estimated by minimizing  $\lVert\mathbf{a} - \rho~ \widetilde{\Pi_{\mathcal{M}}}(\mathbf{a})\rVert_2$, yielding: $\widetilde{\rho} \coloneqq \frac{\mathbf{a}^\top \widetilde{\Pi_{\mathcal{M}}}(\mathbf{a})}{\left\lVert\widetilde{\Pi_{\mathcal{M}}}(\mathbf{a})\right\rVert^2_2}$. The associated parameters $(\widetilde{\bm{\lambda}},\widetilde{\rho})$ can be used to initialize the iterative scheme of the manifold projection.
	
	\bibliographystyle{elsarticle-num}
	 \newcommand{\noop}[1]{}

\end{document}